\documentclass[10pt,letterpaper]{article}%
\usepackage[letterpaper, top=1in, bottom=1in, left=1in, right=1.1in, includefoot]{geometry}

\usepackage[normalem]{ulem}

\usepackage{latexsym,
amsthm,
color,
amssymb,
url,
bm,
cite,
amssymb,
microtype,
amsmath,
caption,
aliascnt,
colordvi,
amsfonts,
mathtools,
amsthm,
nicefrac,
dsfont}

\usepackage{tikz}
\usepackage[ruled,linesnumbered]{algorithm2e}
\RequirePackage[T1]{fontenc}
\RequirePackage[utf8]{inputenc}
\usepackage[english]{babel}
\usepackage{alphabeta}
\usepackage[inline]{enumitem}
\usepackage[mathscr]{euscript}

\usepackage{ifthen}

\definecolor{capri}{rgb}{0.0, 0.75, 1.0}
\definecolor{internationalorange}{rgb}{1.0, 0.31, 0.0}
\definecolor{crimsonglory}{rgb}{0.75, 0.0, 0.2}
\definecolor{brilliantlavender}{rgb}{0.96, 0.73, 1.0}
\definecolor{brightturquoise}{rgb}{0.03, 0.91, 0.87}
\definecolor{darkorange}{rgb}{1.0, 0.55, 0.0}

 \usepackage[pdftex, plainpages = false, pdfpagelabels, 
 bookmarks = true,
 bookmarksopen = true,
 bookmarksnumbered = true,
 breaklinks = true,
 linktocpage,
 pagebackref,
 colorlinks = true, 
 linkcolor = blue!10!black,
 urlcolor = blue,
 citecolor = red!10!black,
 anchorcolor = green,
 hyperindex = true,
 hyperfigures
 ]{hyperref}

\definecolor{MidnightBlack}{rgb}{0.1,0.1,.34}
\definecolor{MidnightBlue}{rgb}{0.1,0.1,0.43}
\definecolor{Black}{rgb}{0,0, 0}
\definecolor{Blue}{rgb}{0, 0 ,1}
\definecolor{myblue}{rgb}{0, 0 ,1}
\definecolor{Red}{rgb}{1, 0 ,0}
\definecolor{White}{rgb}{1, 1, 1}
\definecolor{grey}{rgb}{.6, .6, .6}
\definecolor{Mygreen}{rgb}{.0, .7, .0}
\definecolor{Yellow}{rgb}{.55,.55,0}
\definecolor{Mustard}{rgb}{1.0, 0.86, 0.35}
\definecolor{applegreen}{rgb}{0.55, 0.71, 0.0}
\definecolor{darkgreen}{rgb}{0,.5,0}
\definecolor{darkturquoise}{rgb}{0.0, 0.81, 0.82}
\definecolor{celestialblue}{rgb}{0.29, 0.59, 0.82}
\definecolor{green_yellow}{rgb}{0.68, 1.0, 0.18}
\definecolor{crimsonglory}{rgb}{0.75, 0.0, 0.2}
\definecolor{darkmagenta}{rgb}{0.30, 0.0, 0.30}
\definecolor{magenta}{rgb}{0.50, 0.0, 0.50}
\definecolor{internationalorange}{rgb}{1.0, 0.31, 0.0}
\definecolor{darkorange}{rgb}{1.0, 0.55, 0.0}
\definecolor{ao}{rgb}{0.0, 0.5, 0.0}
\definecolor{awesome}{rgb}{1.0, 0.13, 0.32}
\definecolor{darkcyan}{rgb}{0.0, 0.50, 0.50}
\definecolor{violet}{rgb}{0.93, 0.51, 0.93}
\definecolor{purple}{rgb}{0.625,0.125,0.9375}
\definecolor{brown}{rgb}{0.65, 0.16, 0.16}
\definecolor{orange}{rgb}{1.0, 0.65, 0.0}
\definecolor{cornflowerblue}{rgb}{0.39, 0.58, 0.93}
\definecolor{darkred}{rgb}{.5, 0, 0}
\definecolor{ouryellow}{rgb}{0.64, .6, 0}

\newcommand{\myblue}[1]{{\color{myblue}#1}}

\newcommand{\red}[1]{{\color{Red}#1}}

\newcommand{\darkgreen}[1]{{\color{darkgreen}#1}}
\newcommand{\darkorange}[1]{{\color{darkorange}#1}}

\newcommand{\purple}[1]{{\color{purple}#1}}

\hypersetup{
 colorlinks,
 linkcolor={red!75!black},
 citecolor={green!75!black},
 urlcolor={blue!75!black},
}

\newcommand{\cref}[1]{\autoref{#1}}

\definecolor{BrilliantRose}{rgb}{1.0, 0.33, 0.64}
\definecolor{denim}{rgb}{0.08, 0.38, 0.74}
\definecolor{amethyst}{rgb}{0.6, 0.4, 0.8}

\newcommand{\remove}[1]{}

\newcommand{\Ccal}{\mathcal{C}}
\newcommand{\Dcal}{\mathcal{D}}

\newcommand{\Fcal}{\mathcal{F}}

\newcommand{\Hcal}{\mathcal{H}}

\newcommand{\Lcal}{\mathcal{L}}
\newcommand{\Mcal}{\mathcal{M}}

\newcommand{\Ocal}{\mathcal{O}}
\newcommand{\Pcal}{\mathcal{P}}
\newcommand{\Qcal}{\mathcal{Q}}

\newcommand{\Nbbb}{\mathbb{N}}

\newcommand{\Sbbb}{\mathbb{S}}

\RequirePackage{stmaryrd}
\usepackage{textcomp}
\DeclareUnicodeCharacter{2286}{\subseteq}
\DeclareUnicodeCharacter{2192}{\ifmmode\to\else\textrightarrow\fi}
\DeclareUnicodeCharacter{2203}{\ensuremath\exists}
\DeclareUnicodeCharacter{183}{\cdot}
\DeclareUnicodeCharacter{2200}{\forall}
\DeclareUnicodeCharacter{2264}{\leq}
\DeclareUnicodeCharacter{2265}{\geq}
\DeclareUnicodeCharacter{8614}{\mathbin{\mapsto}}
\DeclareUnicodeCharacter{8656}{\Leftarrow}
\DeclareUnicodeCharacter{8657}{\Uparrow}
\DeclareUnicodeCharacter{8658}{\Rightarrow}
\DeclareUnicodeCharacter{8659}{\Downarrow}
\DeclareUnicodeCharacter{8669}{\rightsquigarrow}
\newcommand{\eqdef}{\stackrel{{\scriptsize\rm def}}{=}}
\DeclareUnicodeCharacter{8797}{\eqdef}
\DeclareUnicodeCharacter{8870}{\vdash}
\DeclareUnicodeCharacter{8873}{\Vdash}
\DeclareUnicodeCharacter{22A7}{\models}
\DeclareUnicodeCharacter{9121}{\lceil}
\DeclareUnicodeCharacter{9123}{\lfloor}
\DeclareUnicodeCharacter{9124}{\rceil}
\DeclareUnicodeCharacter{2208}{\in}
\DeclareUnicodeCharacter{9126}{\rfloor}
\DeclareUnicodeCharacter{9655}{\triangleright}
\DeclareUnicodeCharacter{9665}{\triangleleft}
\DeclareUnicodeCharacter{9671}{\diamond}
\DeclareUnicodeCharacter{9675}{\circ}
\DeclareUnicodeCharacter{10178}{\bot}
\DeclareUnicodeCharacter{10214}{}
\DeclareUnicodeCharacter{10215}{}
\DeclareUnicodeCharacter{10229}{\longleftarrow}
\DeclareUnicodeCharacter{10230}{\longrightarrow}
\DeclareUnicodeCharacter{10231}{\longleftrightarrow}
\DeclareUnicodeCharacter{10232}{\Longleftarrow}
\DeclareUnicodeCharacter{10233}{\Longrightarrow}
\DeclareUnicodeCharacter{10234}{\Longleftrightarrow}
\DeclareUnicodeCharacter{10236}{\longmapsto}
\DeclareUnicodeCharacter{10238}{\Longmapsto} 
\DeclareUnicodeCharacter{10503}{\Mapsto} 
\DeclareUnicodeCharacter{10971}{\mathrel{\not\hspace{-0.2em}\cap}}
\DeclareUnicodeCharacter{65294}{\ldotp}
\DeclareUnicodeCharacter{65372}{\mid}

\newtheorem{definition}{Definition}[section]

\newaliascnt{lemma}{theorem}
\newtheorem{lemma}[lemma]{Lemma}
\aliascntresetthe{lemma}
\newaliascnt{proposition}{theorem}
\newtheorem{proposition}[proposition]{Proposition}
\aliascntresetthe{proposition}
\newaliascnt{observation}{theorem}
\newtheorem{observation}[observation]{Observation}
\aliascntresetthe{observation}
\newaliascnt{corollary}{theorem}
\newtheorem{corollary}[corollary]{Corollary}
\aliascntresetthe{corollary}

\newcommand{\hh}{\end{document}}

\newcommand{\cupall}{\pmb{\pmb{\bigcup}}}
\newcommand{\eg}{{\sf eg}\xspace}
\newcommand{\tw}{{\sf tw}\xspace}

\newcommand{\bw}{{\sf bw}\xspace}

\newcommand{\p}{{\sf p}\xspace}

\usepackage{thm-restate}

\usepackage[nameinlink]{cleveref}
\crefname{subsection}{Subsection}{Subsections}
\crefname{section}{Section}{Sections}
\crefname{proposition}{Proposition}{Propositions}
\crefname{theorem}{Theorem}{Theorems}
\crefname{lemma}{Lemma}{Lemmas}
\crefname{corollary}{Corollary}{Corollaries}
\crefname{observation}{Observation}{Observations}
\crefname{figure}{Fig.}{Figures}

\usepackage{textpos}

\newcommand{\bound}[1]{\textbf{#1}}

\newcommand{\dist}{{\sf dist}}

\usepackage{tikz}
\usetikzlibrary{backgrounds}
\usetikzlibrary{calc,3d}
\usetikzlibrary{intersections,decorations.pathmorphing,shapes,decorations.pathreplacing,fit,shadows,fadings} 
\tikzset{black node/.style={line width=0.5, draw, circle, fill = black, minimum size = 5pt, inner sep = 0pt}}
\tikzset{white node/.style={draw, circle, fill = white, minimum size = 4pt, inner sep = 0pt}}
\tikzset{gray node/.style={draw=gray, circle, fill = gray, minimum size = 4pt, inner sep = 0pt}}
\tikzset{fusion node/.style={draw=crimsonglory, circle, fill=crimsonglory, minimum size = 4pt, inner sep = 0pt}}
\tikzset{root node/.style={draw=black, circle, fill = red!50!white, minimum size = 4pt, inner sep = 0pt}}
\tikzset{sqnode/.style={draw,rectangle, fill = white, minimum size = 4pt, inner sep = 0pt}}
\usetikzlibrary{patterns}

\tikzset{big black node/.style={line width=1pt, draw, circle, fill = black, minimum size = 4.5pt, inner sep = 0pt}}
\tikzset{big white node/.style={line width=1pt, draw, circle, fill = white, minimum size = 4.5pt, inner sep = 0pt}}
\tikzset{big diamond node/.style={line width=1pt,draw = black, diamond, fill = black, minimum size = 6pt, inner sep = 0pt}}
\tikzset{big triangle node/.style={line width=1pt, draw, regular polygon, regular polygon sides=3, fill = black, minimum size = 6pt, inner sep = 0pt}}

\newbox\mybox
\newcommand{\tikzBox}[2][\mybox]{%
	\sbox#1{\pgfinterruptpicture#2\endpgfinterruptpicture}}

\tikzset{diamond node/.style={line width=0.5pt,draw = black, diamond, fill = black, minimum size = 7pt, inner sep = 0pt}}
\tikzset{rect node/.style={line width=1pt, draw,rectangle, fill = black, minimum size = 6pt, inner sep = 0pt}}
\tikzset{star node/.style={line width=1pt, draw, star, fill = black, minimum size = 9pt, inner sep = 0pt}}
\tikzset{square node/.style={line width=0.5pt, draw, rectangle, fill = black, minimum size = 5pt, inner sep = 0pt}}
\tikzset{triangle node/.style={line width=0.5pt, draw, regular polygon, regular polygon sides=3, fill = black, minimum size = 7pt, inner sep = 0pt}}

\definecolor{f2color}{rgb}{0.65, 0.88, 0.93}
\definecolor{f4color}{rgb}{0.53, 0.51, 0.93}
\definecolor{f3color}{rgb}{0.96, 0.54, 0.56}
\definecolor{f1color}{rgb}{1.0, 0.85, 0.45}

\definecolor{cb1}{rgb}{0.667,0.20,0.467}
\definecolor{cb2}{rgb}{1,0.8,0.8}
\definecolor{cb3}{rgb}{0.867,0.867,0.867}
\definecolor{cb4}{rgb}{0,0.447,0.698}
\definecolor{cb5}{rgb}{0.933,0.8,0.4}
\definecolor{cb6}{rgb}{0.835,0.369,0}
\definecolor{cb7}{rgb}{0.267,0.733,0.6}
\definecolor{cb8}{rgb}{0.333,0.333,0.333}

\begin{document}

\title{\Large\bf Finding irrelevant vertices in linear time\\ on bounded-genus graphs%
\thanks{The first author was supported by the Research Council of Norway via the project BWCA (grant no. 314528). The two last authors were supported by the French-German Collaboration ANR/DFG Project UTMA (ANR-20-CE92-0027). 
The third author was also supported by the project BOBR that is funded from the European Research Council (ERC) under the European Union's Horizon 2020 research and innovation program with grant agreement No. 948057. The first and the last author were also supported by the Franco-Norwegian project PHC AURORA 2024-25. Also, the last author has been supported by the French National Research Agency (ANR) under project GODASse ANR-24-CE48-4377 and under the France 2030 grant reference number
ANR-24-RRII-0002 operated by the Inria Quadrant Program.}\,\,$^{,}$\thanks{\scriptsize Emails:
 \texttt{petr.golovach@uib.no},
 \texttt{sgk@di.uoa.gr},
 \texttt{stamoulis@irif.fr},
\texttt{sedthilk@thilikos.info}.}\,\,$^{,}$\thanks{An extended abstract of this paper has been accepted for publication to SODA 2025.}}

{
\author{\bigskip
Petr A. Golovach%
\thanks{Department of Informatics, University of Bergen, Norway.} 
\and
Stavros G. Kolliopoulos%
\thanks{Department of Informatics and Telecommunications, National and Kapodistrian University of Athens, Greece.} 
\and
Giannos Stamoulis\thanks{Université Paris Cité, CNRS, IRIF, F-75013, Paris, France. Most of the research work for this paper was conducted while G.S. was affiliated with (1) Institute of Informatics, University of Warsaw, Poland (2) LIRMM, Univ Montpellier, CNRS, Montpellier, France, (3) the Department of Mathematics and the Department of Informatics and Telecommunications, National and Kapodistrian University of Athens, Greece, and (4) Inter-university Postgraduate Program ``Algorithms, Logic, and Discrete Mathematics'' (ALMA), Athens, Greece.}
\and
Dimitrios M. Thilikos\thanks{LIRMM, Univ Montpellier, CNRS, Montpellier, France.}}
}
\date{}

\maketitle

\vspace{-.7cm}

\begin{abstract}\noindent 
%

%
\noindent 
We introduce an algorithmic approach that permits the application of the {\em irrelevant vertex technique} on graphs of bounded genus in linear time. This technique 
provides a powerful 
tool for the design of parameterized algorithms for a wide variety of 
problems on graphs.
A common characteristic of these problems, permitting the application of this technique 
on surface-embedded graphs, is the fact that every 
graph of large enough treewidth contains a vertex that is {\sl irrelevant}, 
in the sense that 
its removal yields an equivalent instance of the problem. Removing vertices one-by-one yields, in general, algorithms with running time that is {\sl quadratic} in 
the size of the input graph. This running time is due to the fact that it takes linear time 
to detect one irrelevant vertex and the total number of irrelevant vertices to be detected is linear as well.
Using advanced techniques, sub-quadratic 
algorithms have been designed for particular problems,
even in general graphs. However, designing 
a general framework for linear-time algorithms 
has been open, even for the bounded-genus case.

In this paper we introduce a general framework 
that enables finding in linear time 
an {\sl entire} set of irrelevant vertices whose removal yields 
a bounded-treewidth graph, provided that the input graph has bounded genus. Our technique consists of decomposing 
any surface-embedded graph into a tree-structured collection of bounded-treewidth subgraphs 
where detecting globally irrelevant vertices can be done locally and independently. Our method is applicable to a wide variety
of
known graph containment or graph modification problems where the irrelevant vertex technique applies. Examples include the {\sc (Induced) Minor Folio} problem, the {\sc (Induced) Disjoint Paths} problem, and 
the {\sc $\Fcal$-Minor-Deletion} problem.
\end{abstract}
\medskip
 

\begin{textblock}{20}(12.2, 2.4)
 \includegraphics[width=30px]{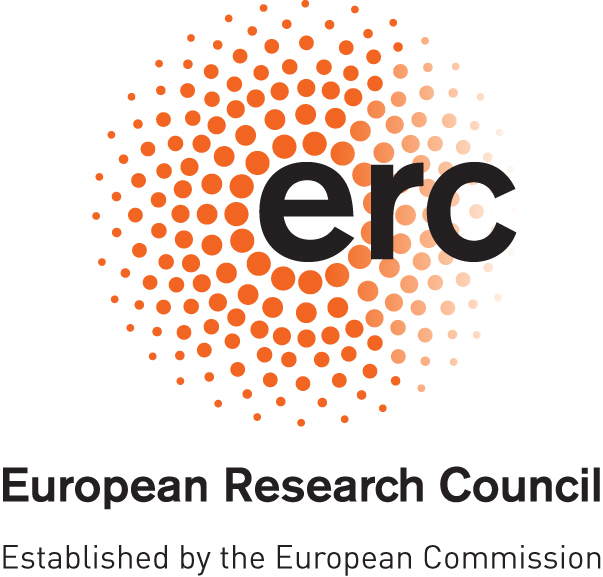}%
\end{textblock}
\begin{textblock}{20}(12.2, 3.0)
 \includegraphics[width=30px]{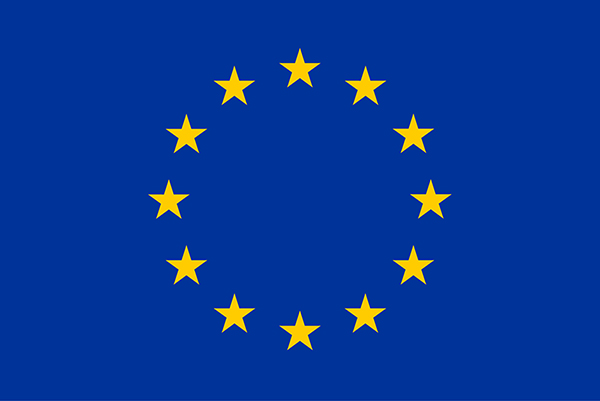}%
\end{textblock}

 \thispagestyle{empty}

\newpage
{\tableofcontents}

\pagenumbering{arabic}
\thispagestyle{empty}
\newpage

\setcounter{page}{1}


\section{Introduction}

The \textsl{irrelevant vertex technique} was introduced by Robertson and Seymour in \cite{RobertsonS95b} for deriving a polynomial algorithm for the {\sc Disjoint Paths} problem. Also, the correctness of this application was proven in \cite{RobertsonS09XXI,RobertsonSGM22}. 
This technique has nowadays evolved to a standard algorithmic paradigm for solving graph containment or graph modification problems~\cite{GolovachKMT17thep,JansenLS14anea,AdlerGK08comp,MarxS07obta,KK09algori,GolovachKPT09indu,KaminskiT12cont,KawarabayashiKM10link,KociumakaP19dele,GroheKMW11find}.
The general idea behind the technique is to exploit certain structural characteristics 
of the input that make possible to find a vertex of the input graph that is \textsl{problem-irrelevant}: its removal from an instance creates a new instance that is equivalent to the old one. 

The archetypical example of the applicability
of this technique is the {\sc Planar Disjoint Paths} problem where, given a planar graph $G$ and $k$ pairs of terminals $\{(s_{i},t_{i})\mid i\in\{1,\ldots,k\}\},$ the 
question is whether $G$ has pairwise vertex-disjoint paths $P_{1},\ldots,P_{k}$ where $P_{i}$ is a path from $s_{i}$ to $t_{i},$ for $i\in\{1,\ldots,k\}.$ For this problem, the core structural characteristic is the ``linkage theorem'' proved by Robertson and Seymour in {\cite{RobertsonS09XXI,RobertsonSGM22}}. This result implies the existence of a function $\ell:\Nbbb\to\Nbbb$ (called \emph{the {linkage} function}) such that the existence 
of a sequence of $\ell(k)$ pairwise disjoint nested cycles in a plane 
embedding of $G$ where all terminals 
are outside the ``outer cycle'' of this sequence implies that
all paths of a solution
of the {\sc Planar Disjoint Paths} problem
can be rerouted away from the ``inner cycle'' of this sequence.
This in turn permits to safely discard from $G$ all 
vertices inside the inner cycle 
and create an equivalent instance of {\sc Planar Disjoint Paths}.
Intuitively, this sequence of nested cycles ``insulates'' the terminals from the vertices declared irrelevant.

Given that every graph of large enough treewidth contains a subdivision of a wall and, in planar graphs, such a wall certifies the aforementioned insulation sequence, as long as the treewidth of the input graph is above a certain threshold, we detect and discard irrelevant vertices until
the treewidth becomes bounded. Then one may solve the problem in linear time using dynamic programming techniques.
While this idea appears to be quite simple and intuitive, to prove that the above rerouting is indeed possible is non-trivial even in the case of planar graphs (see \cite{AdlerKKLST17irre,GolovachST23comb,AdlerK19alower}). 
When it comes to problems on general graphs, the planarity condition is substituted by suitable notions of flatness that 
still permit a rerouting argument based on sufficiently big ``insulation'' \cite{RobertsonS95b,RobertsonS09XXI,RobertsonSGM22}, while 
additional machinery is required in order to find irrelevant vertices for ``non-flat'' instances.

In problems such as {\sc Minor Containment} and ``induced'' or ``rooted'' variants of it, the above insulation property is not enough for declaring the vertices within the inner cycle irrelevant.
Additionally, a ``big enough'' flow between vertices of the inner and the outer cycle is required. In this paper, we call the union of the insulating cycles and the paths of a flow traversing them a \emph{railed nest}. Our results concern all problems where insulation by a railed nest is sufficient for declaring irrelevance.
We refer to this problem property as the \textsl{insulation property}.

A third family of problems that have the insulation property
concerns graph modification problems where 
the question is to find a set of vertices whose removal may yield some particular minor-closed property, such as planarity (see~\cite{KociumakaP19dele,MarxS07obta,BasteST19hittIV,SauST21kapiII,MorelleSST24fast,LokshtanovPSSXZ25Crosssing,GahlawatRZ25Almost}).

Actually, there are families of problems 
where for the applications of the irrelevant vertex technique the insulation property is not enough. The reason is that 
the existence of a railed nest implies that only 
a \textsl{subset} of the vertices inside the inner cycle is irrelevant and 
extra algorithmic effort is required 
in order to detect them. The most representative problem 
of this category is {\sc Topological Minor Containment} \cite{GroheKMW11find,KaminskiT12cont} (see also \cite{FominLPSZ20hitt,GolovachST23hitt,GolovachKMT17thep,GolovachST23comb} for other results of this kind). In fact, there are situations where the irrelevant vertex argument relies on combinatorial conditions that extend the insulation argument and the method is combined with other advanced techniques in graph algorithm design \cite{FominLST18line,MarxS07obta,GroheKMW11find,KawarabayashiR2007compu,AdlerGK08comp,FominGKS23comp,FominLPSZ20hitt}. Finally, there were recent attempts to derive general meta-algorithmic conditions for the 
applicability of the irrelevant vertex technique \cite{GolovachST22model,FominGSST23comp,FominGST22anal,SchirrmacherSST24mode,SauST2024parame}.

\paragraph{The sub-quadratic issue.} As already indicated above, 
the vertex-by-vertex application of the irrelevant vertex argument 
requires time $\Omega(n^2)$. One can achieve time $\Ocal(n)$ for detecting and removing
an irrelevant vertex; this should be repeated 
as long as an irrelevant vertex can be detected, that is $\Ocal(n)$ times. 
We stress that in general, the detection of the irrelevant 
vertex in $\Ocal(n)$ time is not always a straightforward task. Even in the classic case of {\sc Disjoint Paths} and {\sc Minor Containment},
the original results of Robertson and Seymour in \cite{RobertsonS95b} required time $\Ocal(n^{2})$ to detect the 
irrelevant vertex. This was later improved in linear time by Kawarabayashi, Kobayashi, and Reed in \cite{KawarabayashiKR11thed}. Nowadays, most algorithmic 
or meta-algorithmic applications of the irrelevant vertex technique 
require time $\Theta(n^{2})$ (see \cite{SauST2024parame,GolovachST22model,FominGSST23comp}).
However achieving sub-quadratic 
implementations of the irrelevant vertex technique remains 
an open challenge. 
This has been achieved for 
particular problems and, in all of them, the challenge that was met was to detect ``many'' irrelevant vertices at once, instead of finding them one by one. The first problem for which 
a linear-time algorithm was derived 
was {\sc Planar Disjoint Paths} and its extension {\sc Planar Disjoint Connected Subgraphs} by Reed, Robertson, Schrijver, and Seymour in \cite{ReedRSS91find} (see also \cite{Reed95rooted}). Also, for these particular problems, the results of \cite{ReedRSS91find,Reed95rooted} can easily be extended to every class of surface-embeddable graphs.
Next, Mohar, in~\cite{Mohar99alin}, used the irrelevant vertex technique in order to check whether a graph is embeddable 
in some particular surface in linear time. 
The algorithm of \cite{Mohar99alin} was also one of the main ingredients of the linear algorithm of Kawarabayashi and Reed \cite{KawarabayashiR2007compu} for 
checking whether a graph can be embedded in the plane with at most $k$ crossings. The algorithm of \cite{KawarabayashiR2007compu} made use of the irrelevant vertex technique by combining 
the algorithm of \cite{Mohar99alin} with results from discrete geometry.
Later, a simpler algorithm for surface-embeddability 
was proposed by Kawarabayashi, Mohar, and Reed in~\cite{KawarabayashiMR08asim}, where 
irrelevant vertices where found by successively ``shrinking'' the problem instance
to an equivalent one by contracting (big) induced matchings. Kawarabayashi in \cite{Kawarabayashi09plan} used 
a similar approach in order to solve, in linear time, the {\sc Planarizer} problem, 
asking whether the removal 
of at most $k$ vertices can make a graph planar. To our knowledge, \cite{Kawarabayashi09plan} is the only result 
that presents a linear-time implementation of the irrelevant vertex technique for a graph modification problem.

Recently Korhonen, Mi.~\!Pilipczuk, and Stamoulis \cite{korhonen2024minor} gave an \textsl{almost linear} algorithm for solving the {\sc Rooted Minor Folio} problem, 
asking for the set of all rooted minors of some specific size in a graph. 
The techniques in \cite{korhonen2024minor} 
combined the irrelevant vertex technique
with successive instance shrinking and the use of advanced dynamic 
algorithm techniques of Korhonen, Majewski, Nadara, Pilipczuk, and 
Sokołowski in \cite{KorhonenMNPS23dyna}.
As a result of \cite{korhonen2024minor}, the quadratic bound from \cite{KawarabayashiKR11thed} for the {\sc Minor Containment} problem has been reduced for the first time to almost linear.

\paragraph{Our results.}
In this paper we provide an algorithmic framework for surface-embeddable graphs
that, when applicable, can find, in linear time, 
a {\sl set} of irrelevant vertices whose removal yields 
a bounded-treewidth graph. 
We now informally introduce the property that conditions the applicability of our results (for the formal definitions, see \cref{ssiop}). We deal with problems on \textsl{rooted} graphs embedded on surfaces of bounded Euler genus. The input graph comes with a bounded-size set $X$ of distinguished vertices, called \emph{roots}.
We say that such a problem $\Pi$ 
has the \emph{insulation} property
if for every instance $(G,X)$ of $\Pi$ such that
\begin{itemize}
\item $G$ contains a sequence ${\cal C}$ of sufficiently many nested and pairwise vertex-disjoint contractible cycles, 
\item $G$ contains a collection $\Pcal$ of sufficiently many pairwise vertex-disjoint paths between the outer cycle and the inner cycle of $\Ccal,$ and 
\item all roots in $X$ are embedded outside the outer cycle of ${\cal C}$ 
\end{itemize}
the following holds:
if $S$ is the set of vertices inside the inner cycle of ${\cal C},$
then for every vertex $v\in S,$
$(G,X)$ and $(G-v,X)$ are equivalent instances of $\Pi.$
This expresses that fact that 
all vertices in $S$ are \textsl{irrelevant} for the instance $(G,X)$ of $\Pi.$ We refer to the above union of the cycles in ${\cal C}$ and the paths 
in ${\cal P}$ as a \emph{railed nest} insulating $S$ from the roots in $X.$

Typical problems that have the insulation property and for which our results apply include {\sc (Induced) Disjoint Paths}, {\sc (Rooted/Induced) Minor Containment}, and {\sc Planarizer}.
This list covers only a small sample. In order to demonstrate 
the potential of our technique, we dedicate \cref{gen_esxopro} to an abstract description of problems that satisfy the insulation property.
\medskip

Our main result is that for every problem on rooted surface-embedded 
graphs that has the insulation property
there exists a linear-time algorithm that reduces every instance 
to an equivalent one whose graph has bounded treewidth (we state this result as \cref{main_th}). As a corollary (see \cref{our_col}) we obtain for the first time linear-time algorithms 
for a wide family of well-studied graph containment/modification problems on surface-embedded graphs.

\paragraph{Our technique.} In this paper, we follow an approach that deviates%
 significantly 
from the ``shrinking'' techniques applied in \cite{KawarabayashiMR08asim,korhonen2024minor,Kawarabayashi09plan} 
where algorithms eliminate sufficiently many irrelevant vertices to replace the instance with an equivalent one of strictly smaller size. Instead, our strategy identifies multiple railed nests in the input graph and removes vertices that are globally irrelevant across all of them by processing each structure locally. We show that, after applying this procedure, the resulting graph has bounded treewidth. Notably, our technique is self-contained and can be used as a black-box for any problem satisfying the insulation property.

To make this possible, we decompose, in linear time, the graph into a tree-structured collection of subgraphs. These subgraphs are called \emph{slices} and each of them has bounded treewidth. The algorithm processes each slice sequentially. To identify the slices in linear time, we introduce the concept of \textsl{radial distance decomposition}. 
For each slice 
we certify locally \textsl{all} railed nests in it and we prove that this permits the detection and elimination
of all globally-irrelevant vertices relative to this slice. 
This certification is implemented in linear time 
using Courcelle's theorem on some enhanced version of the slice
that, as we prove, is of bounded treewidth as well.
We prove that after all slices have been locally processed, an equivalent instance of bounded treewidth is produced and that this procedure can be implemented in linear time. 
If in addition the problem can be expressed in Counting Monadic Second Order logic (\textsf{CMSO}), it immediately follows that it can be solved in linear time.
A formal description of our results is given in \cref{ssiop} and an outline of our technique is presented in \cref{outline_pr}.

\paragraph{Related work.}
 The approach of producing an equivalent instance of bounded treewidth in linear time has already been employed in the aforementioned works by Reed, Robertson, Schrijver, and Seymour~\cite{ReedRSS91find} and Reed~\cite{Reed95rooted}. Recent works that are based on the ideas of~\cite{ReedRSS91find,Reed95rooted} are the ones of Cho, E. Oh, and S. Oh~\cite{ChoOhOh2023PlanarDisjointPaths} and Włodarczyk and Zehavi~\cite{WlodarczykZehavi2023PlanarDisjointPaths} (see also~\cite{JansenPvL21,JansenW25lossy}).
In particular, Cho, E. Oh, and S. Oh~\cite{ChoOhOh2023PlanarDisjointPaths} 
give a $2^{\Ocal(k^2)}\cdot n$-time algorithm for the $k$-\textsc{Planar Disjoint Paths} problem, while Włodarczyk and Zehavi~\cite{WlodarczykZehavi2023PlanarDisjointPaths} give a polynomial kernel for the same problem parameterized by $k+\tw$, i.e., a polynomial preprocessing algorithm 
that produces an equivalent instance of size $(k+ {\sf tw}(G))^{\Ocal(1)}.$
The idea behind the treewidth reduction in~\cite{Reed95rooted,ChoOhOh2023PlanarDisjointPaths,WlodarczykZehavi2023PlanarDisjointPaths} is that every solution to the $k$-\textsc{Planar Disjoint Paths} problem can be rerouted so as to avoid every vertex that is ``radially-far'' from the terminals, i.e., the radial distance between such vertices and the terminals is big enough. In our terminology, they~\cite{Reed95rooted,ChoOhOh2023PlanarDisjointPaths,WlodarczykZehavi2023PlanarDisjointPaths} identify a single slice (rooted on the terminals) and declare irrelevant every vertex that is not in this slice. 
For our purposes, i.e., problems where irrelevancy is guaranteed in the presence of a railed nest rather than a nested collection of cycles, extra work is needed.

As mentioned above and further explained in~\cref{outline_pr}, our method consists in identifying which slices can ``host'' railed nests: only such slices allow to declare certain sets of vertices as irrelevant. We implement this part algorithmically using Courcelle's theorem, which brings immense parametric dependencies in the running time of our obtained algorithms.
Giving an alternative implementation of this part with better parametric dependencies would readily give a linear-time treewidth reduction with better parametric dependencies.
Such improvements, combined with improved dynamic programming procedures for bounded treewidth instances, would yield improved running times for the problems we consider for the problems considered in this work. See~\cref{sec_open} for a discussion on this direction.

Similar ideas were also used by Jansen, Ma.~\!Pilipczuk, and Van Leeuwen~\cite{JansenPvL21} to give a polynomial kernel for the \textsc{Planar Vertex Multiway Cut} problem. They used \emph{outerplanar decompositions} (which are essentially the same as radial distance decompositions) in order to get an equivalent instance for their problem with bounded radial diameter. While they also identify parts of the instance that can be avoided by an (optimal) solution, their approach is eventually quite different from ours, because of the different nature of the problems. Also, the technique of~\cite{JansenPvL21} was later used by Jansen and Włodarczyk~\cite{JansenW25lossy} to get approximate kernelization for the \textsc{Planarizer} problem. Their work applies some of the aforementioned ideas but instead of exploiting
the full power of the insulation property it rather relies on topological-based arguments, particular to the \textsc{Planarizer} problem.

\section{Formal presentation of the result}
\label{ssiop}

We use $\Nbbb$ for the set of all nonnegative integers. 
Given two integers $p,q$ with $p<q,$
we use $[p,q]$ to denote the set $\{p,p+1,\ldots,q\}$ and $[p,q]^{\sf even}$ to denote the set of all even integers in $[p,q].$
Given a $k\in\mathbb{N}$ we set $[k]:=[1,k]$ and $\Nbbb_{\ge k}:=\Nbbb\setminus[0,k-1]$ and we use $\Nbbb_{\ge k}^{\sf even}$ to denote the set of all even integers in $\Nbbb_{\ge k}.$
If ${\cal S}$ is a collection of objects where the operation $\cup$ is defined, then we use $\cupall{\cal S}$ to denote $\bigcup_{X\in{\cal S}}X.$

\subsection{Basic concepts on graphs}

All graphs in this paper are without loops (but possibly with multiedges), undirected, and finite.
Given a graph $G,$ we denote its vertex and edge set by $V(G)$ and $E(G)$ respectively. We set $|G|:=|V(G)|.$
Given some $S\subseteq V(G),$ we denote by $G\setminus S$ the graph 
obtained by removing from $G$ the vertices in $S,$ together with their incident edges. For $v\in V(G),$ we use $G\setminus v$ to denote the graph $G\setminus\{v\}.$ 
Also, the {\em subgraph of $G$ induced by $S$}, denoted by $G[S],$ is the graph $G\setminus (V(G)\setminus S).$
Given a set $J\subseteq E(G),$ we denote by $G[J]$ the graph $(V_{J},J)$ where $V_{J}$ is the set of all endpoints of the edges in $J.$ 
If $G'$ is a graph where $V(G')\subseteq V(G)$ and $E(G')\subseteq E(G[V(G')])$ then we say that $G'$ is {\em a subgraph of $G$}.

Given a graph $G$ and a set $S\subseteq V(G)$ (resp. $J\subseteq E(G)$) we say that $S$ (resp. $J$) is \emph{connected}
if $G[S]$ (resp. $G[J]$) is a connected graph.
Given a connected set $J$ of edges we denote by $G/J$ 
the graph obtained from $G$ after the contraction of all edges of $J$ to a single vertex.

\paragraph{Embeddings.}
In this paper we deal with graphs embedded in a surface $\Sigma,$
where an embedding of $G$ in $\Sigma$ is a drawing in $\Sigma$ without crossings, see \cite{MoharT01grap} for the formal definition.
When we refer to a {\em $\Sigma$-embedded} graph $G,$ 
we make the implicit assumption that $G$
is accompanied with an embedding of it in $\Sigma.$
For notational convenience, 
we do not distinguish a vertex/edge of $G$ from the points
corresponding to its embedding. For instance, given a $\Sigma$-embedded graph $G,$ if $H$ is a subgraph of $G$ then we denote by $\Sigma\setminus H$ the set of points of $\Sigma$ that are not points of the embedding of $H.$
Also, if $Γ\subseteq Σ,$ then $Γ\cap V(G)$ consists of all vertices 
of $G$ that are embedded to points of $Γ.$
Given a cycle $C$ of $G$ we say that $C$ is {\em contractible}
if one of the two (arcwise) connected
components of $\Sigma\setminus C$ is an open disk.
We use $\Sbbb^{2}\coloneqq\{(x,y,z)\in\mathbb{R}^3\mid x^2+y^2+z^2=1\}$ to denote the sphere. Therefore, $\Sbbb^{2}$-embedded graphs 
are planar graphs. We use $\eg(\Sigma)$ to denote the Euler genus of
the surface $Σ$ and $\eg(G)$ for the minimum Euler genus of a surface where $G$ can be embedded.

\paragraph{Treewidth.}

A graph is \emph{$k$-chordal} if it has no cycle of length $≥4$
as induced subgraph
and does not contain a complete graph on more than $k+1$ vertices. The \emph{treewidth} of a graph $G,$ 
denoted by $\tw(G)$ is the minimum $k$ for which $G$ is a subgraph of some $k$-chordal graph. The parameter of treewidth
is important for our algorithms, however we do not make any use of its original definition (given by Robertson and Seymour in \cite{RobertsonS84GMIII}). Instead we use only results around treewidth and, for this reason, we resort to the above definition for brevity.

\paragraph{Rooted graphs.}
A \emph{rooted graph} is a triple $\bound{G} = (G,R,\rho)$ where $G$ is a graph, $R \subseteq V(G),$ and
$\rho : R \to [|R|]$ is a
bijection. We refer to $R$ as the \emph{boundary} of ${\bf G}$ and 
to its vertices as the {\em roots} of ${\bf G}$ and we denote them by $R({\bf G}).$ 
We say that $\textbf{G}_{1}=(G_{1},R_{1},\rho_{1})$ and $\textbf{G}_{2}=(G_{2},R_{2},\rho_{2})$
are {\em isomorphic} if there is an isomorphism from $G_{1}$ to $G_{2}$
that extends the bijection $\rho_{2}^{-1}\circ \rho_{1}.$
A {rooted graph} ${\bf G}$ is a \emph{$t$-rooted graph} if $t=|R({\bf G})|.$
Notice that the notion of a rooted graph is extending the 
one of a graph as every graph can be seen as a $0$-rooted graph.
A rooted graph $(G,R,\rho)$ is $\Sigma$-embedded if $G$ is $\Sigma$-embedded, i.e.,
is accompanied with a $\Sigma$-embedding of it.
We also use $V({\bf G})$ instead of $V(G)$ and we define $\overline{V}({\bf G}):=V({\bf G})\setminus R({\bf G}).$
Also, if $S\subseteq \overline{V}(G)$ we use ${\bf G}-S$ 
to denote the rooted graph $(G-S,R,\rho).$ Similarly, 
we define ${\bf G}-v$ for some $v\in V(G)\setminus R.$
We set $|{\bf G}|:=|G|$ and, in general, if $\p$ is a graph parameter (such as $\eg$ or $\tw$)
and ${\bf G}=(G,B,\rho),$ we define its rooted graph extension as $\p({\bf G}):=\p(G).$

In this paper, we see every decision problem on rooted graphs as a set $\Pi$ of rooted graphs, i.e., the set of yes-instances of the problem,
whose number of roots is bounded by some fixed number.
Also we adopt the convention that all problems that we consider 
concern rooted graphs with at least one root. 
In the case of simple graphs, i.e., $0$-rooted graphs,
we arbitrarily choose some vertex as a root. This convention 
is useful for the uniformity of our presentation.

\paragraph{Railed nests.} Let ${\bf G}$ be a $\Sigma$-embedded rooted graph and let $r\in\Nbbb_{≥1}.$

\begin{figure}[ht]
 \begin{center}
 \scalebox{.669}{\includegraphics{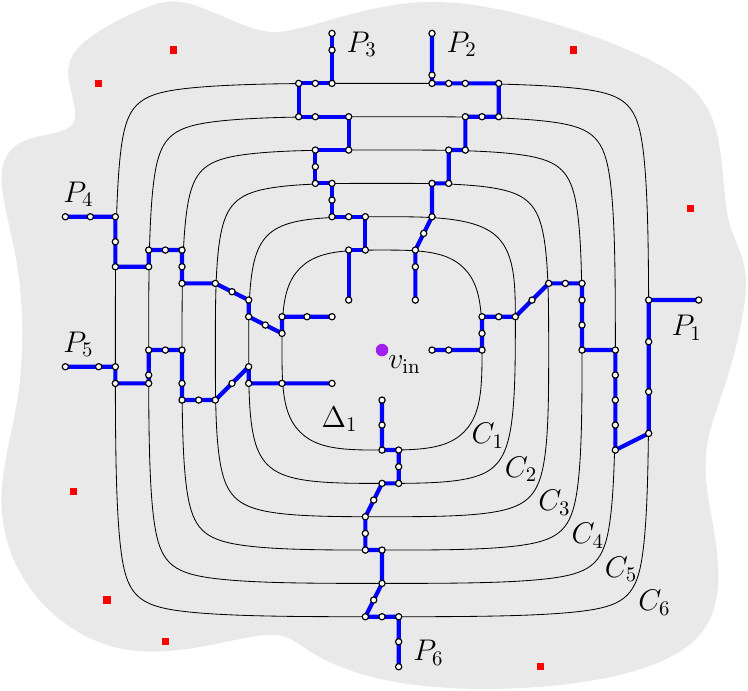}}
 \end{center}
 \caption{A railed nest $W$ with 6 cycles (in black) and 4 paths (in blue) as a subgraph of a graph $G.$ Apart from $W,$ we depict 
 only the vertex $v_{\rm in}$ in \purple{purple} and the terminals in \red{red}.}
 \label{todrodop}
\end{figure}%

\begin{definition} 
\label{def_otho}
An {\em $r$-railed nest} of ${\bf G}$ is the union $W$ of a collection $\Ccal=\{C_{1},\ldots,C_{r}\}$ of pairwise vertex disjoint cycles of $G$ and a collection $\Pcal=\{P_{1},\ldots,P_{r}\}$
of pairwise vertex-disjoint paths of $G$ such that
\begin{enumerate}
\item %

 for every $i\in\{1,\ldots,r\}$ the cycle $C_{i}$ bounds an open disk $Δ_{i}$ of the surface where 
all the cycles among $C_{1},\ldots,C_{i-1}$ are embedded, 
\item\label{it_orhogonality} for every $i,j\in[r]$, the graph $C_{i}\cap P_{j}$ is a (possibly trivial) path, 
\item all roots of ${\bf G}$ are embedded in $\Sigma\setminus \Delta_{r},$ i.e., 
$R({\bf G})\subseteq V(G)\cap (\Sigma\setminus \Delta_{r}),$ and 
\item at least one vertex $v_{\rm in}$ of $G$ is embedded in $\Delta_{1}.$
\end{enumerate}
We refer to $C_{r}$ (resp. $C_{1}$) as the \emph{outer} (resp. \emph{inner}) cycle of $W.$ The \emph{interior} (resp. \emph{exterior}) of $W$ is the open disk $\Delta_{1}$ (resp. the open set $\Sigma\setminus \Delta_{r}$) and is denoted by ${\sf int}(W)$ (resp. ${\sf ext}(W)$).
We also refer to $\Ccal$ and $\Pcal$ as the \emph{cycle collection} and the \emph{path collection} of the $r$-railed nest $W.$
\end{definition}

The combinatorial importance of treewidth resides in the 
fact that it certifies the existence of a railed nest (see also \cref{thi_oplos_s}).

\begin{proposition}[]
\label{proog_tw}
There is a universal constant $c$ such that, for every $r\in \Nbbb_{≥1}$ and $t\in\Nbbb,$
if ${\bf G}$ is a $t$-rooted graph where $\tw({\bf G})>c\cdot r \cdot\sqrt{t+1}\cdot (\eg({\bf G})+1)$
then ${\bf G}$ contains an {\em $r$-railed nest}. Moreover, such an $r$-railed nest can be found in time $\Ocal_{\eg({\bf G})+t+r}(|{\bf G}|).$\footnote{Given two functions $\chi,\psi\colon \mathbb{N}\rightarrow \mathbb{N},$ we write $\chi(n)=\Ocal_{x}(\psi(n))$ to denote that there exists a computable function $f\colon\mathbb{N} \rightarrow \mathbb{N}$ such that $\chi(n)\le f(x)\cdot \psi(n).$}
\end{proposition}

The above proposition is a consequence of the results of \cite{DemaineHT06thebidim}, 
as explained in \cref{subsec_treewidth}.

An \emph{annotated graph} is a tuple $(G,R_{1},\ldots,R_{k})$ for some $k\ge 0$ where each $R_{i}$ is either a vertex (resp. edge) subset of $V(G)$ (resp. $E(G)$).
The sets $R_{1},\ldots,R_{k}$ are called \emph{annotation sets}. 
\emph{Monadic Second Order logic} (\textsf{MSO}) is a basic tool to express properties in (rooted and/or annotated) graphs.
The syntax of \textsf{MSO} includes logical connectives $\land,$ $\lor,$ $\neg,$ $\Leftrightarrow,$ $\Rightarrow,$ variables for vertices, edges, vertex sets, and edge sets, quantifiers $\exists,$ $\forall$ over these variables, the relations 
$u\in V$ when $u$ is a vertex variable and $U$ is a vertex set variable (or a unary relation symbol interpreted as an annotation set);
$e\in E$ when $e$ is an edge variable and $E$ is a edge set variable (or a binary relation symbol interpreted as an annotation set);
${\rm adj}(u,v)$ when $u$ and $v$ are vertex variables, with the interpretation that $u$ and $v$ are adjacent;
${\rm inc}(u,e)$ when $u$ is a vertex variable and $e$ is an edge variable, with the interpretation that $e$ is incident to $u$;
and equality of variables representing vertices, edges, vertex sets, edge sets and relation symbols representing annotation sets.
\emph{Counting Monadic Second Order logic} (\textsf{CMSO}) extends \textsf{MSO} by including atomic sentences testing whether the cardinality of a set is equal to $q \pmod r,$ where $r\in\Nbbb_{\ge 2}$ and $q\in[0,r-1].$ A \emph{$\mathsf{CMSO}$-sentence on annotated graphs} is a formula of $\mathsf{CMSO}$ without free variables that is evaluated on annotated graphs (the formula has access to the annotation sets of the input structure). Given an annotated graph $(G,X)$ and a $\mathsf{CMSO}$-sentence $\varphi$ on annotated graphs of this form, we write $(G,X)\models\varphi$ to denote that $\varphi$ is satisfied in $(G,X).$

The algorithmic importance of treewidth resides, among others, in the next result, known as Courcelle's theorem. It has appeared in different versions and proofs in \cite{Courcelle90them,Courcelle97,Courcelle92,BoriePT92auto,ArnborgLS91easy}. In this paper we make use of the following optimization version (see \cite{ArnborgLS91easy}); we denote by $|\varphi|$ the size of the formula $\varphi$.

\begin{proposition}
\label{ow_dec_dource}
There is an algorithm that, given an {\sf CMSO}-sentence $\varphi$ on annotated graphs and an $n$-vertex graph $G,$ outputs a maximum size set $X\subseteq V(G)$
where $(G,X)\models\varphi,$ in time $\Ocal_{\tw(G),|\varphi|}(n).$
\end{proposition}

The decision version of \cref{ow_dec_dource} asserts that for every \textsf{CΜSO} sentence $\psi$ on rooted graphs, there is an algorithm deciding whether ${\bf G}\models \psi$ in $\Ocal_{\tw({\bf G})}(|{\bf G}|)$-time.
A rooted graph problem $\Pi$ is {\em {\sf CMSO}-expressible} 
if there is a \textsf{CMSO}-sentence $\psi$ on rooted graphs such that ${\bf G}\in\Pi$ iff ${\bf G}\models \psi.$ 
Notice that \cref{proog_tw}, readily implies that every {\sf CMSO}-expressible rooted graph problem with the insulation property, can be solved in quadratic time on bounded-genus bounded-boundary rooted graphs:
indeed, to see this we apply \cref{proog_tw} in order to detect some big enough railed nest 
and we discard all vertices inside its inner cycle from $G$. After applying 
this step a linear number of times, the treewidth will become bounded by some function of $g$ and $t$
and it will be possible to apply the decision version of Courcelle's theorem in order to solve the problem in quadratic time. 
The main contribution of our paper is that any such problem can be solved in linear time.

\subsection{Formal description of our result}

\paragraph{The insulation property.}
Let $\Pi$ be a rooted graph problem.
Given a rooted graph ${\bf G},$ a vertex $v\in \overline{V}({\bf G})$ is \emph{$\Pi$-irrelevant} if ${\bf G}\in\Pi$ if and only if ${\bf G}-v\in\Pi.$

We say that a rooted graph problem $\Pi$ has the \emph{insulation property} if 
\begin{quote}
\textsl{there exists some constant $c_{\Pi}$ such that, for every $\Sigma$-embedded rooted graph ${\bf G}$
with a $c_{\Pi}$-railed nest $W,$ if $R({\bf G})\subseteq {\sf ext}(W)\cap V({\bf G})$ then every vertex in ${\sf int}(W)\cap V({\bf G})$ is $\Pi$-irrelevant. }
\end{quote}
Notice that the constant $c_{\Pi}$ does not depend on the choice of the surface $\Sigma.$
We refer to this constant as $c_{\Pi}$ and we call it \emph{the insulation constant of $\Pi$}.
Given two rooted graphs ${\bf G}$ and ${\bf G}'$ we say that they are \emph{$\Pi$-equivalent} if 
${\bf G}\in\Pi\iff{\bf G}'\in\Pi.$

Our main result is the following.

\begin{restatable}{theorem}{mainthm}
\label{main_th}
There exists an algorithm that, given a non-negative integer $c,$ and an $n$-vertex $b$-rooted graph ${\bf G}$ where $\eg({\bf G})≤g,$ outputs a set $I\subseteq V({\bf G})$
such that
\begin{itemize}
\item $\tw({\bf G}-I)=\Ocal(c\cdot (g+1)^{5/2}\cdot (b+1)^{3/2})$ and 
\item for every rooted graph problem $\Pi$ 
that has the insulation property with insulation constant at most $c,$ it holds that ${\bf G}$ and ${\bf G}-I$ are $\Pi$-equivalent. 
\end{itemize}
Moreover, the algorithm runs in time $\Ocal_{c}(n).$
\end{restatable}

Using~\cref{main_th} and the decision version of \cref{ow_dec_dource} we get the following corollary.

\begin{corollary}
\label{our_col}
There is an algorithm such that, for every rooted graph problem $\Pi$ that has the insulation property with insulation constant $c$ and is {\sf CMSO}-expressible by a formula $\varphi$ and every $n$-vertex $b$-rooted graph ${\bf G}$ where $\eg({\bf G})≤g,$ decides whether ${\bf G}$ belongs to $\Pi$ in time $\Ocal_{g,b,c,|\varphi|}(n).$
\end{corollary}

\subsection{Outline of the proof of \texorpdfstring{\cref{main_th}}{\cref{main_th}}}
\label{outline_pr}

We provide here the main ideas of the proof of \cref{main_th}. The proof requires several concepts on embedded graphs that 
we only present informally here. All formal definitions, statements, and proofs are available in \cref{aligned_cycles,lamnitw_section,slice_section,removing_irr_sec,algo_section}.

\paragraph{Radial distance decompositions} We first present how the algorithm of \cref{main_th} works in the special case where $G$ is a plane graph, i.e., it is $\Sbbb^{2}$-embedded and where there is only one root vertex $u_{\rm root}.$ 
We use $(G,u_{\rm root})$ in order to denote problem instances of this type.
The root $u_{\rm root}$
is useful in order to orient cycles of $G$ that do not meet it: each cycle 
defines two open disks and its \emph{exterior} is the open disk that contains $u_{\rm root}$ while its \emph{interior} is the open disk that does not contain $u_{\rm root}.$ We also need the concept of a \emph{radial} graph $R_{G}$ of $G$ that is a bipartite graph whose vertices are the vertices and the faces of $G$
and where adjacency expresses incidence between the corresponding faces and vertices. The definition of $R_{G}$ naturally 
defines an embedding of $R_{G}$ in $\Sbbb^{2}$ 
and can be extended for graph embeddings on any fixed surface. 
A path in $R_{G}$ defines a 
\emph{radial path} in $G$ between the corresponding vertices or faces.
Accordingly, the distance in $R_{G}$ between two vertices defines the \emph{radial distance} in $G$ between the corresponding 
vertices or faces.

We next introduce the \emph{radial distance decomposition} of $(G,u_{\rm root})$ as the triple $(T,t_{0},\chi)$ where $(T,t_{0})$ is a rooted tree, the ``bags'' $\{\chi(t): t\in V(T)\}$ form a partition of $V(R_{G})$ where $\chi(t_{0})=\{u_{\rm root}\},$ for every $t\in V(T)$ in distance 
$r$ from $t_{0},$ in $T,$ all vertices in $\chi(t)$ are in radial distance $r$ from $u_{\rm root}$ in $G.$ We also require that if two vertices $v,v'$ or faces $f,f'$ 
have the same, say $r,$ radial distance from $u_{\rm root}$ and there is a 
radial path between them all whose vertices/faces
have radial distance $≥r$ from $u_{\rm root},$ then both $v,v'$ (resp. $f,f'$) belong to the same bag.
See~\cref{fig_rddtree} for an illustration.
The construction of the radial distance decomposition of $(G,v)$ can be done based on a BFS transversal of ${R}_{G},$ in linear time. We see all edges of the rooted tree $(T,t_{0})$ as being directed away from the root, i.e., it is an out-arborescence. A {\em vertex-face edge} of $T$ is an edge $(v,f)$ 
where $\chi(v)$ consists of vertices and $\chi(f)$ consists of faces of $G.$ Radial distance decompositions offer the guiding underlying structure for our algorithm. Similar schemes for decomposing $V(G)$ in layers based on their radial distance from a root have already been employed several times in the algorithmic study of planar graphs; see e.g.~\cite{Reed95rooted,ChoOhOh2023PlanarDisjointPaths,WlodarczykZehavi2023PlanarDisjointPaths,JansenPvL21,JansenW25lossy}.
Our definition is essentially corresponds to \emph{outerplanar decompositions}, defined in~\cite[Definition 5.1]{JansenPvL21} (see also~\cite[Definition~6.7]{JansenW25lossy}).

\begin{figure}[ht]
	\centering
	\scalebox{.9}{\begin{tikzpicture}[ scale=0.24]
		
		\node[big diamond node, fill= cb1,
		 ] (A) at (-3,5) {};
		\node[big diamond node, fill= cb1,
		 ] (B) at (0,20) {};
		\node[big triangle node, fill= cb1,
		 ] (C) at (0,5) {};
		\node[big black node, fill= cb1,
		] (D) at (1,13) {};
		\node[big black node,
		fill= cb1,
		] (E) at (2,7.5) {};
		\node[big black node, 
		fill= cb1,
		] (F) at (5,10) {};
		\node[big white node, label = {right:$u_{\mathrm{root}}$}
		] (G) at (5,5) {};
		\node[big diamond node,
		fill= cb1,
		] (H) at (9,13) {};
		\node[big diamond node,fill= cb1,
		] (I) at (10,0) {};
		\node[big black node,
		fill= cb4, 
		] (J) at (13,15) {};
		\node[big black node,
		fill= cb4, 
		] (J1) at (11,17) {};
		\node[big black node,
		fill= cb4, 
		] (J2) at (14,17.5) {};
		\node[big black node,
		fill= cb4,
		] (J3) at (11.5,10) {};
		\node[big black node,
		fill= cb4, 
		] (J4) at (14.5,12) {};
		
		\node[big diamond node, 
		fill= cb4,
		] (K) at (17,16) {};
		\node[big diamond node,
		fill= cb4, 
		] (L) at (17,13) {};
		\node[big diamond node, 
		fill= cb4,
		] (M) at (15,2) {};
		\node[big black node, 
		fill= cb6,
		] (N) at (19,10) {};
		\node[big black node, 
		fill= cb6,
		] (O) at (19,6.5) {};
		\node[big diamond node, 
		fill= cb8,
		] (P) at (21,8) {};
		\node[big black node,
		fill= cb6,
		] (Q) at (22,5) {};
		\node[big black node, 
		fill= cb6, 
		] (R) at (21,1) {};
		\node[big diamond node, 
		fill= cb4, 
		] (S) at (24,-2) {};
		\node[big black node,
		fill= cb6, 
		] (T) at (24.5,3) {};
		\node[big black node,
		fill= cb6, 
		] (U) at (24,9) {};
		\node[big diamond node,
		fill= cb4, 
		] (V) at (29,10) {};
		\node[big diamond node,
		fill= cb6, 
		] (X) at (25,15) {};
		\node[big diamond node,
		fill= cb6,
		] (W) at (21,16) {};
		\node[big diamond node, 
		fill= cb6,
		] (Y) at (24,17.5) {};
		\node[big diamond node,
		fill= cb4,
		] (Z) at (26,20) {};
		\node[big black node, 
		fill= cb6,
		] (UQ) at (24,6.5) {};
		
		\node[
			] (f1) at (4,5.5) {};
		\node[
		] (f2) at (4,13) {};
		\node[
		] (f3) at (8,6) {};
		\node[
		] (f4) at (3,9) {};
		\node[
		] (f5) at (8,15) {};
		\node[
		] (f6) at (18,18) {};
		\node[
		] (f7) at (13,5) {};
		\node[
		] (f8) at (1,1) {};
		\node[
		] (f9) at (12.7,15.2) {};
		\node[
		] (f10) at (13,11) {};
		\node[
		] (f11) at (21,17.5) {};
		\node[
		] (f12) at (28,12.5) {};
		\node[
		] (f13) at (24.7,16.2) {};
		\node[
		] (f14) at (26.5,7.5) {};
		\node[
		] (f15) at (18,2) {};
		\node[
		] (f16) at (17.5,8) {};
		\node[
		] (f17) at (20,6.5) {};
		\node[
		] (f18) at (22.5,5.5) {};
		\node[
		] (f19) at (22.5,2) {};

		\begin{scope}[on background layer]
		
		\shade[inner color=cb3, outer color=white] (0,20) circle (3cm);
		\shade[inner color=cb3, outer color=white] (-3,5) circle (3.1cm);
		\shade[inner color=cb3, outer color=white] (24,-2) circle (3cm);
		\shade[inner color=cb3, outer color=white] (10,0) circle (3cm);
		\shade[inner color=cb3, outer color=white] (29,10) circle (3.1cm);
		\shade[inner color=cb3, outer color=white] (26,20) circle (3cm);
		
		\fill[cb3] plot[smooth] coordinates {(-3,5) (0,20) (26,20) (29,10) (24,-2) (10,-1) };

		\tikzBox{%
			\tikz\shade[bottom color=cb3, top color=white] (0,20) -- (26,20) -- (26,23) -- (0,23) -- cycle;%
		}
		\path[rotate=0,transform shape]
		(13,21.5) node{\usebox\mybox};

		\tikzBox{%
			\tikz\shade[bottom color=cb3, top color=white] (0,20) -- (15,20) -- (15,23) -- (0,23) -- cycle;%
		}
		\path[rotate=78.1,transform shape]
		(12,5.6) node{\usebox\mybox};
		
		\begin{scope}
		\tikzBox{%
			\tikz\shade[bottom color=cb3, top color=white] (0,20) -- (14,20) -- (14,23) -- (0,23) -- cycle;%
		}
		\path[rotate=159,transform shape] (-2.5,-2) node{\usebox\mybox};
		\end{scope}
		
		\begin{scope}
		\tikzBox{%
			\tikz\shade[bottom color=cb3, top color=white] (0,20) -- (14,20) -- (14,23) -- (0,23) -- cycle;%
		}
		\path[rotate=172,transform shape] (-17,0.2) node{\usebox\mybox};
		\end{scope}
		
		\begin{scope}
		\tikzBox{%
			\tikz\shade[bottom color=cb3, top color=white] (0,20) -- (12.5,20) -- (12.5,23) -- (0,23) -- cycle;%
		}
		\path[rotate=247.5,transform shape] (-13.6,24.5) node{\usebox\mybox};
		\end{scope}
		
		\begin{scope}
		\tikzBox{%
			\tikz\shade[bottom color=cb3, top color=white] (0,20) -- (10.3,20) -- (10.3,23) -- (0,23) -- cycle;%
		}
		\path[rotate=-73,transform shape] (-6.3,32.25) node{\usebox\mybox};
		\end{scope}

		\filldraw[fill=cb2] (-3,5) -- (0,20) -- (9,13) -- (5,5) -- (5,10) -- (1,13) -- (2,7.5) -- (0,5) -- cycle;

		\filldraw[fill=cb3] (1,13) -- (2,7.5) -- (5,10) -- cycle;
		
		\filldraw[fill=cb2] (-3,5) -- (0,5) -- (2,7.5) -- (5,10) -- (5,5) -- (10,0) -- cycle;
		
		\filldraw[fill=cb2] (10,0) -- (5,5) -- (9,13) -- (10,0) -- cycle;
		
		\filldraw[fill=cb3] (0,20) -- (11,17) -- (13,15) -- (11.5,10) -- (10,0) -- (9,13) -- cycle;
		
		\filldraw[fill=cb3] (0,20) -- (26,20) -- (17,16) -- (14,17.5) -- (11,17) -- cycle;
		
		\filldraw[fill=cb5] (13,15) -- (11,17) -- (14,17.5) -- cycle;
		
		\filldraw[fill=cb5] (11.5,10) -- (14.5,12) -- (13,15) -- cycle;
		
		\filldraw[fill=cb5] (17,16) -- (26,20) -- (24,17.5) -- (21,16) -- cycle;
		
		\filldraw[fill=cb5] (26,20) -- (29,10) -- (17,16) -- (21,16) -- (25,15) -- (24,17.5) -- cycle;
		
		\filldraw[fill=cb7] (25,15) -- (24,17.5) -- (21,16) -- cycle;
		
		\filldraw[fill=cb3] (11.5,10) -- (14.5,12) -- (13,15) -- (14,17.5) -- (17,16) -- (29,10) -- (17,13) -- (15,2) -- (24,-2) -- (10,0) -- cycle;
		
		\filldraw[fill=cb5] (17,13) -- (19,10) -- (19,6.5) -- (15,2) -- cycle;
		
		\filldraw[fill=cb5] (17,13) -- (29,10) -- (24,-2) -- (21,1) -- (24.5,3) -- (22,5) -- (24,9) -- (19,10) -- cycle;
		
		\filldraw[fill=cb5] (15,2) -- (19,6.5) -- (22,5) -- (21,1) -- (24,-2) -- cycle;
		
		\filldraw[fill=cb7] (19,10) -- (19,6.5) -- (22,5) -- (21,8) -- cycle;
		
		\filldraw[fill=cb7] (19,10) -- (24,9) -- (24,6.5) -- (22,5) -- (21,8) -- cycle;
		
		\filldraw[fill=cb7] (22,5) -- (24.5,3) -- (21,1) -- cycle;
		
		\end{scope}

		\end{tikzpicture}}%
		 \hspace{4mm} \scalebox{1.0}{
 \begin{tikzpicture}[scale=0.17]
\node[white node, label={$t_{0}$}] (S0) at (-5,6) {};
 \node[square node, fill=cb2,
 ] (F0) at (0,6) {};
 \node[line width=1pt,black node,
 fill=cb1, 
 ] (A1) at (5,10) {};
 \node[line width=1pt,diamond node,
 fill=cb1, 
 ] (A2) at (5,4.5) {};
 \node[line width=1pt,triangle node,
 fill=cb1, 
 ] (A3) at (5,1) {};
 \node[square node,
 fill=cb3, 
 ] (B1) at (10,11) {};
 \node[square node,
 fill=cb3, 
 ] (B2) at (10,4.5) {};
 \node[line width=1pt,diamond node,
 fill=cb4, 
 ] (C1) at (15,7) {};
 \node[line width=1pt,black node,
 fill=cb4, 
 ] (C2) at (15,2) {};
 \node[square node,
 fill=cb5, 
 ] (D1) at (20,10) {};
 \node[square node,
 fill=cb5, 
 ] (D2) at (20,5) {};
 \node[square node,
 fill=cb5, 
 ] (D3) at (20,1) {};
 \node[square node,
 fill=cb5, 
 ] (D4) at (20,-2.5) {};
 \node[line width=1pt,black node,
 fill=cb6, 
 ] (E1) at (25,13) {};
 \node[line width=0.5pt,diamond node, 
 fill=cb6,
 ] (E2) at (25,3.5) {};
 \node[square node,
 fill=cb7, 
 ] (F1) at (30,14.5) {};
 \node[square node, 
 fill=cb7,
 ] (F2) at (30,11) {};
 \node[square node, 
 fill=cb7,
 ] (F3) at (30,2) {};
 \node[line width=1pt,diamond node,
 fill=cb8, 
 ] (G) at (35,10) {};

 \path (1,-10) -- (0,-10);
 \begin{scope}[on background layer]
	\draw[line width=2.5pt] (S0.center) -- (F0.center);
 	\draw[black] (F0) -- (A1) (F0) -- (A2) (F0) -- (A3);
 	\draw[line width=2.5pt] (A1.center) -- (B1.center) (A2.center) -- (B2.center);
 	\draw[black] (B2) -- (C1) (B2) -- (C2);
 	\draw[line width=2.5pt] (C1.center) -- (D1.center) (C1.center) -- (D2.center) (C2.center) -- (D3.center) (C2.center) -- (D4.center);
 	\draw[black] (D1) -- (E1) (D2) -- (E2);
 	\draw[line width=2.5pt] (E1.center) -- (F1.center) (E1.center) -- (F2.center) (E2.center) -- (F3.center);
 	\draw[black] (F2) -- (G);
 \end{scope}

 \end{tikzpicture}}
	\caption{\textbf{Top}: An example of a pair $(G,u_{\mathrm{root}})$ and an illustration of the bags of the radial distance decomposition of it. Vertices/faces of the same color have the same radial distance from $u_{\mathsf{root}}$ in $G.$ In this example, same colored faces that share an edge are attributed to the same bag of the radial distance decomposition. Also, vertices depicted with the same color and the same style belong to the same bag of the radial distance decomposition.
 \textbf{Bottom}: The rooted tree $(T,t_{0})$ of the radial distance decomposition of the graph on the left. Every node $t$ of $T$ is associated with a set $\chi(t)$ of vertices or faces of $G$ and is depicted as a square if $\chi(t)$ consists of faces of $G$ and as a diamond/disk/triangle if $\chi(t)$ consists of vertices of $G.$
 The color/style of each node $t$ of $T$ matches the color/style of the corresponding vertices and faces in $G$ that {belong to} $\chi(t).$ The bold edges of $T$ are its vertex-face edges.}
	\label{fig_rddtree}
\end{figure}

\begin{figure}[ht]
	\centering
	\includegraphics[width=4.3cm]{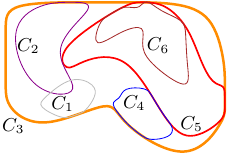}
	\caption{The collection $\{C_{1},\ldots,C_{6}\}$ is not laminar. However, it becomes laminar if we remove from it either $C_{1}$ or $C_{2}.$ Also, $\{C_{2},\ldots,C_{6}\}$ is not strongly laminar but it becomes strongly laminar if we remove from it $C_{5}$ or $C_{6}.$}
	\label{fig_subg}
\end{figure}

As can be observed in \cref{fig_rddtree}, and formally proved in \cref{aligned_cycles}, 
each vertex-face edge $e$ of $T$ corresponds to a cycle $C_{e}$ of $G$ (\cref{segments}).
Moreover, an important property of a radial distance decomposition
is that the $r$ vertex-face edges of a (directed) path of length $2r$ in $(T,t_{0})$
correspond to a sequence of nested and pairwise vertex-disjoint cycles of $G$ (\cref{pensions}).
Similar statements have been proven in~\cite{ChoOhOh2023PlanarDisjointPaths,WlodarczykZehavi2023PlanarDisjointPaths} (see also~\cite[Lemma~5.3]{JansenPvL21}). In this paper, we prove our own versions in order to specify that the obtained cycles correspond to vertex-face edges of the decomposition.
These cycles are called \emph{aligned} and will be the ``vital space'' where our algorithm will look for disjoint paths that will form the railed nests where the insulation property will be applied. 
Another property of radial tree decompositions that we need is that every two aligned cycles, i.e., cycles corresponding to vertex-face edges of $(T,t_{0}),$ are \emph{laminar}, i.e., no two points of the one cycle are separated by the other (see~\cref{subsec_laminar} for a formal definition of laminarity and \cref{fig_subg} for an illustration). 

\begin{figure}[ht]
 \begin{center}
 \scalebox{.77}{\includegraphics{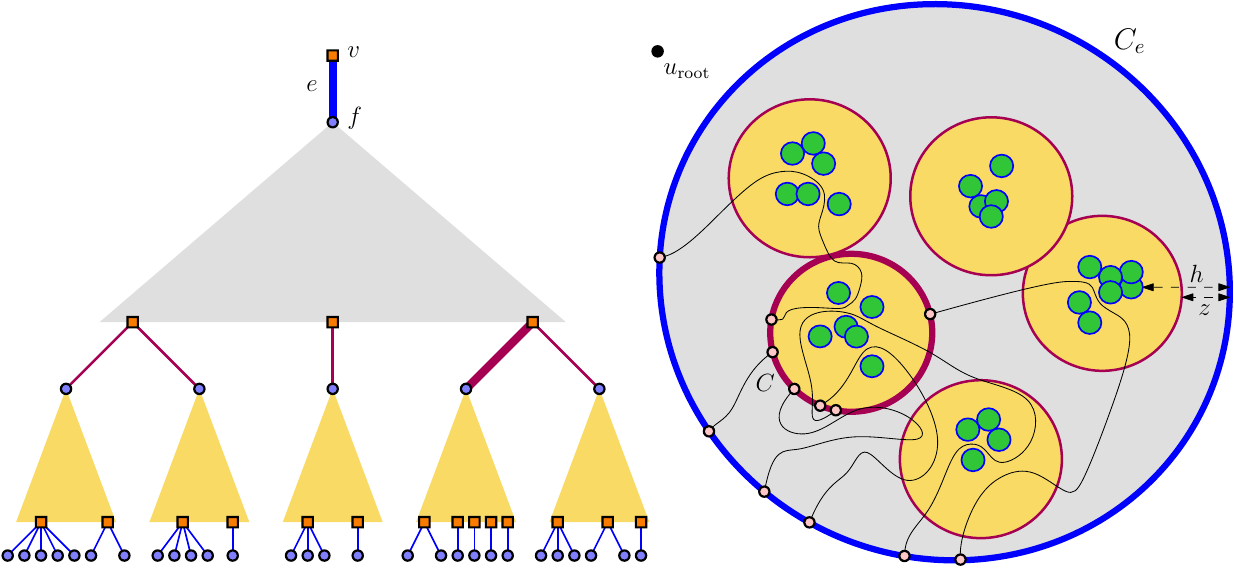}}
 \end{center}
 \caption{{\bf Left}: Some part of the radial distance decomposition of a planar single-rooted graph. The nodes corresponding to vertex sets are depicted as \darkorange{orange} squares and the nodes corresponding 
 to sets of faces are depicted as blue disks. Each vertex-face edge corresponds to a same-color cycle of $G.$ {\bf Right}: The visualization of an embedding of ${\sf Slice}_{G}^{h}(C_{e})$ in a pseudo-disk $\Psi$ where $C_{e}$ is the bold blue cycle on the outer boundary of $\Psi$ ($\mathsf{Slice}_G^h(C_e)$ is the union of the gray and the yellow regions). The cycles in $\mathsf{Cycles}(C_{e},z)$ are depicted in \red{red} while $\mathsf{Slice}_{G}^z(C_e)$ is the gray region.
 The cycles $\mathsf{Cycles}(C_{e},h)$ are depicted in \myblue{blue} and are bounding the \darkgreen{green} regions, which are their interiors.
 The black paths are disjoint paths in $G$ from the vertices of some cycle $C$ of $\mathsf{Cycles}(C_{e},z),$ depicted in bold red to the vertices of $C_e.$ As asserted by \cref{more_reroute_linkage}, all these paths can be assumed to be embedded inside $\Psi,$ i.e., they do not meet the vertices of the interiors of the cycles in $\mathsf{Cycles}(C_{e},h)$ (depicted in \darkgreen{green}).}
 \label{torodop}
\end{figure}

\paragraph{Slices.} Let $e=(v,f)$ be
some vertex-face edge of $T$ where $v\neq t_{0}.$ For some $h\in\mathbb{N}_{≥2}^{\mathsf{even}},$ 
we define $\mathsf{Cycles}(C_{e},h)$ as the set containing every cycle $C_{e'}$ of $G$ 
corresponding to a vertex-face edge $e'=(v',f')$
where the distance in $T$ between $v$ and $v'$ is $h.$
As all cycles in $\{C_{e}\}\cup\mathsf{Cycles}(C_{e},h)$ are pairwise laminar we may consider a subset of the plane, called \emph{pseudo-disk}, obtained by removing all open interiors of the cycles in $\mathsf{Cycles}(C_{e},h)$ from the closed interior of $C_{e}.$ We define ${\sf Slice}_{G}^{h}(C_{e})$ as the part of the embedding of $G$ that is embedded inside this pseudo-disk (see~\cref{torodop}).

By considering the set $E_{h}$ of vertex-face edges $(v,f)$ of $T$ 
where $\dist_{T}(u_{\rm root},v)\equiv 2 \pmod h,$ we may 
decompose $G$ into the set of slices $\{{\sf Slice}_{G}^{h}(C_{e})\mid e\in E_{h}\}.$
Notice that the union of these slices is the graph $G$ without the root vertex $u_{\rm root}.$
Let us recall at this point that our plan is to sketch how to solve the problem for instances consisting of planar single-rooted graphs. 
Later in the end of this section we describe how to reduce to this special case the general case where we have many roots and surface-embedded graphs.
In that approach, a general instance is viewed as a planar single-rooted instance $(G,u_{\rm root})$ that is the ``contracted version'' of the original rooted graph where $u_{\rm root}$ represents the contracted territory.

A key idea of the algorithm for the planar single-rooted case is to process each of the slices separately and certify the existence inside them of big-enough 
railed nests that justify the elimination of (globally) irrelevant vertices from each individual slice. For a visualization of six slices 
and the way they correspond to subtrees of the radial tree decomposition, see \cref{torodop}.

\paragraph{Flows and railed nests.} Assume that the problem $\Pi$ 
that we want to solve has the insulation property for some constant $c_{\Pi}.$ In our arguments the choice of $h$ will depend on $c_{\Pi}.$
We set $z:=2(c_{\Pi}-1).$
Let $e=(v,f)$ be a vertex-face edge of $T$ and $B:={\sf Slice}_{G}^{z}(C_{e}).$ In \cref{torodop} $C_{e}$ is depicted in bold blue and the cycles in $\mathsf{Cycles}(C_{e},z)$ are depicted in red. Let $C\in \mathsf{Cycles}(C_{e},z)$ and recall that $B$ contains 
a nested sequence ${\cal C}=\{C_{1},\ldots,C_{c_{\Pi}}\}$ of cycles where $C=C_{1}$ and $C_{e}=C_{c_{\Pi}}.$ If additionally $G$ contains a collection ${\cal P}=\{P_{1},\ldots,P_{c_{\Pi}}\}$ of $c_{\Pi}$-many pairwise vertex-disjoint paths from $V(C_{e})$ to $V(C),$ then $W:=\cupall\Ccal\cup\cupall\Pcal$ forms a $c_{\Pi}$-railed 
nest justifying that all vertices in the open interior of $C$ are $\Pi$-irrelevant. Our plan is to find, if it exists, such a collection of $c_{\Pi}$-many pairwise vertex-disjoint paths 
for every cycle $C\in\mathsf{Cycles}(C_{e},z).$ Clearly, these paths are not necessarily paths of $B$ and it would cost too much time to look for these disjoint 
paths in \textsl{the whole} of $G.$ Instead, we prove that it is enough to
look for disjoint paths in $D={\sf Slice}_{G}^{z+w}(C_{e})$
where $w=2c_{\Pi}$

.
We would like to stress that, while a flow (in $G$) from $C_e$ to $C$ can obviously be assumed to not enter the open interior of $C,$ it \textsl{cannot} be assumed to avoid the open interiors of any \textsl{other} cycle $C'\in\mathsf{Cycles}(C_{e},z)$ different from $C$ (denoted by thin red in~\cref{torodop}). However, we can ensure that such a flow can avoid the \textit{open interior of every cycle in $\mathsf{Cycle}(C',w)$} (denoted in thin blue in~\cref{torodop}).
Indeed, here we make use of the fact that between any cycle $C'$ in $\mathsf{Cycles}(C_{e},z)$ other than $C$ and the cycles of $\mathsf{Cycles}(C',w)$
there is always a nested sequence of $c_\Pi$ pairwise vertex-disjoint cycles. The existence of this nested sequence of cyclescombined with Menger's theorem shows that if there is a $c_{\Pi}$-flow from $C$ to $C_{e}$ in $G$ then there is such a flow also \textsl{inside} $D.$ Let ${\cal Q}$ be the set of cycles in $\mathsf{Cycles}(C_{e},z)$
that send a $c_{\Pi}$-flow to $C_{e}.$

For every individual $C\in{\cal Q},$ 
we immediately know that all vertices in the interior of $C$ are $\Pi$-irrelevant and we certainly could safely remove them.
However, discarding \textsl{only one} such set of $\Pi$-irrelevant vertices is not enough for deriving a sub-quadratic algorithm.%
At this point, one might be tempted to remove 
the interiors of \textsl{all} cycles in $\Qcal.$
However, this is unsafe as the vertices that we remove from the interior of some particular $C\in\Qcal$
might 
 be vertices of the $c_{\Pi}$-flow from $C'$ to $C_{e}$
 certifying the inclusion of some $C'\neq C$ in $\Qcal.$
In other words, discarding the interior 
of $C$ might entail that the vertices in the interior of $C'$ lose the property of being $\Pi$-irrelevant.

Our way out of this is to discard the vertices in the union $\Lambda$ of the open interiors of all cycles in $\mathsf{Cycles}(C,w),$ for every $C\in\Qcal.$
This is safe as all $|\Qcal|$ aforementioned collections of paths, as argued above, avoid 
$\Lambda.$
 We will argue that this elimination is ``massive''.

\paragraph{Discarding irrelevant vertices in slices.}
Let $h:=z+w.$
We just argued that all collections of $c_{\Pi}$-many pairwise vertex-disjoint paths in $G$ 
between $C_{e}$ and the cycles of $\mathsf{Cycles}(C_{e},z)$
also exist in $D={\sf Slice}_{G}^{h}(C_{e})$ and that this permits us to safely 
discard the interiors (drawn in green in \cref{torodop}) of the cycles in $\mathsf{Cycles}(C,w),$
for every $C\in\Qcal.$ Our next step is to find ${\cal Q}$ in time that is linear in the size of the slice $D.$ For this, we exploit 
 the fact that $D={\sf Slice}_{G}^{h}(C_{e})$ is a planar graph of bounded radial diameter, therefore its treewidth is $\Ocal(h)$ (see e.g., \cite{Eppstein2000DiameterTreewidth,DemaineHajiaghayi2004DiameterTreewidthRevisited,Bodlaender1988SomeClasses}).
 We next add in $D,$ for every $C\in \mathsf{Cycles}(C_{e},z)\cup\{C_{e}\},$ a new vertex $v_{C}$ adjacent to all vertices of $C$ and, in \cref{lamnitw_section}, we prove that this enhancement does not significantly increase the treewidth of the resulting graph $D^{*}.$ Then 
we ask for the maximum size of a set $S\subseteq \{v_{C}\mid C\in \mathsf{Cycles}(C_{e},z)\}$ such that $v_{C_{e}}$ sends a flow of $c_{\Pi}$-many pairwise vertex-disjoint paths to all the vertices of $S.$ As this question can be expressed as an optimization query in \textsf{CMSO}, \cref{ow_dec_dource} implies that one
 may compute the set $S$ in time $\Ocal_{c_{\Pi}}(|D|)$ (\cref{lemma_courcelle}). As every vertex in $S$ corresponds to a cycle in $\mathsf{Cycles}(C_e,z),$ from $S$ we can easily derive the collection $\mathcal{Q}.$ Knowing $\mathcal{Q},$ we can safely discard from $G$ all vertices that are embedded inside $\Lambda.$

\paragraph{Bounding the treewidth.}
We apply the above procedure for all slices of depth $h$ 
corresponding to the vertex-face edges of the above-defined set $E_h,$

starting from the edges of $E_h$ that are closer to the root. Recall that a vertex of a slice can appear multiple times only if it belongs to the ``boundaries'' of different slices. An elementary use of Euler's formula implies that the total running time for processing all the slices is linear in the size of $G.$ This procedure produces a set $I$ of irrelevant vertices whose removal produces a 
$\Pi$-equivalent instance $(G-I,u_{\rm root}).$ We use 
$(T,t_{0},\chi)$ for the (remaining) radial distance decomposition. Our next step is 
to observe that the new instance $(G-I,u_{\rm root})$ satisfies the following property: \textsl{for every directed path of $T$ with vertices $v_{1},f_{1},\ldots,v_{2h},f_{2h}$
there are no $c_{\Pi}$ disjoint paths between $C_{(v_{1},f_{1})}$ and $C_{(v_{2h},f_{2h})}$}. The proof of this property is in \cref{bstwso9o} and, intuitively, is based on the fact that such paths would contain
vertex-face edges corresponding to cycles that should have already been eliminated according to the above procedure. Next, again in \cref{bstwso9o}, we prove that this property implies the desired bound on the treewidth of $G-I.$

\paragraph{From single-rooted and planar to multi-rooted and surface embedded.}
What we just sketched is the proof that \cref{main_th} holds for single-rooted and planar instances.
This does not directly extend to graphs embedded in surfaces of higher Euler genus. In particular, when the graph is not planar, we cannot ensure the cycle structure given by the radial distance decomposition. We next explain how to reduce from (general) surface-embedded graphs to planar graphs. This comes with additional tricks.

Assume now that ${\bf G}$ is a $t$-rooted graph embedded in some surface $\Sigma$ where $\eg(\Sigma)≤g.$ We next prove 
that one may find, in linear time, a collection ${\cal P}$ of $\Ocal(g+t)$ shortest paths in the radial graph $R_{G}$ of $G$ 
meeting all roots of ${\bf G}$ and such that if we contract them to a single vertex in $R_{G},$ the resulting graph is planar. This is proved in \cref{main_tw} using as a departure point the algorithmic results of Cabello, Colin de Verdière, and Lazarus in \cite{CabelloCdVL12algo}.
Using these radial paths of $G,$ we find, also in linear time, a connected set $J$ of edges of $G$
such that if we contract them to a single vertex $u_{\rm root}$ the 
resulting graph $G':=G/J$ is planar. Next we 
 apply the aforementioned irrelevant-vertex elimination procedure on $(G',u_{\rm root})$ and we prove that the set $I$ of irrelevant vertices for $(G',u_{\rm root})$ is also a set of irrelevant vertices for ${\bf G}.$
Intuitively, this holds because all railed-nests that justified the inclusion
of vertices in $I,$ while processing $G',$ are also present in $G.$ What remains is to 
prove that the bound on the treewidth of $G'-I$ also implies a bound on the treewidth of $G-I.$ For this, we use the fact (guaranteed by the choice of $J$) that all edges in $J$
that have been contracted towards creating $G'$ have endpoints 
that are within 
radial distance two from some of the vertices of the $\Ocal(g+t)$ radial shortest paths in $\Pcal.$
Based on this fact and making use of the results of Demaine, Hajiaghayi, and Kawarabayashi in \cite{DemaineHK11cont}, we prove in \cref{main_tw} that 
the contraction of the edges $J$ does not decrease the treewidth of 
$G-I$ by more than a constant, depending on $t$ and $g.$
This implies the bound of \cref{main_th} for general surface-embedded rooted graphs.

\section{Radial distance decompositions and nested cycle sequences}
\label{aligned_cycles}

In this section we give the formal definition of radial distance decompositions of pairs $(G,u)$ and show that each vertex-face edge of the underlying tree corresponds to a cycle of $G,$ which we call \emph{$u$-aligned}.

\subsection{Radial distance decompositions}
\label{radial_dec}

We start by giving the formal definition of radial graphs of surface-embedded graphs.
In the rest of the paper, we always assume that the considered embedding of every given $\Sigma$-embedded connected graph $G$ is a \emph{$2$-cell embedding}, i.e., every face is homeomorphic to an open disk. 
A \emph{bridge} in a graph $G$ is an edge $e$ such that 
$G-e$ has more connected components than $G.$ A graph is \emph{$2$-edge connected} if the removal of any edge leaves the graph connected. Equivalently, a graph is 2-edge connected if it is connected and has no bridges.

\paragraph{Radial graphs.}
Let $\Sigma$ be a surface and let $G$ be a $\Sigma$-embedded connected graph.
We define the {\em radial graph} of $G,$ denoted by $R_{G},$ as the 
graph whose vertex set is $V(G) \cup F(G)$ and whose edge set is defined as follows:
for every $f\in F(G)$ we consider the closed walk of $G$ defined by the boundary 
of $f$ and we make $f$ adjacent to all vertices
in this walk (we permit multiple edges as a vertex can appear many times in the walk).

Note that $R_{G}$ is a $\Sigma$-embedded graph that is bipartite and connected. 
A \emph{radial path} in $G$ is a path between two vertices
in $R_{G}.$ Notice that each endpoint of such a path may correspond either to a vertex or a face of the embedding of $G.$
The \emph{radial distance} of two disjoint cycles $C,C'$ of $G$ is the length of the shortest radial path with one endpoint in $V(C)$ and one endpoint in $V(C').$

We give the following straightforward observation for radial graphs of connected and bridgeless graphs embedded on $\Sbbb^{2}.$

\begin{observation}\label{obs_squareface}
	Let $G$ be a $\Sbbb^{2}$-embedded graph. If $G$ is connected and bridgeless, then every face of $R_{G}$ is incident to exactly four vertices. 
\end{observation}

\paragraph{Tree distance decompositions.}
Before presenting the definition of tree distance decompositions and radial distance decompositions, let us introduce some additional notation and definitions.
Given two vertices $x,y$ of a graph $G,$ we denote by $\dist_{G}(x,y)$ the minimum number of edges of a path in $G$ with endpoints $x$ and $y$; if there is no such path, we set $\dist_{G}(x,y)=\infty.$
Also, an \emph{out-arborescence} $(T,t_{0})$ is a directed graph $T$ and $t_{0}$ is a vertex of $T$ such that $t_{0}$ has in-degree zero and each vertex different than $t_{0}$ has in-degree exactly one. In other words, $T$ is a tree with some distinguished vertex $t_{0}$ where all edges of $T$ are ``oriented away'' from $t_{0}.$ For every edge $(t,t')\in E(T),$
we set $\mathsf{parent}(t')=t.$

Let $G$ be a connected graph and $u\in V(G).$
A {\em tree distance decomposition of $(G,u)$} is a triple $(T,t_{0},\chi),$
where $(T,t_{0})$ is an out-arborescence
and $\chi:V(T)\to 2^{V(G)},$ that satisfies the following conditions.
\begin{itemize}
 \item $\{\chi(t): t\in V(T)\}$ is a partition of $V(G),$ with $\chi(t_{0})=\{u\},$
 \item %

 for every $v\in V(G),$ if $v\in \chi(t),$ there is a $z\in \chi(\mathsf{parent}(t))$ so that $vz\in E(G)$ and $\mathsf{dist}_{G}(u,v)=\mathsf{dist}_{G}(u,z)+1,$ and
 \item for every $v,w\in V(G),$ if $\mathsf{dist}_{G}(u,v) = \mathsf{dist}_{G}(u,w)$ and there is a $(v,w)$-path in $G[\{z\in V(G)\mid \mathsf{dist}_{G}(u,z)\geq \mathsf{dist}_{G}(u,v)\}],$ then there is a $t\in V(T)$ such that $\{v,w\}\subseteq \chi(t).$
\end{itemize}

Tree distance decompositions (using a different formulation) have been introduced in \cite{YamazakiBFT99isom}.
Let us note that tree distance decompositions are obtained by the BFS-layering of graphs (with respect to some fixed root) after ``grouping'' vertices in every layer corresponding to connected components in a bottom-up fashion.

Observe that for every connected graph $G$ and every vertex $u\in V(G),$ there is a unique triple $(T,t_{0},\chi)$ satisfying the above conditions. This allows us to refer to $(T,t_{0},\chi)$ as \textsl{the} (unique) tree distance decomposition of $(G,u).$

\begin{lemma}
	\label{lem_comp_rdd}
	Given a connected graph $G$ and a vertex $u\in V(G),$ we can compute the tree distance decomposition of $(G,u)$ in linear time.
\end{lemma}

\begin{proof}
The tree distance decomposition of $(G,u)$ can be computed by producing a BFS-tree for $G$ rooted at $u$ and recursively storing which vertices of the same level of the search tree are connected in the graph induced by the vertices of the current subtrees. It is easy to observe that the recursive storage of this information maintains linearity of BFS.
\end{proof}
 
\paragraph{Rooted embeddings and their radial distance decompositions.} 
A \emph{rooted embedding} is every pair $(G,u),$ where 
$G$ is a $2$-edge-connected graph, accompanied with an embedding 
of it in the sphere, and $u$ is a vertex of $G.$ We refer to $u$ 
as the ``reference point'' of $(G,u).$ The $2$-edge-connected assumption in rooted embeddings facilitates our presentation
and we explain why it does not hurt generality in \cref{make_e_edge}.

From now on, we will only consider tree distance decompositions 
of $(R_{G},u)$ where $(G,u)$ is a rooted embedding.
If $(T,t_{0},\chi)$ is the tree distance decomposition of $(R_{G},u),$ we refer 
to it as the \emph{radial distance decomposition} of the embedding pair $(G,u).$
See~\cref{fig_rddtree} for an illustration of the radial distance decomposition of a rooted embedding.

Our definition is essentially the same as the one of outerplanar decompositions defined in~\cite[Definition 5.1]{JansenPvL21} and \cite[Definition~6.7]{JansenW25lossy}. Similar decomposition schemes based on radial distances have been presented in~\cite{ChoOhOh2023PlanarDisjointPaths,WlodarczykZehavi2023PlanarDisjointPaths,JansenPvL21,JansenW25lossy}. Also, in~\cite{DemaineHK11cont}, the authors define \emph{radial colorings} of graphs that correspond to BFS-layerings of their radial graphs. Their definition (as well as the ones in~\cite{ChoOhOh2023PlanarDisjointPaths,WlodarczykZehavi2023PlanarDisjointPaths}) differs from the definition of radial distance decomposition in the fact that our radial colorings do not “group” vertices in the same “bag” when they are connected in the ``suffix'' of the decomposition, i.e., radial colorings do not demand the last condition of the definition of a tree distance decomposition.

\paragraph{Vertex and face layers of a radial distance decomposition.}
Let $(T,t_{0},\chi)$ be the radial distance decomposition of 
a rooted embedding $(G,u).$
Note that since $R_{G}$ is bipartite,
for every $t\in V(T)$ that is in odd (resp. even) distance from $t_{0}$ in $T,$
it holds that $\chi(t)\subseteq F(G)$ (resp. $\chi(t)\subseteq V(G)$).

For this reason we call the set of nodes of $T$ with odd distance from $t_{0}$ a \emph{face layer} of $T$ and the set of nodes of $T$ with even distance from $t_{0}$ a \emph{vertex layer} of $T.$ We say that an edge $(v,f)$
of $T$ is a \emph{vertex-face edge} of $T$ if $v$ is in a vertex layer of $T.$ A vertex-face edge of $T$ that is incident to the root $t_{0}$ is called \emph{root edge}.
In~\Cref{app_resp} we show that there is exactly one root edge, but this is only used in the proofs in~\Cref{app_resp}.

\subsection{Aligned cycles}
\label{subsec_alinged_cycles}

In this subsection we show that, given a rooted embedding $(G,u)$ together with its radial distance decomposition $(T,t_{0},\chi),$ every non-root vertex-face edge of $T$ corresponds to a cycle $C_{e}$ of $G.$
Moreover, this cycle bounds an open disk containing all vertices and faces of $G$ that are in bags of nodes in the subtree of $T$ rooted at this edge.
In order to state this result formally, let us give some additional definitions.

Let $G$ be a $\Sbbb^{2}$-embedded graph and let $u\in V(G).$
Let also $C$ be a cycle of $G$ disjoint from $u.$
Note that $\Sbbb^{2}\setminus C$ consists of two open disks.
The \emph{interior} of $C,$ denoted by $\Delta_{C},$ is the open disk bounded by $C$ that does not contain $u.$
The \emph{open exterior} of $C,$ denoted by $\overline{\Delta}_{C},$ is the open disk bounded by $C$ that contains $u.$ In other words $u$ serves as a ``reference vertex''
that orients all cycles that it does not intersect, that way
$\Delta_{C}$ is the disk that is ``away from $u$'' and $\overline{\Delta}_{C}$ is the disk containing $u.$

Let $(T,t_{0},\chi)$ be the radial distance decomposition of a rooted embedding $(G,u).$
Given an edge $e=(t,t')$ of $T$ (in this case $t'$ is a child of $t$),
we use $\mathsf{suffix}_{\chi}(e)$ to denote the union of all $\chi(z),$ where $z$ is either $t'$ or a descendant of $t'$ in $(T,t_{0}).$

\begin{restatable}{lemma}{lemonecycle}
	\label{segments}
	Let $(T,t_{0},\chi)$ be the radial distance decomposition of a rooted embedding $(G,u).$ 
	For every non-root vertex-face edge $e=(v,f)$ of $T,$ there is a unique cycle $C_{e}$ in $G$ such that $V(C_{e})\subseteq\chi(v)$ and $V(G)\cap \mathsf{suffix}_{\chi}(e)$ is the set of vertices of $G$ in the interior of $C_{e}.$
\end{restatable}

The proof of~\cref{segments} follows from the fact that each (non-root) vertex-face edge of $T$ partitions the set of faces of $G$ into two parts, so that the union of (the closure of) all faces in each part is a disk. This can be shown by induction on the tree $T,$ using the definition of the radial distance decompositions as well as some easy observations on $\Sbbb^{2}$-embedded graphs (see also~\cite[Lemma 5.3]{JansenPvL21}); a full proof of~\cref{segments} is given in~\Cref{app_resp}.

\paragraph{Root-aligned cycles.}
Let $(T,t_{0},\chi)$ be the radial distance decomposition of 
a rooted embedding $(G,u).$
Given a non-root vertex-face edge $e$ of $T,$ we denote by $C_{e}$ the unique cycle of $G$ corresponding to it, given by \cref{segments}.
Also, we say that a cycle $C$ of $G$ is \emph{$u$-aligned} if there is a vertex-face edge $e$ of $T$ such that $C=C_{e}.$

The next observation about the relation between $u$-aligned cycles of $G$ will be used later in~\cref{removing_irr_sec}.
It follows directly from~\cref{segments} and the fact that for every two nodes $t,t'$ of the radial distance decomposition of a rooted embedding, the sets $\chi(t)$ and $\chi(t')$ are disjoint.

\begin{observation}
	\label{obs_interior}
	Let $(G,u) $ be a rooted embedding and let $C,C'$ be two $u$-aligned cycles of $G.$
	If $C'$ intersects the interior $\Delta_{C}$ of $C,$ then $C'$ is embedded in $\Delta_{C}.$
\end{observation}

Next, we show that in linear time we can dispose the set of all $u$-aligned cycles of $G$ of a given rooted embedding $(G,u).$

\begin{lemma}
	\label{lem_comp_align}
Let $(T,\chi)$ be the radial distance decomposition of 
a rooted embedding $(G,u).$ There is a function that maps each non-root vertex-face edge $e$ of $T$ to the cycle $C_{e}.$
Moreover, this function can be computed in linear time.
\end{lemma}

\begin{proof}
Due to~\cref{segments}, for every non-root vertex-face edge $e=(v,f)$ of $T,$ the cycle $C_{e}$ corresponds to the graph induced by the edges of $G$ that are incident to one face in $\chi(f)$ and one face in the union of $\chi(f'),$ for all nodes $f'\neq f$ of $T$ that are neighboring to $v$; every edge is incident to exactly two faces because $G$ is $2$-edge-connected. Therefore, the function that associates each non-root vertex-face edge $e=(v,f)$ of $T$ to the cycle $C_{e}$ can be computed by local queries on the neighborhood of the faces in $\chi(f)$ inside the union of $\chi(f'),$ for all nodes $f'\neq f$ of $T$ that are neighboring to $v,$ which in total can be done in linear time.
\end{proof}

\paragraph{Nested cycles.} Let $G$ be a planar graph
and let ${\cal C}=[C_{1},\ldots,C_{r}], r\geq 2$ be a sequence of cycles in $G.$
Given a vertex $u\in V(G)$ that is not a vertex of any of the cycles of $\Ccal,$ we call ${\cal C}$ \emph{$u$-nested}, if its cycles are pairwise vertex-disjoint and
for every $i\in[r-1],$ the interior (with respect to $u$) of $C_i$ is contained in the interior of $C_{i+1}$ 
(see~\cref{dnsamfnamsd} for an example).
We call $C_{1}$ the \emph{inner cycle} of $\Ccal$ and $C_{r}$ the \emph{outer cycle} of $\Ccal,$ always with respect to the ``orientation point'' $u.$ We also say that $\Ccal$ is \emph{nested} if 
it is $u$-nested for some $u\in V(G).$

 \begin{figure}[ht]
	\centering
	\scalebox{0.85}{\begin{tikzpicture}[scale=0.5]
	\node[black node] (A) at (7,0) {};
	\node[black node] (B) at (5,3) {};
	\node[black node] (C) at (8,3) {};
	\node[black node] (D) at (9,4.5) {};
	\node[black node] (E) at (12,4.5) {};
	\node[black node] (F) at (0,5.5) {};
	\node[black node] (G) at (7,5) {};
	\node[black node] (I) at (3,6) {};
	\node[black node] (J) at (5,6) {};
	\node[black node] (K) at (8,5.5) {};
	\node[black node] (L) at (10,5.5) {};
	\node[black node] (M) at (11,6) {};
	\node[black node] (N) at (10,7) {};
	\node[black node] (O) at (7,7) {};
	\node[black node] (P) at (6,7) {};
	\node[black node] (R) at (8,8) {};
	\node[black node] (Q) at (5,8) {};
	\node[black node] (S) at (9,8.5) {};
	\node[black node] (T) at (6,9) {};
	\node[black node] (U) at (4,9) {};
	\node[black node] (V) at (1,10) {};
	\node[black node] (W) at (7,10) {};
	\node[black node] (X) at (11,10.5) {};
	\node[black node] (Y) at (6,11) {};
	\node[black node] (Z) at (5,12) {};
	
	\draw[-,ultra thick] (A) -- (E) -- (X) -- (Z) -- (F) -- (A);
	\draw[gray] (F) -- (V) -- (Z);
	\draw[ultra thick] (I) -- (B) -- (D) -- (S) -- (W) -- (U) -- (I);
	\draw[gray] (Z) -- (Y) -- (U);
	\draw[gray] (F) -- (I);
	\draw[gray] (A) -- (C) -- (D);
	\draw[gray] (E) -- (L) -- (M) -- (N) -- (X) -- (W);
	\draw[gray] (U) -- (B) -- (G) -- (D);
	\draw[gray] (R) -- (S);
	\draw[gray] (R) -- (W) -- (T) -- (Q);
	\draw[-,ultra thick] (Q) -- (R) -- (K) -- (J) -- (Q);
	\draw[gray] (R) -- (P) -- (O) -- (R);
	\draw[gray] (K) -- (P) -- (O) -- (K);
	\begin{scope}[on background layer]
 \fill[fill=orange,opacity=0.2] (7,0) -- (12,4.5) -- (11,10.5) -- (5,12) -- (0,5.5) -- cycle;
 \fill[fill=red,opacity=0.3] (3,6) -- (5,3) -- (9,4.5) -- (9,8.5) -- (7,10) -- (4,9) -- cycle;
 \fill[fill=cyan,opacity=0.6] (5,8) -- (8,8) -- (8,5.5) -- (5,6) -- cycle;
	\end{scope} 
	\end{tikzpicture}}
	\caption{An example of a $\Sbbb^{2}$-embedded graph $G$ and a sequence ${\cal C}$ of 3 nested cycles in $G.$ The outer/inner cycle is the one bounding the yellow/blue disk.}	
	\label{dnsamfnamsd}
\end{figure}

By repeatedly applying~\cref{segments}, we show that a sequence of vertex-face edges appearing in a path of the radial distance decomposition gives rise to a sequence of nested cycles. Statements similar to the following have been proved in the full versions of~\cite{ChoOhOh2023PlanarDisjointPaths,WlodarczykZehavi2023PlanarDisjointPaths}.

\begin{lemma}
	\label{pensions}
Let $(T,t_{0},\chi)$ be the radial distance decomposition of 
a rooted embedding $(G,u).$
Let $e=(v,f)$ and $e'=(v',f')$ be two non-root vertex-face edges of $T$ where $v'$ is a descendant of $f.$ 
If $\mathsf{dist}_{T}(v,v')=2r,$ for some integer $r\ge 1,$
then there is a $u$-nested sequence $\Ccal$ of $r+1$ many $u$-aligned cycles of $G$ 
where the inner cycle of $\Ccal$ is 
$C_{e'}$ and the outer cycle of $\Ccal$ is $C_{e}.$
\end{lemma}

\begin{proof}
Let	$v_{r+1},f_{r+1},\ldots,v_{1},f_{1}$ be the vertices of the path connecting $v$ and $f'$ in $T,$ ordered in the ancestor-descendant relation, i.e., $v_{r+1}=v$ and $f_{1}=f'.$
For every $i\in[r+1],$ let $e_{i}$ denote the vertex-face edge $(v_{i},f_{i})$ of $T.$
We use $C_{i}$ to denote the cycle $C_{e_i}$ and $\Delta_{i}$ to denote the open disk $\Delta_{C_{e_i}}.$
Due to \cref{segments}, for every $i\in[r+1]$ it holds that $V(C_{i})\subseteq\chi(v_{i})$ and 
$V(G)\cap \mathsf{suffix}_{\chi}(e_{i})$ is the set of vertices of $G$ in the interior of $C_{i}.$
Since for every $i\in[r]$ we have $\chi(v_{i})\subseteq \mathsf{suffix}_{\chi}(e_{i+1}),$
we get that $C_{i}\subseteq \Delta_{i+1}.$ 
Therefore, $\Ccal=[C_{1},\ldots,C_{r+1}]$ is a $u$-nested sequence of $u$-aligned cycles of $G$ 
where $C_{1}=C_{e'}$ and $C_{r+1}=C_{e}.$ 
\end{proof}

\section{Bounding the treewidth of laminar stellations}
\label{lamnitw_section}

An important element of our approach, as explained in~\Cref{outline_pr}, is to show that the treewidth of a $\Sbbb^{2}$-embedded graph does not increase too much after extending it to what we define in \cref{subsec_laminar} as its \textsl{laminar stellation}.
This corresponds to the graph obtained from a given graph and a
collection of strongly laminar cycles (also defined in \cref{subsec_laminar}) by introducing one new vertex per cycle and making this vertex adjacent to all vertices of the corresponding cycle.
In~\Cref{subsec_laminar}, we also show that the treewidth of laminar stellations is a linear function of the treewidth of the original graph.
Before this, in~\Cref{subsec_treewidth}, we provide some preliminary facts about treewidth and surface-embedded graphs. 
We will use $P_{r}$ to denote the path on $r$ vertices and $K_{1,r}$ to denote the \emph{star graph}, i.e., the tree consisting of one vertex that is adjacent to $r$ leaves.

\subsection{Treewidth and and walls in surfaces}
\label{subsec_treewidth}

Let $G_{1}$ and $G_{2}$ be graphs. The \emph{Cartesian product} of $G_{1}$ and $G_{2},$ denoted by $G_{1} \square G_{2},$ is the graph with vertex set $V(G_{1} \square G_{2}) = V(G_{1}) \times V(G_{2}),$ where two vertices $(u,v)$ and $(u’,v’)$ are adjacent if and only if either $u=u’$ and $vv’\in E(G_{2}),$ or $v=v’$ and $uu’\in E(G_{1}).$ We use $G_{1}\square G_{2}$ to denote the Cartesian product of the graphs $G_{1}$ and $G_{2}.$
We need the following known result about treewidth products (see e.g., \cite[Lemma 19]{HickingbothamW23stru}). 

\begin{proposition}
\label{cart_pr_tw}
If $G_{1}$ and $G_{2}$ are graphs, then $\tw(G_{1}\square G_{2})≤
|G_{1}|\cdot (\tw(G_{2})+1)-1$ 
\end{proposition}

 \begin{figure}[ht]
	\centering
	\scalebox{0.9}{
\pgfmathsetmacro{\size}{8} 
\pgfmathsetmacro{\qq}{.07} 
\pgfmathsetmacro{\ww}{.12} 
\pgfmathsetmacro{\ee}{2} 
\pgfmathsetmacro{\uu}{1.8pt} 
\pgfmathsetmacro{\showl}{0} 
\pgfmathsetmacro{\oddlayercolor}{"darkorange!100"} 
\pgfmathsetmacro{\evenlayercolor}{"green!50!black"} 
\pgfmathsetmacro{\branchingcolor}{"capri!100"} 
\pgfmathsetmacro{\subdivisioncolor}{"crimsonglory!50"} 
\pgfmathsetmacro{\regularcolor}{"black!100"} 
\pgfmathsetmacro{\scale}{.65} 
\pgfmathsetmacro{\rotate}{0} 

\newcommand{\perimetry}[8]{
\begin{scope}[xscale=#7,shift={(#1,#2)},rotate=#3]
\pgfmathsetmacro{\tmy}{1-mod(#6,2)}
\draw[line width=#4, color=#8] (0,0) -- (#5,0);
\draw[line width=#4, color=#8] (\tmy,#6) -- (\tmy+#5,#6);
\pgfmathsetmacro{\tbb}{#6-1}
\foreach \y in {0,1,...,\tbb} {
\draw[line width=#4, color=#8] (0,\y) -- (1,\y);
}
\ifthenelse{\equal{#6}{\one}}{}{%
\foreach \y in {0,...,\tbb} {
\pgfmathsetmacro{\tmy}{mod(\y,2)}
\draw[line width=#4, color=#8] (\tmy,\y) -- (\tmy,\y+1);
}}
\ifthenelse{\equal{#6}{\one}}{}{%
\foreach \y in {0,...,\tbb} {
\pgfmathsetmacro{\tmy}{mod(\y,2)}
\draw[line width=#4, color=#8] (#5+\tmy,\y) -- (#5+\tmy,\y+1);
}}
\ifthenelse{\equal{#6}{\one}}{}{%
\foreach \y in {1,2,...,\tbb} {
\draw[line width=#4, color=#8] (#5,\y) -- (#5+1,\y);
}}
\ifthenelse{\equal{#6}{\one}}{
\draw[line width=#4, color=#8] (0,0) -- (0,1);
\draw[line width=#4, color=#8] (#5,0) -- (#5,1);
}{}
\end{scope}
}

\newcommand{\subdedgedots}[7]{
\ifthenelse{\ee > 0}{
\pgfmathsetmacro{\xa}{#1}
\pgfmathsetmacro{\ya}{#2}
\pgfmathsetmacro{\xb}{#3}
\pgfmathsetmacro{\yb}{#4}
\pgfmathsetmacro{\k}{#5}
\pgfmathsetmacro{\t}{\k+1}
\pgfmathsetmacro{\tt}{\t-1}
\tikzstyle{subdivisionvertex}=[thick, fill=\subdivisioncolor]
\foreach \s in {1,...,\tt} {
\pgfmathsetmacro{\xxx}{\s/\t*(\xb-\xa)}
\pgfmathsetmacro{\yyy}{\s/\t*(\yb-\ya)}
\pgfmathsetmacro{\rnd}{100*rnd}
\ifthenelse{#7>\rnd}{\draw[subdivisionvertex] (\xa+\xxx,\ya+\yyy) circle(#6)}{};
}
}{}
}

\newcommand{\subdedgelines}[7]{
\pgfmathsetmacro{\xa}{#1}
\pgfmathsetmacro{\ya}{#2}
\pgfmathsetmacro{\xb}{#3}
\pgfmathsetmacro{\yb}{#4}
\pgfmathsetmacro{\k}{#5}
\pgfmathsetmacro{\t}{\k+1}
\pgfmathsetmacro{\tt}{\t-1}
\foreach \s in {1,...,\t} {
\pgfmathsetmacro{\xxx}{\s/\t*(\xb-\xa)}
\pgfmathsetmacro{\yyy}{\s/\t*(\yb-\ya)}
\pgfmathsetmacro{\xxxx}{(\s-1)/\t*(\xb-\xa)}
\pgfmathsetmacro{\yyyy}{(\s-1)/\t*(\yb-\ya)}
\draw[color=\regularcolor] (\xa+\xxxx,\ya+\yyyy) -- (\xa+\xxx,\ya+\yyy);
}
}

\ifthenelse{\size>3}{\pgfmathsetmacro{\size}{\size-mod(\size,2)}}{\pgfmathsetmacro{\size}{2}}
\pgfmathsetmacro{\a}{\size}
\pgfmathsetmacro{\b}{\size}

\begin{tikzpicture}[rotate=\rotate,thick,scale=\scale]
\pgfmathsetmacro{\corr}{(\a-2)/4}
\pgfmathsetmacro{\lay}{\a/2-1-\corr}
\pgfmathsetmacro{\start}{\a-1}
\pgfmathsetmacro{\a}{2*\a}
\pgfmathsetmacro{\bb}{\b-1}

\tikzstyle{noire}=[fill=\branchingcolor]

\pgfmathsetmacro{\one}{1}
\pgfmathsetmacro{\zero}{0}
\pgfmathsetmacro{\my}{1-mod(\b,2)}

\pgfmathsetmacro{\aa}{\a-1}
\pgfmathsetmacro{\bb}{\b-1}

\foreach \x in {0,...,\aa} {
\subdedgelines{\x}{0}{\x+1}{0}{\ee}{\qq}{50}
}

\ifthenelse{\equal{\b}{\one}}{}{%
\foreach \x in {0,...,\aa} {
\subdedgelines{\my+\x}{\b}{\my+\x+1}{\b}{\ee}{\qq}{50}
}
}

\foreach \y in {0,...,\bb} {
\foreach \x in {0,2,...,\a} {
\pgfmathsetmacro{\my}{mod(\y,2)}
\subdedgelines{\my+\x}{\y}{\my+\x}{\y+1}{\ee}{\qq}{50}
}
}

\ifthenelse{\equal{\b}{\one}}{}{
\foreach \y in {1,...,\bb} {
\subdedgelines{\a}{\y}{\a+1}{\y}{\ee}{\qq}{50}
}
}

\ifthenelse{\equal{\b}{\one}}{}{
\foreach \y in {1,...,\bb} {
\foreach \x in {0,...,\aa} {
\subdedgelines{\x}{\y}{\x+1}{\y}{\ee}{\qq}{50}
}}
}

\ifthenelse{\showl = 1}{
\foreach \x in {0,...,\lay} {
\perimetry{\x*4}{\x*2}{0}{\uu}{\a-8*\x}{\b-4*\x}{1}{\oddlayercolor}
}
\pgfmathsetmacro{\clay}{mod(\size,4)/2}
\pgfmathsetmacro{\llay}{\lay-\clay}
\ifthenelse{\NOT \equal{\size}{2}}{
\foreach \x in {0,...,\llay} {
\perimetry{-2-4*\x}{\start-2*\x}{180}{\uu}{\a-4-\x*8}{\b-2-4*\x}{-1}{\evenlayercolor}
}
}{}
}{}

\ifthenelse{\equal{\b}{\one}}{}{%
\foreach \y in {1,...,\bb} {
\draw[noire] (\a+1,\y) circle(\ww);
}
}

\foreach \x in {0,...,\a} {
\draw[noire] (\my+\x,\b) circle(\ww);
}

\foreach \y in {0,...,\bb} {
\foreach \x in {0,...,\a} {
\draw[noire] (\x,\y) circle(\ww);
}
}

\pgfmathsetmacro{\aa}{\a-1}
\pgfmathsetmacro{\bb}{\b-1}

\foreach \x in {0,...,\aa} {
\subdedgedots{\x}{0}{\x+1}{0}{\ee}{\qq}{50}
}

\ifthenelse{\equal{\b}{\one}}{}{%
\foreach \x in {0,...,\aa} {
\subdedgedots{\my+\x}{\b}{\my+\x+1}{\b}{\ee}{\qq}{50}
}
}

\foreach \y in {0,...,\bb} {
\foreach \x in {0,2,...,\a} {
\pgfmathsetmacro{\my}{mod(\y,2)}
\subdedgedots{\my+\x}{\y}{\my+\x}{\y+1}{\ee}{\qq}{50}
}
}

\ifthenelse{\equal{\b}{\one}}{}{
\foreach \y in {1,...,\bb} {
\subdedgedots{\a}{\y}{\a+1}{\y}{\ee}{\qq}{50}
}
}

\ifthenelse{\equal{\b}{\one}}{}{
\foreach \y in {1,...,\bb} {
\foreach \x in {0,...,\aa} {
\subdedgedots{\x}{\y}{\x+1}{\y}{\ee}{\qq}{50}
}}
}

\end{tikzpicture}}
	\caption{A 9-wall. The subdivision vertices are depicted in \red{red}.} 
	\label{dnsamsfsnamsd}
\end{figure}

The \emph{elementary $k$-wall}, for $k≥3,$ is obtained from 
$P_{k}\square P_{2k}$ by first deleting a perfect matching formed by ``vertical edges'', i.e., edges whose endpoints have the same second coordinate, and then remove the vertices of degree 1 that appear (see \cref{dnsamsfsnamsd}). 
 A \emph{$k$-wall}~$W$ is a graph isomorphic to a subdivision of the elementary $k$-wall. Notice 
that every $k$-wall has a unique (up to homeomorphism) embedding in the sphere $\Sbbb^{2}$ and its \emph{perimeter} is the boundary 
of the unique face of this embedding that has more than six degree-3 vertices.

All results on treewidth in this paper are based on the following 
result that can be seen as a surface-embedded version of the ``grid minor theorem'' of Robertson and Seymour \cite{RobertsonS84GMIII,RobertsonST94quic} (see also \cite{ChuzhoyT21towar}). 

\begin{proposition}
\label{thi_oplos_s}
There is a universal constant $c$ 
such that if a $\Sigma$-embedded graph $G$ has treewidth bigger than $c\cdot r\cdot \sqrt{t+1}\cdot (\eg(\Sigma)+1)$ and $B$ is a subset of $V(G)$ of size $t$, then $G$ contains as a subgraph an $r$-wall $W$ whose perimeter is a contractible cycle bounding a closed disk that is disjoint from $B$ and where the entire $W$ is embedded. 
\end{proposition}

The above result follows from \cite[Theorem 4.8]{DemaineHT06thebidim} (see also \cite[Lemma 4]{GeelenRS04embe}). 

The

$\Theta(\sqrt{t+1})$-overhead
permits us to consider a partition of wall to $t$ pairwise vertex-disjoint walls and then pick one that is embedded in a closed disk bounded by its perimeter that does not contain any of the vertices of $B.$
To prove \cref{proog_tw} from \cref{thi_oplos_s} it remains to observe that a big enough wall can give rise to the required $r$-railed nest. The algorithmic part of \cref{proog_tw} follows by applying the linear-time algorithm of \cite{AdlerDFST12fastm}.

In the next section, we will also use the fact that the treewidth of a graph is upper-bounded by a linear function of its \emph{radial radius}, which is defined as follows. 
Given a graph $G$ and a vertex $v\in V(G),$ the \emph{eccentricity} of $v$ in $G$ is the maximum distance between $v$ and some vertex of $G.$ The \emph{radius} of $G$ is the minimum eccentricity of its vertices. Given a $\Sbbb^{2}$-embedded graph, we define its \emph{radial radius} as the radius of its radial graph. We need the following result that follows easily from \cref{thi_oplos_s}. For a proof with improved constants, see \cite[2.1]{RobertsonS84GMIII}.

\begin{proposition}
\label{rad_tw_rad}
If $G$ is a $\Sbbb^{2}$-embedded graph of radial radius $≤k,$
then $\tw(G)=\Ocal(k).$
\end{proposition}

A graph is called \emph{sub-cubic} if it has no vertex of degree bigger than three. We say that $H$ is a \emph{contraction} of $G$ if $H$ can be obtained from $G$ after contracting edges.
A vertex $v\in V(G)$ is a \emph{cut-vertex} of $G$ 
if $G-v$ has more connected components than $G.$
A graph is \emph{2-connected} if it is connected and has no cut-vertices.
A \emph{block} of a graph $G$ is a maximal 2-connected subgraph of $G$ that is not isomorphic to $K_{2}.$

\begin{lemma}
\label{easyconse}
Every planar graph $G$ is the contraction of a planar sub-cubic graph $G^{+}$ where $\tw(G^{+})=\Ocal( \tw(G)).$
\end{lemma}

\begin{proof}
We examine the non-trivial case where $G$ is not a forest.
We assume that $G$ is 2-connected, otherwise we consider all its blocks of $G$ separately and join them by adding edges, an operation that does not increase the treewidth of a graph that is not a forest. The parameter of \textsl{branch-width} was defined by Robertson and Seymour in \cite{RobertsonS91GMX} and can be seen as an alternative for treewidth.
We do not give the definition of branch-width here. We only 
need two properties of it. The first is that for every graph $\bw≤ \tw(G)+1
\leq \lfloor \frac{3}{2}\bw(G)\rfloor+2,$ because of \cite[(5.2)]{RobertsonS91GMX}. The second is that every 2-connected planar 
graph is a contraction of some sub-cubic planar graph of the same branch-width, because of \cite[Lemma 3.4]{FominT06domin}.
It is now easy to derive the bound of the lemma from these two properties.
\end{proof}

\subsection{Laminar stellations}
\label{subsec_laminar}

Let $G$ be a $\Sbbb^{2}$-embedded graph and let 
$\Ccal$ be a collection of cycles of $G.$
We say that $\Ccal$ is \emph{laminar} if for every 
$C,C'\in \Ccal,$ if $\Delta$ and $\Delta'$ are open disks bounded by the embeddings 
of $C$ and $C'$ respectively, then either one of the disks 
in $\{\Delta,\Delta'\}$ is a subset of the other or the two disks in $\{\Delta,\Delta'\}$ are disjoint.
We call $\Ccal$ \emph{strongly laminar} if it is laminar and there is no cycle $C\in \Ccal$
such that both \textsl{closed} disks bounded by $C$ contain some cycle in $\Ccal-\{C\}$ 
. 
Given a $\Sbbb^{2}$-embedded graph $G$ and a strongly laminar collection $\Ccal$ of cycles of $G,$
the \emph{laminar stellation of $G$ with respect to $\Ccal$}, which we denote by $G_{\Ccal},$ is the graph obtained 
by introducing a new vertex $v_{C}$, for each $C \in \Ccal,$ and joining it to
all the vertices of $C.$ 
Notice that laminar stellations are not necessarily planar graphs.

\begin{lemma}
\label{cycles_col_tw}
Let 
$G$ be a planar graph $G$ and 
$\Ccal$ be a strongly laminar collection of cycles of $G.$
Then $\tw(G_{\Ccal})=\Ocal(\tw(G)).$
\end{lemma} 

\begin{proof}

We first apply \cref{easyconse} on $G$ and consider a $\Sbbb^{2}$-embedded sub-cubic graph $G^{+}$ such that $G$ is a contraction of $G^+$ and 
\begin{eqnarray}
\tw(G^{+}) & = & \Ocal( \tw(G)).\label{eq_dpdother}
\end{eqnarray}

The fact that $G$ is a contraction of $G^{+}$ implies that 
for each cycle $C\in\Ccal$ there is a cycle in $G^{+},$ which we denote by $C^{+},$ such that the set $\Ccal^{+}:=\{C^{+}\mid C\in\Ccal\}$ 
is a strongly laminar collection of cycles in $G^{+}.$
We now consider the subgraph $Z^+=\cupall \Ccal^{+}$ of $G^+$ 
Clearly, $Z^+$ is also sub-cubic and $\Sbbb^{2}$-embedded.
Also, as $\Ccal^{+}$ is strongly laminar, each cycle of $\Ccal^{+}$ corresponds to some face of $Z^+.$ Also keep in mind that because $G$ is a contraction of $G^{+},$ it also holds that $G_{\Ccal}$ is a contraction of $G_{\Ccal^+}^{+}.$
We denote by $M_{Z^+}$ the \emph{map graph} of $Z^+$ that is the graph whose vertices are the faces of $Z^+$ and where two vertices are adjacent if the corresponding faces have some common incident vertex. Map graphs have been introduced by Chen, Grigni, and Papadimitriou in \cite{ChenGP02mapg} and in general, map graphs of planar graphs are not necessarily planar. However, it is easy to see that map graphs of \textsl{sub-cubic} planar graphs are also planar. Using the four-color theorem, we may color the vertices of $M_{Z^+}$ using four colors. This means that the faces of $Z^+,$ and therefore also the cycles of $\Ccal^{+},$ can be colored with four colors in a way that 
no two faces (resp. cycles of $\Ccal^{+}$) with a common incident vertex have the same color. We partition $\Ccal^{+}$ to four same-colored classes of cycles $\Ccal_{1}^{+},\ldots,\Ccal_{4}^{+}.$ 
We next consider the 
graph $Q:=G^{+}\square K_{1,4}$ and keep in mind that, from \cref{cart_pr_tw}, 
\begin{eqnarray}
\tw(Q) & = & O(\tw(G^{+})).\label{eq_f8iiooa}
\end{eqnarray}
Clearly $Q$ 
contains five disjoint copies of $G$: a copy $G_{0}$ corresponding 
in the non-leaf vertex of $K_{1,4}$ and 
four copies $G_{1},\ldots,G_{4},$ each corresponding to some of the four leaves of $K_{1,4}.$ 
Let $Q^-$ be the subgraph of $Q$ obtained after removing from $Q$ all vertices
of $G_{1}\cup \cdots \cup G_{4}$ that do not correspond vertices of cycles of $\Ccal^+.$

\begin{figure}[ht]
	\centering
	\includegraphics[height=3.5cm]{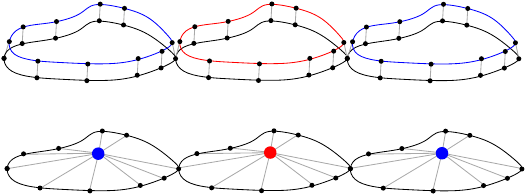}
	\caption{The contractions in the proof of \cref{cycles_col_tw}. the \myblue{blue} cycles are disjoint cycles of $G_{1}$ and the \red{red} cycle
	is one of the cycles of $G_{2}$
	and they are all copies of cycles in $\Ccal^+$ in $Q=G^{+}\square K_{1,4}.$ These cycles are contracted to the fat \myblue{blue} and \red{red} vertices that play the roles of the new vertices $v_{C},$ for each $C \in \Ccal^+,$ in $G^+_{\Ccal^+}.$}
	\label{fig_ssubg}
\end{figure}

Notice now that for each color $i\in[4],$ the copies of the cycles in $\Ccal_{i}^{+}$ inside $G_{i}$ do not have common vertices. 
We now consider a minor of $Q^-$ as follows:
for every $i\in[4],$ we contract in $Q^-$ every copy of a cycle of $\Ccal_{i}^{+}$ in $G_{i}$ to a single vertex and remove all vertices of the copies of the cycles of $\cupall(\{\Ccal_{1}^{+},\ldots,\Ccal_{4}^{+}\}\setminus \{\Ccal_{i}^{+}\})$ in $G_{i}.$ Observe that the resulting graph is the graph $G_{\Ccal^+}^{+}$: for every $C^{+}\in\Ccal^{+},$ colored $i,$
the vertex $v_{C^+}$ is the vertex created because of the contraction of the copy of $C^{+}$ in $G_{i}.$ Recall now that $G_{\Ccal}$ is a contraction of $G_{\Ccal^+}^{+}$ which, as we have just seen, is a contraction of $Q^-$ (see \cref{fig_ssubg} for an visualization of these contractions).
As $Q^-$ is an induced subgraph of $Q,$ we have that 
 $G_{\Ccal}$ is a minor of $Q,$ therefore $\tw(G_{\Ccal})≤\tw(Q).$
 Combining this fact with \eqref{eq_dpdother} and \eqref{eq_f8iiooa}, 
we obtain that $\tw(G_{\Ccal})=\Ocal(\tw(G)),$ as required.

\end{proof}

\section{Finding flows inside slices in linear time}
\label{slice_section}

As explained in~\cref{outline_pr}, the key idea of our algorithm is to split the given rooted embedding to slices and then process each slice separately in order to detect big enough railed nests inside them.
In this section, and in particular in~\Cref{subsec_slices}, we give the definition of slices and we use the results of the previous section in order to bound the treewidth of laminar stellations of slices. Also, in~\cref{subsec_flows_in_slices} using Menger's theorem we show that in sufficiently large slices, in order to detect flows between particular cycles in each slice, it suffices to search locally inside the slice.

%
%

\subsection{Slices and their treewidth}
\label{subsec_slices}

Slices are graphs that are embedded in \emph{pseudo-disks}.
We next define pseudo-disks and associated notions and we prove that the treewidth of graphs embedded in pseudo-disks is upper-bounded by a linear function of their radial depth. 

\paragraph{Graphs embedded in pseudo-disks.}
A \emph{pseudo-disk} $\Psi$ is 
obtained from a closed disk $\Delta$ by removing a finite collection $\Dcal$ of pairwise disjoint open 
disks in the interior of $\Delta.$
We call the boundary of $\Delta$ \emph{outer boundary} of $\Psi$ 
and the boundary of each of the disks of $\Dcal$ an \emph{inner} boundary of $\Psi.$
A \emph{boundary} of $\Psi$ is either the outer boundary of $\Delta$ or the boundary of an open disk in $\Dcal$.

We say that a graph $G$ is \emph{$\Psi$-embedded} for 
some pseudo-disk $\Psi$ if it is accompanied with an embedding $\Gamma$ in 
$\Psi$ where 
each boundary of $\Psi$ is the embedding of some cycle of $G.$
As an \emph{embedding} $\Gamma$ of a graph $G$ in a pseudo-disk $\Psi$ we consider a
a drawing of $G$ in $\Psi$ where no point of $\Psi$ is used more than once. 
The \emph{radial depth} of a pseudo-disk embedding $(\Psi,Γ)$
of a graph $G$ is the minimum $k$ for which, for every vertex $x$ of $G,$ there is a sequence $[v_{1},f_{2},\ldots,v_{k-1},f_{k},v_{k+1}]$
of vertices and faces of $Γ,$ where 

\begin{itemize}
\item faces and vertices alternate, i.e., $v_{1},v_{3},\ldots,v_{k-1},v_{k+1}$ are vertices and $f_{2},\ldots,f_{k}$ are faces,
\item $v_{1}$ is drawn in the outer boundary of $\Psi$ and $v_{k+1}=x,$ and \\
 
\item if $v,f$ or $f,v$ are two consecutive elements of this sequence, then $v$ is incident to the face $f,$ i.e., $v$ is embedded in the boundary of $f.$
\end{itemize}

It is straightforward to observe that every graph embedded in a pseudo-disk that has bounded radial depth also has bounded radial radius.
Thus, using~\cref{rad_tw_rad}, we can upper-bound its treewidth.

\begin{lemma}
\label{pseu_tw}
Let $\Psi$ be a pseudo-disk and let $G$ be a $\Psi$-embedded graph of radial depth at most $k.$
Then $\tw(G)={\Ocal}(k).$
\end{lemma}

\begin{proof}
Notice that $G$ has a $\Sbbb^{2}$-embedding 
 where one, say $f$ of its faces is the outer boundary of $\Psi.$
Notice also that all vertices or faces of $G$ are within radial distance 
$k+2$ from $f.$ This means that the radial radius of $G$ is at most 
$k+2$ and the result follows from \cref{rad_tw_rad}.
\end{proof}

\paragraph{Slices.}
Let $(G,u)$ be a rooted embedding.
Let $C$ be a cycle of $G$ that is $u$-aligned and let $r\in \Nbbb^{\sf even}.$ We denote by 
$\mathsf{Cycles}(C,r)$ the set of all cycles $C'$ of $G$ that satisfy the following properties:
\begin{itemize}
	\item $C'$ is $u$-aligned,
	\item $V(C')\subseteq \Delta_{C},$ and
	\item the radial distance of $C'$ and $C$ is equal to $r.$
\end{itemize}
We also denote by $\mathsf{Interior}(C,r)$ the set $\bigcup_{C'\in \mathsf{Cycles}(C,r)}\Delta_{C'}.$
Keep in mind that $\mathsf{Interior}(C,r)$ is an open set.

We also define ${\sf Slice}_{G}^{r}(C)$ as the graph 
obtained from $G$ if by removing all vertices and edges 
that are disjoint from the closure of $\Delta_{C}$ and all vertices and 
edges that are embedded in $\mathsf{Interior}(C,r).$
The construction of ${\sf Slice}_{G}^{r}(C)$ immediately implies the following.

\begin{observation}
\label{obs_slice_pseudo}
If $h\in\Nbbb_{≥2}^{\sf even},$ every ${\sf Slice}_{G}^{h}(C)$ has a pseudo-disk embedding of radial depth at most $h.$
\end{observation}

As a direct consequence of \cref{obs_slice_pseudo,pseu_tw,cycles_col_tw},
we get the following result.
\begin{lemma}
\label{tw_slice}
Let $(G,u) $ be a rooted embedding and let $C$ be $u$-aligned cycle of $G.$
Let also $h\in\Nbbb_{≥2}^{\sf even}$ and $z\in[0,h]^{\sf even}.$ If $D={\sf Slice}_{G}^{h}(C)$ and $\Ccal=\mathsf{Cycles}(C,z),$ then $\tw(D_{\Ccal})=\Ocal(h).$
\end{lemma}

\begin{proof}
Due to~\cref{obs_slice_pseudo}, $D$ has a pseudo-disk embedding of radial depth at most $h.$ Then,~\cref{pseu_tw} implies that $\tw(D)=\Ocal(h).$
Moreover, observe that the definition of $\Ccal$ implies that $\Ccal$ is strongly laminar.
Therefore,~\cref{cycles_col_tw} implies that $\tw(D_{\Ccal})=\Ocal(h).$
\end{proof}

\subsection{Flows in slices}
\label{subsec_flows_in_slices}

\paragraph*{Big flows between cycles.}
Given a graph $G$ and subgraphs $H,H'$ of $G,$
we use $\mathsf{flow}_{G}(H,H')$ to denote the maximum number of pairwise vertex-disjoint paths in $G$ with one endpoint in $V(H)$ and one endpoint in $V(H').$ 
Let $(G,u) $ be a rooted embedding
and let $C$ be a $u$-aligned cycle of $G.$
Let also $z\in\Nbbb_{≥2}^{\sf even}$ and $q\in\Nbbb.$ 
We denote by ${\sf bfc}_{G}(C,z,q)$ the set containing every cycle $C'$
of $\mathsf{Cycles}(C,z)$ for which
$\mathsf{flow}_{G}(C,C')≥q.$

We show that in order to compute ${\sf bfc}_{G}(C,z,q)$ it suffices to check for which cycles $C'$ in $\mathsf{Cycles}(C,z)$ there is a $q$-flow from $C$ to $C'$ in ${\sf Slice}_{G}^{z+2q}(C).$ This reduces the flow question in the general graph to a ``local'' question in the corresponding slice.

\begin{lemma}
	\label{more_reroute_linkage}
	Let $(G,u) $ be a rooted embedding 
	and let $C$ be a $u$-aligned cycle of $G.$
	Let also $z\in\Nbbb_{≥2}^{\sf even}$ and $q\in\Nbbb.$ 
	Then for every cycle $C'\in {\sf bfc}_{G}(C,z,q),$ it holds that 
	$\mathsf{flow}_{D}(C,C')≥q,$ where $D={\sf Slice}_{G}^{z+2q}(C).$
\end{lemma}

\begin{proof}
Let $(T,t_{0},\chi)$ be the radial distance decomposition of $(G,u)$ and let $e=(v,f)$ be the vertex-face edge of $T$ such that $C=C_{e}.$
Also, let $E$ be the set of all vertex-face edges $\tilde{e}=(\tilde{v},\tilde{f})$ of $T$ such that $\mathsf{dist}(v,\tilde{v})=h$. We have $\bigcup_{\tilde{e}\in E}\{C_{\tilde{e}}\}=\mathsf{Cycles}(C,h).$
Consider an arbitrary ordering $e_{1},\ldots,e_{p}$ of the edges in $E.$
We set $G_{0}:=G$ and for every $i\in[p],$ we set $G_{i}$ to be the graph obtained from $G_{i-1}$ after removing all vertices of $G$ in the interior of $C_{e_i}.$
Note that $D=G_p.$

Next, fix a cycle $C'\in {\sf bfc}_{G}(C,z,q),$ i.e., a cycle $C'$ of $\mathsf{Cycles}(C,z)$ such that $\mathsf{flow}_{G}(C,C')\ge q.$
We show inductively that for every $i\in[p],$ $\mathsf{flow}_{G_i}(C,C')\ge q.$

We fix some $i> 0$ and we assume that there is a collection of at least $q$ pairwise vertex-disjoint paths in $G_{i-1}$ with one endpoint in $V(C)$ and one endpoint in $V(C').$ 
We aim to prove that there is such a collection of paths also in $G_{i}.$ 

We start with a series of observations.
First, we can assume that $C_{e_i}$ is not in the interior of $C',$ since otherwise the paths of $G_{i-1}$ give directly the paths in $G_i.$
We use $(v_{i},f_{i})$ to denote the vertex-face edge $e_{i}$ and $e'=(v',f')$ to denote the edge of $T$ where $v'$ is an ancestor of $v_{i}$ and $\mathsf{dist}(v',v_{i})=2(q-1).$
Due to~\Cref{pensions}, there is a $u$-nested sequence $\Ccal$ of $q$ many $u$-aligned cycles of $G$ where the inner one is $C_{e_i}$ and the outer one is $C_{e'}.$
Note that $C_{e'}\in\mathsf{Cycles}(C,z+2),$ since $h=z+2q,$ and the cycles of $\Ccal$ are also cycles of $G_i.$

Suppose towards a contradiction that $\mathsf{flow}_{G_i}(C,C')< q.$ By Menger's theorem, this means that there is a set $S\subseteq V(G_{i}$ of less than $q$ vertices such that there is no path from $C$ to $C'$ in $G_i\setminus S.$ Since $\Ccal$ consists of $q$ many pairwise vertex-disjoint cycles, there should be a cycle $\tilde{C}$ of $\Ccal$ in $G_i\setminus S.$ 
Also, among the at least $q$ paths in $G_{i-1}$ certifying that $\mathsf{flow}_{G_i}(C,C')\ge q,$ there should be one, call it $\tilde{P}$ that does not intersect $S.$ Now let $Q$ be the subgraph of $G_{i}\setminus S$ induced by $V(\tilde{C})$ and $V(\tilde{P})\cap V(G_i).$ In $Q,$ there is a path from $C$ to $C'.$ Indeed, since the endpoints of $\tilde{P}$ are in the exterior of $\tilde{C}$ and the interior of $C_{e_i}$ is contained in the interior of $\tilde{C},$ if $\tilde{P}$ intersects $C_{e_i}$ it also intersects $\tilde{C}.$
The existence of such a path in $Q$ (and therefore in $G_i\setminus $) from $C$ to $C'$ contradicts the assumption that $S$ separates $C$ from $C'$ in $G_i.$
\end{proof}

We conclude this section by showing that we can compute the set of all aligned cycles in a slice (in an appropriate radial distance) that certify the existence of sufficiently large railed nests. The ingredients used in this proof are the following. First,~\cref{more_reroute_linkage} allows to restrict our attention to the given slice. 
The existence of large flows can be expressed in \textsf{MSO} in the appropriate laminar stellation of the slice, which has bounded treewidth because of~\Cref{tw_slice}. Therefore, we can compute these flows in total linear time using~\Cref{ow_dec_dource}.

\begin{lemma}
\label{lemma_courcelle}
There is an algorithm that given a rooted embedding $(G,u),$ a $u$-aligned cycle $C$ of $G,$
two integers $z\in \Nbbb_{≥2}^{\sf even}$ and $q\in\Nbbb,$ and
the set of cycles $\mathsf{Cycles}(C,z),$
computes ${\sf bfc}_{G}(C,z,q),$
in time $\Ocal_{z+q}(|D|),$
where $D={\sf Slice}_{G}^{h}(C)$ and $h=z+2q.$
\end{lemma}

\begin{proof}
Our aim is to compute ${\sf bfc}_{G}(C,z,q)$ using~\Cref{ow_dec_dource}.
By~\cref{more_reroute_linkage}, it suffices to check for which cycles $C'$ of $G$ in $\mathsf{Cycles}(C,z)$ it holds that $\mathsf{flow}_{D}(C,C')\ge q.$
We use $\Ccal$ to denote $\mathsf{Cycles}(C,z).$
In linear time, we construct $D$ using the algorithm of~\cref{lem_comp_rdd} and from this we get the graph $D_{\Ccal}$ (by introducing a vertex $v_{C'}$ for every cycle $C'\in\Ccal$ and making it adjacent to all vertices of $C$).
Also, we color the set of all newly introduced vertices by a color, say $R.$
Notice that there is an \textsf{MSO}-formula $\varphi$ that given a set $S\subseteq R,$ expresses, for every $s\in S,$ the existence of a collection $\Pcal_{s}$ of $q$ paths in $D_{\Ccal}$ from $s$ to the vertices of $V(C)$ that have $s$ as a common endpoint and no other common vertex. 
A maximum size such set $S\subseteq R$ corresponds to the collection of cycles $\mathsf{bfc}(C,z,q).$
Therefore, using~\Cref{ow_dec_dource}, we can compute the maximum size set $S\subseteq R$ such that $(G,S)\models\phi,$ from which we can also get $\mathsf{bfc}(C,z,q)$ (by taking the neighborhood of each $s\in S$ in $D_{\Ccal}$). Following~\cref{ow_dec_dource}, this can be done in time $\Ocal_{q,\tw(D_{\Ccal})}(|V(D_{\Ccal})|).$
Due to~\Cref{tw_slice}, $\tw(D_{\Ccal})=\Ocal(h)$ and therefore we get the claimed running time. 
\end{proof}

%


\section{Removing all irrelevant vertices in linear time}
\label{removing_irr_sec}

In this section, we show how to split a planar single-rooted instance into slices and apply the algorithm of~\Cref{lemma_courcelle} in order to find railed nests inside each slice that certify the irrelevance of sets of vertices.

Since we want to argue about irrelevance in the original instances, the results in this section are stated for general surface-embedded multiple-rooted graphs.
However, radial distance decompositions and slices are defined for rooted embeddings, which are single-rooted $2$-edge-connected $\Sbbb^{2}$-embedded graphs.
We will next work with \emph{reduced embedding pairs},
which are rooted embeddings obtained from a surface-embedded graph $G$ after the contraction of a connected edge set $J$ that make it planar and then adding some edges in the obtained $\Sbbb^{2}$-embedded graph $G/J$ to make it $2$-edge-connected.
The corresponding definitions and results about achieving $2$-edge-connectivity and the notion of reduced embedding pairs are given in~\Cref{make_e_edge}. The specifications of the set $J$ to be contracted as well as the algorithm that finds it is presented in~\Cref{algo_section}.

The set of vertices to be declared irrelevant for each slice as well as the set of all irrelevant vertices are described in~\Cref{subsec_deducing_irr}, where we also show that their removal from the instance indeed gives an equivalent instance.
Then, in~\Cref{subsec_algo_irr}, we give the algorithm that actually detects this set.
We conclude this section with~\Cref{bstwso9o}, where we show that, after the removal of the set of irrelevant vertices, the obtained (equivalent) instance has bounded treewidth.

\subsection{Making a sphere-embedded graph \texorpdfstring{$2$}{2}-edge-connected}
\label{make_e_edge}

In order to show how to get a $2$-edge-connected supergraph of a given $\Sbbb^{2}$-embedded graph, we define the operation of \emph{digon completion}.

\paragraph{Digon completions.}
A $2$-edge-connected component of a graph $G$ is a vertex-maximal $2$-edge-connected subgraph of $G.$ 
Note that if $H$ and $H'$ are two distinct $2$-edge-connected components of a graph then $V(H)\cap V(H')=\emptyset$ and if there is an edge $e\in E(G)$ with one endpoint in $V(H)$ and one endpoint in $V(H'),$ then $e$ is a bridge of $G.$

Let $G$ be a $\Sbbb^{2}$-embedded graph with $|V(G)|\ge 2.$
The operation of \emph{digon completion} in $G$ constructs a new graph $G'$ as follows. 
Let $H_{1},H_{2}$ be two distinct $2$-edge-connected components of $G$ that share a face $f$ of $G.$
We distinguish two cases, depending whether $H_{1}$ and $H_{2}$ are connected by a bridge, or not.
If $H_{1}$ and $H_{2}$ are connected by a bridge, say with endpoints $v_{1},v_{2},$ then $G'$ is obtained by introducing a new edge between $v_{1}$ and $v_{2},$ which we embed inside $f.$
If there is no bridge between $H_{1}$ and $H_{2},$
we pick arbitrary vertices $v_{1}\in V(H_{1})$ and $v_{2}\in V(H_{2})$ and we construct $G'$ by introducing two new edges between $v_{1}$ and $v_{2},$ both embedded inside $f.$
We say that $G'$ is obtained by a \emph{digon completion} from $G.$
We call the newly introduced edges \emph{digon edges}.
Note that $G'$ is also a $\Sbbb^{2}$-embedded graph.

Let $G$ be a $\Sbbb^{2}$-embedded graph.
We say that a $\Sbbb^{2}$-embedded graph $G'$ is a \emph{$2$-edge-connected digon-completion} of $G$ if $G'$ is $2$-edge-connected and either $G=G'$ or there is a sequence $G_{1},\ldots,G_{d}$ of graphs, where $G_{1}=G$ and $G_{d}=G'$ such that
for every $i\in[d-1],$ $G_{i}$ is not $2$-edge-connected and
$G_{i+1}$ is obtained by a digon completion from $G_{i}.$

\begin{lemma}
	\label{lemma_digons}
	Let $G$ be a $\Sbbb^{2}$-embedded graph and let $G'$ be a $2$-edge-connected digon-completion of $G.$
	Also, let $u$ be a vertex of $G.$
	Then for every $u$-aligned cycle $C$ of $G',$
	we have that either $E(C)$ contains no digon edge or the interior of $C$ is a face of $G'.$ 
\end{lemma}	

\begin{proof}
Assume that $G\neq G',$ since otherwise $G'$ contains no digon edges.
Let $G_{1},\ldots,G_{d}$ be the sequence of graphs where $G_{1}=G$ and $G_{d}=G'$ certifying that $G'$ is a $2$-edge-connected digon-completion of $G.$

Let $(T,t_{0},\chi)$ be the radial distance decomposition of $(G',u)$ and let $(v,f)$ be the vertex-face edge of $T$ such that $C=C_{(v,f)}.$
Suppose towards a contradiction that $E(C)$ contains a digon edge and the interior of $C$ is not a face of $G'.$
Let $i$ be the maximum integer in $\{1,\ldots,d-1\}$ for which there is a digon edge $e\in E(C)$ introduced when obtaining $G_{i+1}$ from $G_{i}.$
Observe that maximality of $i$ implies that $C$ is a cycle in $G_{i+1}$ and therefore both endpoints of $e$ should {belong to} the same connected component of $G_{i}.$
Therefore, by definition of the digon completion operation, $E(C)\setminus\{e\}$ should contain only one edge $e'$ which should also be a bridge in $G_{i-1}.$ Since $E(C)$ consists of the two edges $e$ and $e',$ again by the definition of the digon completion operation we have that the interior of $C$ bounds a face of $G',$ a contradiction.
\end{proof}

\paragraph*{Reduced embedding pairs.}
Let $\Sigma$ be a surface, let $G$ be a $\Sigma$-embedded graph, and let 
$J$ be a connected set of edges of $G$ such that $G/J$ is planar.
We consider 
a rooted embedding $(G',u_{J})$ where $G'$
is a $2$-edge-connected digon-completion of $G/J,$ accompanied with 
the resulting embedding of $G/J$ in $\Sbbb^{2},$ and $u_{J}$ is the result of the contraction of the edges in $J$ in $G.$ We refer to $(G',u_{J})$ as the \emph{reduced embedding pair} of $(G,J).$
We say that a cycle $C$ of $G$ is \emph{$J$-aligned} if $C$ is a $u_{J}$-aligned cycle of $G'$ (with respect to $(G',u_{J})$). 

\subsection{Deducing irrelevance from flows}
\label{subsec_deducing_irr}

Let $(G,u) $ be a rooted embedding and let $C$ be a $u$-aligned cycle of $G.$
Given a $q\in\mathbb{N}_{\ge 2},$
we define the \emph{$q$-irrelevant set for $C$} as the vertex set $$\mathsf{Irr}_{G}(C,q):=V(G)\cap \bigcup_{C'\in \Ccal}\mathsf{Interior}(C',2q),$$
where $\Ccal={\sf bfc}_{G}(C,2(q-1),q)$.
In words, the $q$-irrelevant set for $C$ is the set of all vertices of $G$ that are embedded in the interior $\mathsf{Interior}(C',2q)$ of all cycles that are in radial distance $2q$ from $C',$ for all aligned cycles $C'$ that are in radial distance $2(q-1)$ from $C$ and send a $q$-flow to $C.$

We next show that given an instance $\mathbf{G}$ of a rooted graph problem $\Pi$ that has the insulation property, if $J$ is a connected set of edges that spans the roots $R(\mathbf{G})$
and whose contraction makes $G$ planar, then for every $J$-aligned cycle $C$ of $G$ 
the removal of the $c_{\Pi}$-irrelevant set of $C$ yields an equivalent instance for $\Pi.$

\begin{lemma}
\label{equiv_lem}
Let $\Pi$ be a rooted graph problem
that has the insulation property.
Also, let $\mathbf{G}$ be a $\Sigma$-embedded rooted graph and let $J$ be a connected set of edges of $G$ such that $G/J$ is planar and $R(\mathbf{G})\subseteq V(J).$
Then, for every $J$-aligned cycle $C$ of $G,$
we have that
$\mathbf{G}$ and $\mathbf{G}-\mathsf{Irr}_{G}(C,c_{\Pi})$ are $\Pi$-equivalent.
\end{lemma}

\begin{proof}
	Let $(G',u_{J})$ be a reduced embedding pair of $(G,J)$ and let $(T,t_{0},\chi)$ be the radial distance decomposition of $(G',u_{J}).$
	By definition, the fact that $C$ is $J$-aligned implies that there is a non-root vertex-face edge $e=(v,f)$ of $T$ such that $C=C_{e}.$
	Also, note that because of~\Cref{lemma_digons}, all $u_{J}$-aligned cycles of $G'$ that correspond to vertex-face edges $e'=(v',f')$ of $T$ for which $f'$ is not a leaf, are also cycles of $G.$ 

	Consider an arbitrary ordering $C_{1},\ldots,C_{p}$ of the cycles in $\mathsf{bfc}(C,2(c_{\Pi}-1),c_{\Pi}).$
	We set $\mathbf{G}_{0}:=\mathbf{G}$ and for every $i\in[p],$ we set $\mathbf{G}_{i}$ to be the rooted graph obtained from $\mathbf{G}_{i-1}$ by removing all vertices in $\mathsf{Interior}(C_{i},2c_{\Pi}).$ Keep also in mind that $\mathsf{Interior}(C_{i},2c_{\Pi})$ is a subset of the interior of $C_{i}.$
	
	Because of~\Cref{more_reroute_linkage}, for every $i\in[0,p]$ and every $j>i,$ $\mathsf{flow}_{G_i}(C,C_{j})\ge c_{\Pi}.$ In other words, given that $G$ contains $c_{\Pi}$ pairwise vertex-disjoint paths between $C$ and $C_{j},$ such paths also exist in $G_{i}.$ This allows us to consider, for every $i\in[p],$ a collection $\Pcal_{i}$ of $c_{\Pi}$ pairwise vertex-disjoint paths from $V(C)$ to $V(C_{i})$ in $G_{i-1}.$
	Moreover, since for every $i\in[p],$ $C_{i}\in \mathsf{Cycles}(C,2(c_{\Pi}-1)),$
	by~\Cref{pensions}, there is a nested sequence $\Ccal$ of $c_{\Pi}$ cycles of $G'$ such that the inner cycle of $\Ccal$ is $C_{i}$ and the outer cycle of $\Ccal$ is $C.$

Notice now that, for every $i\in[p],$ $W_i := \cupall \Ccal_i \cup \cupall \Pcal_i$ 
is a $c_{\Pi}$-railed nest of $G'$ such that $u_J\in \mathsf{ext}(W_i)$ and
\(\mathsf{int}(W_i)\) is the interior of \(C_i\).
To justify the last assertion,
observe that all requirements in \Cref{def_otho} for an \(r\)-railed nest are
immediate, except possibly the orthogonality condition
\Cref{it_orhogonality}. The latter can be imposed by a standard rerouting
argument for the paths; see, e.g.,~\cite[Lemma~4.11]{GorskyPW2026Quickly},~\cite[Lemma~6.15]{JansenW25lossy} or the full version of~\cite{WlodarczykZehavi2023PlanarDisjointPaths}.

	Observe also that the fact that $J$ is connected implies that $W_{i}$ is also a $c_{\Pi}$-railed nest of $\mathbf{G},$ where the endpoints of all edges in $J$ {belong to} $\mathsf{ext}(W)$ and $\mathsf{int}(W)$ is the interior of $C_{i}.$ Furthermore, since $R(\mathbf{G})$ is a subset of $V(J),$ it is also a subset of $\mathsf{ext}(W)\cap V(\mathbf{G}).$
	Therefore, by the insulation property, we have that for every $i\in[p],$ $\mathbf{G}_{i}$ and $G_{i-1}$ are $\Pi$-equivalent. Thus, $\mathbf{G}_{0}=\mathbf{G}$ and $\mathbf{G}_{i}$ are $\Pi$-equivalent for every $i\in[p],$ which in particular means that $\mathbf{G}$ and $\mathbf{G}_{p}=\mathbf{G}-\mathsf{Irr}_{G}(C,c_{\Pi})$ are $\Pi$-equivalent.
\end{proof}

\paragraph{Global irrelevant sets.}

Let $(G,u) $ be a rooted embedding and let $q\in\mathbb{N}_{\ge 2}.$
We set $h\coloneqq 2(q-1)+2q.$
Let $\Ccal_{h}$ be the set of all $u$-aligned cycles of $G$ whose radial distance from $u$ is equal to $2 \pmod h$ and we set $I:=\bigcup_{C\in\Ccal_{h}}\mathsf{Irr}_{G}(C,q).$
We call $I$ the \emph{global $q$-irrelevant set} of $(G,u).$
It is easy to see that $2$-edge-connectivity of $G$ implies that $G-I$ is also $2$-edge-connected.

By repeatedly applying~\Cref{equiv_lem} for all cycles in $\Ccal_{h},$ we can show that the removal of the global $q$-irrelevant set from an instance gives an equivalent one.

\begin{lemma}
	\label{eq_irrel}
	Let $\Pi$ be a rooted graph problem
	that has the insulation property.
	Also, let $\mathbf{G}$ be a $\Sigma$-embedded rooted graph and let $J$ be a connected set of edges of $G$ such that $G/J$ is planar and $R(\mathbf{G})\subseteq V(J).$
	If $I$ is the global $c_{\Pi}$-irrelevant set for the reduced embedding pair $(G',u_{J}),$
	then $\mathbf{G}$ and $\mathbf{G} - I$ are $\Pi$-equivalent.
	\end{lemma}
	
	\begin{proof}
	Let $(T,t_{0},\chi)$ be the radial distance decomposition of $(G',u_{J}).$
	Keep in mind that, because of~\Cref{lemma_digons}, all $u_{J}$-aligned cycles of $G'$ that correspond to vertex-face edges $e'=(v',f')$ of $T$ where $f'$ is not a leaf are also cycles of $G.$ 

 We set $h:=2(c_{\Pi}-1)+2c_{\Pi}.$
	Let $E_{h}$ be the set of all vertex-face edges $e=(v,f)$ of $T$ so that the distance between $t_{0}$ and $v$ is equal to $2\pmod h.$ Also, consider a linear ordering $e_{1},\ldots,e_{d}$ of $E_{h}$ such that if $e_{i}=(v_{i},f_{i}),$ $e_{j}=(v_{j},f_{j}),$ and $v_{j}$ is a descendant of $v_{i},$ then $i<j.$
	The latter property implies that if $C_{e_j}$ is in the interior of $C_{e_i},$ then $i<j.$

	We set $\mathbf{G}_{0}:=\mathbf{G}$ and for every $i\in[p],$ we set $\mathbf{G}_{i}$ be the rooted graph obtained from $\mathbf{G}_{i-1}$ after the removal of all vertices in $\mathsf{Irr}_{G}(C_{e_i},c_{\Pi}).$

	In order to show that $\mathbf{G}_{i}$ and $\mathbf{G}_{i-1}$ are $\Pi$-equivalent, we distinguish two cases.
	If there is a $j<i$ such that
	$C_{e_i}$ contains a vertex of $\mathsf{Irr}_{G}(C_{e_j},c_{\Pi}),$ then by~\Cref{obs_interior} 
	$C_{e_i}\subseteq \mathsf{Irr}_{G}(C_{e_j},c_{\Pi}).$
	This in turn implies $\mathsf{Irr}_{G}(C_{e_i},c_{\Pi})\subseteq \mathsf{Irr}_{G}(C_{e_j},c_{\Pi})$ and therefore $\mathbf{G}_{i}=\mathbf{G}_{i-1}.$
	If for every $j<i,$ we have that $C_{e_i}$ is disjoint from $\mathsf{Irr}_{G}(C_{e_j},c_{\Pi})$ then the interior of $C_{e_i}$ is also disjoint from $\mathsf{Irr}_{G}(C_{e_j},c_{\Pi})$ for every $j<i$ because of the definition of the ordering. Therefore, the $c_{\Pi}$-irrelevant set of $C_{e_i}$ in $G$ and in $G_{i-1}$ is the same, i.e., 
	$\mathsf{Irr}_{G}(C_{e_i},c_{\Pi}) = \mathsf{Irr}_{G_i}(C_{e_i},c_{\Pi}).$ By applying~\cref{equiv_lem} for the rooted graph $\mathbf{G}_{i-1}$ and the cycle $C_{e_i},$ we get that $\mathbf{G}_{i}$ and $\mathbf{G}_{i-1}$ are $\Pi$-equivalent. 
	Thus, $\mathbf{G}$ and $\mathbf{G}_{i}$ are $\Pi$-equivalent for every $i\in[p],$ which in particular means that $\mathbf{G}$ and $\mathbf{G}_{p}=\mathbf{G}-I$ are $\Pi$-equivalent.
	\end{proof}

\subsection{Finding the irrelevant vertices}
\label{subsec_algo_irr}

We next show how to compute the $q$-irrelevant set for each slice and the global $q$-irrelevant set in time that is linear in the size of the split and in the size of the instance, respectively.

\begin{lemma}
	\label{algo_irrelevant_set}
	There is an algorithm that given a rooted embedding $(G,u),$ a $u$-aligned cycle $C$ of $G,$ an integer $q\in\mathbb{N}_{\ge 2},$ and the set of cycles $\mathsf{Cycles}(C,h),$ where $h= 2(q-1)+2q,$ computes the $q$-irrelevant set for $C$ in time $\Ocal_{q}(|D|),$ where $D=\mathsf{Slice}_{G}^{h}(C).$ 
\end{lemma}	
	
\begin{proof}
	The algorithm first computes $\mathsf{bfc}(C,2(q-1),q)$ in time $\Ocal_{q}(|H|)$ using the algorithm of~\Cref{lemma_courcelle}. Then, it outputs the union of sets of vertices of $G$ in $\mathsf{Interior}(C',2q)$ for every $C'\in \mathsf{bfc}(C,2(q-1),q).$
\end{proof}

Let us now describe an algorithmic procedure that, given a rooted embedding $(G,u)$ and an integer $q\in\mathbb{N}_{\ge 2},$ outputs a vertex set $I.$ After presenting this procedure, we show that the output is the $q$-global irrelevant set and that it runs in linear time.

\begin{tabbing}
~~~~~~~~\=~~~~~~~~~~~\=~~~~~\=~~~~~\=~~~~~\=~~~~~\=\\

Algorithm {\bf global\_trim}$(G,u,q)$\\

{\sf Input}: \= a rooted embedding $(G,u)$ and a $q\in\Nbbb_{≥2}.$
 \\
 {\sf Output}: \= a vertex set $I\subseteq V(G).$\\

 1. \> Compute the radial distance decomposition $(T,t_{0},\chi)$ of $(G,u)$ and\\
 \>\> a function that maps every non-rooted vertex-face edge $e$ of $T$ to the cycle $C_{e}.$\\

2. \> Let $h:=2(q-1)+2q$ and let $I=\emptyset.$\\

3. \> Let $E_{h}$ be the set containing every vertex-face edge $e=(v,f)$ of $T$ \\
\>\> where the distance in $T$ between $t_{0}$ and $v$ is $2\pmod h.$ \\

4. \> 
Consider a linear ordering $\Lcal:=[e_{1},\ldots e_{d}]$ of $E_{h}$ such that \\

\>\> if $e_{i}=(v_{i},f_{i}),$ $e_{i'}=(v_{i'},f_{i'})$ and $v_{i'}$ is a descendant of $v_{i}$
then $i<i'.$\\

5. \> Let $e$ be the first edge of $\Lcal.$\\

6. \> Let $G:=G-\mathsf{Irr}_{G}(C_{e},q).$\\

7. \> Remove from $\Lcal$ the edge $e$ and every edge $e'$ where $V(C_{e'})\cap \mathsf{Irr}_{G}(C_{e},q)\neq\emptyset.$\\

8. \> Let $I:=I\cup \mathsf{Irr}_{G}(C_{e},q).$\\

9. \> If $\Lcal$ is non-empty, go to Step 5, otherwise output $I.$
\end{tabbing}

We now show that the above algorithm outputs in linear time the global $q$-irrelevant set.

\begin{lemma}
\label{alg_basic_trim}
The algorithm {\bf global\_trim}$(G,u,q)$
outputs the global $q$-irrelevant set $I$ in
time $\Ocal_{q}(|G|).$
\end{lemma}

\begin{proof}
Let $E$ be the subset of $E_{h}$ consisting of all edges $e\in E_{h}$ for which $\mathsf{Irr}_{G}(C_{e},q)$ was added in Step 8.
We claim that the union of $\bigcup_{e\in E}\mathsf{Irr}_{G}(C_{e},z)$ for all $e\in E$ is the $z$-irrelevant set for $(G,u),$ i.e., $\bigcup_{e\in E_h}\mathsf{Irr}_{G}(C_{e},z)=\bigcup_{e\in E}\mathsf{Irr}_{G}(C_{e},z).$

To show this, we argue that for every $e'\in E_{h},$ there is an $e\in E$ such that $\mathsf{Irr}_{G}(C_{e'},z)\subseteq \mathsf{Irr}_{G}(C_{e},z).$
Indeed, note that an edge $e'\in E_{h}\setminus E$ was not considered in Step 8 because it was discarded in Step 7 from $\Lcal.$
Therefore, there is some $e\in E$ such that $V(C_{e'})\cap \mathsf{Irr}_{G}(C_{e},q)\neq \emptyset.$
Since $V(C_{e'})\cap \mathsf{Irr}_{G}(C_{e},q)\neq \emptyset,$ we have that $C_{e'}$ intersects the interior of a cycle $C'\in\mathsf{Cycles}(C,h).$
By~\Cref{obs_interior}, every vertex in the interior of $C_{e'}$ is also in the interior of $C'.$
Therefore every vertex of $G$ in the interior of $C_{e'}$ is contained in $\mathsf{Irr}_{G}(C_{e},q).$
Since $\mathsf{Irr}_{G}(C_{e'},q)$ is a subset of the vertices in the interior of $C_{e'},$ we get $\mathsf{Irr}_{G}(C_{e'},q)\subseteq \mathsf{Irr}_{G}(C_{e},q).$

We next argue that the algorithm {\bf global\_trim}$(G,u,q)$ terminates in linear time.
Note that because of~\Cref{lem_comp_rdd}, we can compute the radial distance decomposition $(T,t_{0},\chi)$ of $(G,u)$ in linear time.
Also, because of~\Cref{lem_comp_align}, in linear time we can get a function that maps every non-rooted vertex-face edge of $T$ to the cycle $C_{e}.$
Now, for each $e\in E,$ we set $D_{e}:=\mathsf{Slice}_{G}^{h}(C_{e}).$
Observe that the set $\mathsf{Irr}(C_{e},q)$ can be computed in time $\Ocal_{q}(|V(D_{e})|),$ using the algorithm of~\Cref{algo_irrelevant_set}.
The claimed bound on the running time of {\bf global\_trim}$(G,u,q)$ follows by observing that $\Sigma_{e\in E}|V(D_{e})|\le 2 |V(G)|.$ To see why the latter inequality holds, note that for every two distinct $e,e'\in E,$ the cycles $C_{e},C_{e'}$ do not intersect and $V(D_{e})\setminus V(C_{e})$ and $V(D_{e'})\setminus V(C_{e'})$ are disjoint sets. Thus $\Sigma_{e\in E}|V(D_{e})|= \Sigma_{e\in E}|V(D_{e})\setminus V(C_{e})| + \Sigma_{e\in E}|V(C_{e})|\le 2|V(G)|.$
\end{proof}

\subsection{Bounding the treewidth}
\label{bstwso9o}

As mentioned in~\Cref{outline_pr}, we show that the removal of the global irrelevant set gives a graph of bounded treewidth. In fact, we show that in a rooted embedding where there is no large nested sequence of aligned cycles with a large flow from the inner one to the outer one, the treewidth is bounded. Let us formally define the aforementioned property.

Let $(G,u)$ be a rooted embedding and let $r\in\mathbb{N}_{\ge 1}$ and $q\in\Nbbb_{≥2}.$
We say that $(G,u)$ is \emph{$(r,q)$-thin} if for every two $u$-aligned cycles $C,C'$ of $G$ where $V(C')$ is embedded in the interior of $C$ and their radial distance is at least $2r$ it holds that ${\sf flow}_{G}(C,C')<q.$

The next lemma states that if $(G,u)$ is $(r,q)$-thin, then the treewidth of $G$ is bounded by $\mathcal{O}(r+q).$ Before proceeding to the proof, let us sketch its main ideas.
Supposing towards a contradiction that the treewidth of $G$ is large enough, we get a large railed nest. While the cycles in $\Ccal$ are not guaranteed to be $u$-aligned, we show that $u$-aligned cycles are ``roughly parallel'' to the cycles of $\Ccal.$
This is based on the observation that if a $u$-aligned cycle $C$ crosses many cycles of $\Ccal,$ then the same will happen to an other $u$-aligned cycle $C'$ in an appropriate radial distance. In this case, (sub)paths of cycles of $\Ccal$ will give a large flow between $C$ and $C',$ contradicting thinness. Therefore, many $u$-aligned cycles are ``roughly parallel'' to the cycles $\Ccal$ of the railed nest, and in this case the paths $\Pcal$ of the nest certify the existence of a large flow, contradicting thinness.

\begin{lemma}
\label{thin_tw}
For every rooted embedding $(G,u),$
if $(G,u)$ is $(r,q)$-thin, then $\tw(G)=\Ocal(r+q).$
\end{lemma}

\begin{proof}
We set $d:=2q+3r+1.$

Assume towards a contradiction that 
$G$ has treewidth at least $c\cdot d,$ where $c$ is the constant of \cref{proog_tw}. Then, from \cref{proog_tw}, applied on the rooted graph $(G,u)$ and for $\Sigma=\Sbbb^{2},$
we have that $(G,u)$ contains some $d$-railed nest $W$ whose 
cycle collection is $\Ccal=\{C_{1},\ldots,C_{d}\},$ whose path collection is $\Pcal=\{P_{1},\ldots,P_{d}\},$ and such that 
the interior of $W$ contains some vertex $v.$ For $i\in[d],$
we denote by $\Delta_{i}$ the open disk bounded by $C_{i}$
where $v$ is embedded.

Observe that the radial distance between $v$ and $u$
should be at least $2(d+1)$, because there are at least $d$
cycles separating $v$ and $u.$
Let $(T,t_{0},\chi)$ be the radial distance decomposition of $(G,u).$
Let $t\in V(T)$
such that $v\in \chi(t)$ and consider the 
path in $T$ between $t_{0}$ and $t.$ 
This path has length at least $2(d+1)$ and contains 
at least $d+1$ vertex-face edges. Among them let $e_{1},\ldots,e_{d'},$ for some $d'≥d,$
be those for which (1) the cycle $C_{e_1}$ is not embedded in $\Delta_{1},$ (2) 
are as close as possible to the vertex $t$ in $T,$ and (3) are 
ordered so that those that are closer to $t$ appear earlier. 
Also we consider $\Qcal=[Q_{1},\ldots,Q_{d'}]$ be the sequence of nested cycles so that $Q_{i}=C_{e_{i}}, i\in[d'].$ By the definition of the radial distance decomposition, if $1≤i≤j≤d',$ then the radial distance between $Q_{i}$ and $Q_{j}$ is $2(j-i).$

For every cycle $Q\in\Qcal$
we denote by ${\sf ind}(Q)$ the set $\{h\mid V(C_{h})\cap V(Q)\neq \emptyset\}$
and by ${\sf range}(Q)$ the set $\max {\sf ind}(Q)- \min {\sf ind}(Q).$
Intuitively, ${\sf ind}(Q)$ encodes the set of all cycles of $\Ccal$ that $Q$ intersects and ${\sf range}(Q)$ is the difference (in indices) between the first and the last cycle of $\Ccal$ that $Q$ intersects.
Observe the following
\begin{eqnarray}
\forall i\in[d']\ \min {\sf ind}(Q_{i}) & ≤ & i, \label{iok2o}\\
\mbox{if~} 1≤i≤j≤d, \mbox{~then~} \max {\sf ind}(Q_{i}) & ≤ & \max {\sf ind}(Q_{j}), \text{ and} \label{secf}\\
\mbox{if~} 1≤i≤j≤d, \mbox{~then~} \min {\sf ind}(Q_{j})& ≤ & \min {\sf ind}(Q_{i})+j-i.\label{iopl}
\end{eqnarray}
Let us comment how to get these observations:
\eqref{iok2o} follows because the radial distance between $v$ and $Q_{i}$
is $2i.$ 
\eqref{secf} follows because $v$ is embedded in $\Delta_{Q_{i}}$ and $\Delta_{Q_{i}}\subseteq \Delta_{Q_{j}}.$
To see \eqref{iopl} it is enough to observe that for every vertex $x$ of $Q_{i}$ there is a vertex of $Q_{j}$ within radial distance $2(j-i)$ from $x$ and this, in turn, implies that $\min {\sf ind}(Q_{j})-\min {\sf ind}(Q_{i})$ is upper-bounded by the half of the radial distance between $Q_{i}$ and $Q_{j}.$

We next show that the range of each cycle $Q_{i}$ in $\Qcal$ is small.\medskip

\noindent{\sl Claim}: for every $i\in[d-(r+q+1)],$
it holds that ${\sf range}(Q_{i})≤r+q.$ 
\smallskip

\noindent{\sl Proof of claim}: Suppose that ${\sf range}(Q_{i})≥r+q+1.$
We set $\ell:=\min {\sf ind}(Q_{i}).$
Due to~\eqref{iok2o}, $\ell≤i.$
Observe that because of \eqref{secf} and \eqref{iopl}, for every $1≤i≤j≤d,$ it holds that ${\sf range}(Q_{j})\geq {\sf range}(Q_{i})-(j-i).$
This in turn implies that each of the $r+1≥r$ cycles in $\{Q_{i},\ldots,Q_{i+r}\}$ intersect each of the $q$ pairwise disjoint cycles in $\{C_{\ell+r+1},\ldots,C_{\ell+r+q+1}\}.$ This certifies 
that there is a flow of size $q$ from $Q_{i}$ to $Q_{i+r},$ a contradiction to the fact that $(G,u)$ is $(r,q)$-thin, as the radial distance between $Q_{i}$ and $Q_{i+r}$ is $2r.$\hfill$\diamond$
\medskip

We next set $y$ to be minimum such that $\max {\sf ind}(Q_{y})≥r+q+1.$ 
This means that either $y=1,$ or $\max {\sf ind}(Q_{y-1})≤r+q.$ 
From the above Claim, we have that $\min {\sf ind}(Q_{y})≥1$. To see why this holds note that if $y=1,$ then by the fact that $\min {\sf ind}(Q_{y})=\max {\sf ind}(Q_y)-\mathsf{range}(Q_y)\ge r+q+1-\mathsf{range}(Q_y)\ge 1,$ where the last inequality holds because of the above Claim. In case $y\ge 2,$ we use the fact that $\min {\sf ind}(Q_i)\le \min {\sf ind}(Q_j)$ for every $1\le i\le j \le d,$ which holds because $\mathcal{Q}$ is a sequence of nested cycles. From this and the fact that, as we showed just before, $\min {\sf ind}(Q_1)\ge 1,$ we get that $\min {\sf ind}(Q_{y})≥1$ also if $y\ge 2.$

The fact that $\min {\sf ind}(Q_{y})≥1$ implies that the whole $Q_{y}$ is embedded outside the open disk $\Delta_{1}.$ Consider the cycles $\{Q_{y},\ldots,Q_{y+r}\}$
and observe that, from \eqref{iopl}, 
$\min {\sf ind}(Q_{y+r})\leq \min {\sf ind}(Q_{y})+r.$ By the above Claim, we also have $\max {\sf ind}(Q_{y+r}) ≤ \min {\sf ind}(Q_{y+r})+q+r.$
Therefore, 
\begin{eqnarray*}
\max {\sf ind}(Q_{y+r}) & ≤ & \min {\sf ind}(Q_{y})+r+q+r\\
& ≤^{}& \min {\sf ind}(Q_{y-1})+1+q+2r \mbox{~~~(because of \eqref{iopl})}\\
& ≤ & \max {\sf ind}(Q_{y-1})+1+q+2r \mbox{~~~(because of \eqref{secf})}\\
&≤ & r+q+1+q+2r =d. \mbox{~~~(because $\max {\sf ind}(Q_{y-1})≤r+q.$)}
\end{eqnarray*}

We conclude that $Q_{y+r}$ is embedded inside the closure of the disk $\Delta_{d}.$
Then any $q$ of the $d$ paths of $\Pcal$ should intersect all 
cycles $Q_{y},\ldots,Q_{y+r}.$ Therefore there is a flow of size $q$ 
from $Q_{y}$ to $Q_{y+r},$ a contradiction to the fact that $(G,u)$ is $(r,q)$-thin, as the radial distance between $Q_{y}$ and $Q_{y+r}$ is $2r.$
\end{proof}

We next show that the removal of the global $q$-irrelevant set $I$ from $G$ yields an $(h,q)$-thin graph, where $h=2(q-1)+2q.$ Then, due to~\Cref{thin_tw}, it follows that $\tw(G-I)$ is a linear function of $q.$

\begin{lemma}
\label{final_tw}
Let $(G,u)$ be a rooted embedding and let $q\in \Nbbb_{≥2}.$
If $I$ is the global $q$-irrelevant set for $(G,u),$ then $\tw(G-I)=\Ocal(q).$
\end{lemma}

\begin{proof}
 We set $h=2(q-1)+2q.$
 We show that $(G-I,u)$ is $(h,q)$-thin, which together with \cref{thin_tw} implies that $\tw(G-I)=\Ocal(h+q)=\Ocal(q).$
 To see why $(G-I,u)$ is $(h,q)$-thin, suppose towards a contradiction that there exist two $u$-aligned cycles $C,C'$ of $G-I$ with the following properties: $V(C')\subseteq \Delta_{C},$ their radial distance is at least $2h,$ and $\mathsf{flow}_{G}(C,C')\ge q.$
 Then, there exist two $u$-aligned cycles $\tilde{C},\tilde{C}'$ of $G$ such that $\Delta_{C'}\subseteq \Delta_{\tilde{C}'}\subseteq \Delta_{\tilde{C}}\subseteq \Delta_{C}$ and the radial distance from both $\tilde{C}$ and $\tilde{C}'$ to $u$ is equal to $2 \pmod h.$ Moreover, note that $\mathsf{flow}_{G}(C,C')\ge q$ implies that $V(\tilde{C}')\subseteq \mathsf{Irr}_{G}(\tilde{C},q),$ a contradiction to the assumption that $\tilde{C}'$ is a cycle of $G-I.$
\end{proof}

\section{Dealing with surface-embeddable multi-rooted graphs}
\label{algo_section}

In this section, we wrap-up the proof of~\Cref{main_th} by showing how to deal with the general case of surface-embeddable multi-rooted graphs.

We first show how to find a connected set $J$ of edges of a surface embeddable whose contraction yields a planar graph.
We then show how to enhance this set $J$ in order to guarantee that the roots of $\mathbf{G}$ are spanned by $J,$ while maintaining the property that if the treewidth of $G/J$ is bounded then the treewidth of $G$ is also bounded. We conclude this section by assembling the proof of~\Cref{main_th}.

\subsection{From surfaces to spheres}
\label{subsec_surf_sphere}

We will use the following result that is implicit in the proof of \cite[Lemma 8]{CabelloCdVL12algo}.

\begin{proposition}
\label{prop_cabello_et_al}
There is an algorithm that, given a connected graph $G$ embedded in a surface $\Sigma$ and a vertex $s\in V(G),$
outputs, in linear time, a collection $\Pcal$ of shortest paths of $G$ with the following properties:
\begin{itemize}
	\item $|\Pcal|=\Ocal(\eg(G)),$
	\item the paths in $\Pcal$ have $s$ as a common endpoint, and
	\item $\Sigma\setminus \cupall\Pcal$ is an open disk.
\end{itemize}	
\end{proposition}

While~\cite[Lemma 8]{CabelloCdVL12algo} does not explicitly state that $\Sigma\setminus \cupall\Pcal$ is an open disk, this can be easily derived from its proof.

In fact, in the proof of~\cite[Lemma 8]{CabelloCdVL12algo}, the authors construct $\Pcal$ by picking a subgraph of a cut graph $\mathbb{L}$ that is, in turn, a subgraph of $T+X,$ where $(T,T^{*},X)$ is the tree–cotree decomposition of Eppstein~\cite{Eppstein03dyna}.
Since when cutting open the surface $\Sigma$ along $T+X,$ they obtain a disk, it is easy to observe that the removal of $\cupall\Pcal$ from $\Sigma$ also creates an open disk. 

We next enhance the paths of \cref{prop_cabello_et_al}
by few more shortest paths between vertices of a set of $B$ of terminals.
As asserted in the next lemma, such an enhancement can be done so as to construct a collection of shortest paths whose contraction makes the graph planar.

\begin{lemma}
\label{lem_contract_paths}
There is an algorithm that, given a $\Sigma$-embedded connected graph $G$ and a set $B\subseteq V(G),$
outputs, in linear time, a collection ${\cal P}$ of shortest paths of $G$ with the following properties:
\begin{itemize}
	\item $|{\cal P}|=\Ocal(\eg(G)+|B|),$
	\item $\cupall\Pcal$ is a connected subgraph of $G$ whose vertex set contains $B,$ and
	\item $G/ E(\cupall\Pcal)$ is planar.
\end{itemize}	

\end{lemma}

\begin{proof}
The algorithm works as follows.
It first finds a collection $\Qcal$ of $|B|-1$ shortest paths among the vertices of $B,$ by fixing an arbitrary vertex $b\in B$ and computing a BFS tree with $b$ as root.
Then, it invokes the algorithm of~\cref{prop_cabello_et_al} with the graph $G$ and the vertex $b$ as inputs,
which gives a collection $\Pcal$ of shortest paths with $b$ as a common endpoint. The latter algorithm guarantees that $|\Pcal|=\Ocal(\eg(G))$ and that $\Sigma\setminus \cupall\Pcal$ is an open disk.
It is therefore easy to see that $\Qcal\cup\Pcal$ is a collection of shortest paths that satisfies the first two properties in the statement of the lemma.
Let $J$ be the set of edges of all paths in $\Pcal$; we claim that $G/J$ is planar. Indeed, consider the graph $G'$ obtained from $G$ by subdividing once each edge of $G$ that is incident to a vertex of some path in $\Pcal.$ Since $\Sigma\setminus \cupall\Pcal$ is an open disk, $G'$ is embedded in a closed disk $\Delta$ with all the subdivision vertices, i.e., the vertices in $V(G')\setminus V(G)$ lying on the boundary of $\Delta.$ Therefore, by introducing a new vertex that is adjacent to all these vertices in the boundary of $\Delta,$ we get a graph that is planar and isomorphic to $G/J.$ Planarity of $G/J$ implies that if we further contract the edges of all paths in $\Qcal$ we still get a planar graph. 
\end{proof}

As we explain in the next subsection, the above result (\Cref{lem_contract_paths}) will be applied for radial graphs.
Therefore, we need to relate the contraction of paths in $R_{G}$ to the contraction of edges in the original graph $G.$ For this, we have to associate edges of $R_{G}$ that have to be contracted, to edges of $G$ that have to be contracted in order to achieve planarity for $G,$ while not
 changing significantly the radial distances in $G.$ 

\begin{figure}[ht]
	\centering
	\includegraphics[width=6cm]{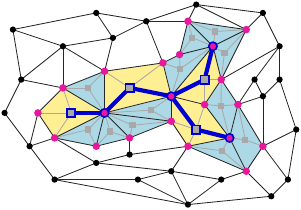}
	\caption{An example of an $\Sbbb^{2}$-embedded graph $G$ (its vertices are depicted as disk-shaped nodes) and a subgraph $H$ of its radial graph $R_{G},$ depicted in blue.
	The faces of $G$ that correspond to vertices of $H$ are also drawn as yellow regions. The set $Z(H)$ consists of the magenta-colored disk-shaped vertices and the gray-colored square-shaped vertices; all vertices are vertices of the radial graph $R_{G}.$
	The faces of $G$ that correspond to vertices of $H$ are also depicted as yellow regions. The faces of $G$ that are incident to vertices of $V(G)\cap V(H)$ are also depicted as cyan regions. The set $J(H)$ is set of gray-colored edges while the set $E_{H}(G)$ is the set of all edges of $G$ that are incident to colored faces.}
	\label{fig_subg_radial}
\end{figure}

Let $G$ be a $\Sigma$-embedded graph and let $H$ be a subgraph of its radial graph $R_{G}.$ 
Let $Z(H)$ be the set of all vertices of $R_{G}$ whose distance from some vertex of $V(G)\cap V(H)$ is at most two. 
We denote by $J(H)$ the set of all edges of $R_{G}$
with both endpoints in $Z(H).$
Also we denote by $E_{H}(G)$ the set of all edges of $G$
where both their endpoints belong to $Z(H)$; in other words $E_H(G)\coloneqq E(G[Z(H)])$.
See~\Cref{fig_subg_radial} for an illustration of the above notions.

\begin{lemma}
	\label{lem_prop_contr}
Let $G$ be $\Sigma$-embedded graph and let $H$ be a subgraph of its radial graph $R_{G}.$ 
Then the following hold:
\begin{enumerate}
\item If $H$ is connected then $E_{H}(G)$ is a connected set of edges in $G.$
\item If $R_{G}/E(H)$ is planar, then $G/E_{H}(G)$ is planar.
\item Let $v$ be a vertex in $V(G)\setminus V(E_{H}(G))$ and let $u$ be a vertex in $V(G)\cap V(H).$ Then the radial distance between $v$ and $u$ is at least four.
\end{enumerate}
\end{lemma}

\begin{proof}
	We use $N_{1}$ to denote the set $V(G)\cap V(H)$ and $N_{2}$ to denote the set of all vertices of $G$ that are in distance two from a vertex of $N_{1}$ in $R_{G}$; note that $\{N_{1},N_{2}\}$ is a partition of $Z(H)\cap V(G).$
	Also, keep in mind that $E_{H}(G) = E(G[N_{1}\cup N_{2}]).$

	We first show that $R_{G}/J(H)$ is the radial graph of $G/E_{H}(G).$
	Let $Q$ be the subgraph of $G$ induced by the edges of $E_{H}(G).$
	Observe that when contracting all edges of $E_{H}(G)$ in $G,$ every connected component of $Q$ becomes a single vertex.
	Each such vertex is incident to the faces of $G$ that were adjacent to some vertex in $N_{2}.$ Similarly, the contraction of the edges of $J(H)$ in the radial graph $R_{G}$ gives a set of new vertices, each adjacent to the vertices of $R_{G}$ in the neighborhood of the vertices of $N_{2}.$ In order to derive that $R_{G}/J(H)$ is the radial graph of $G/E_{H}(G),$ observe that there is a bijection between the vertices of $G$ that are result of the contraction of the edges from $E_{H}(G)$ and the vertices of $R_{G}$ that are result of the contraction of the edges from $J(H).$

	Property (1) follows from the fact that $E_{H}(G) = E(G[N_{1}\cup N_{2}])$ and that if $H$ is connected, then $G[N_{1}\cup N_{2}]$ is connected.
 To see why the latter holds, pick any three distinct vertices $x,y,z$ in $H$ with $xy,yz\in E(H)$ and 
 $x,z$ corresponding to vertices of $G$ (thus $x,z\in N_1$) and $y$ corresponding to a face $f_y$ of $G.$ Note that every vertex of $G$ on the boundary of the face $f_y$ is at distance exactly two from $x$ and $z$ in $R_G$ and thus belongs to $N_2.$ Therefore, by following the edges (in $G$) along the boundary of $f_y,$ we can obtain a path in $G$ from $x$ to $z$ containing only vertices of $N_1\cup N_2.$ 
 Therefore, for any two vertices in $N_1,$ since $H$ is connected, there is a path in $G[N_1\cup N_2]$ connecting them.
 In the same way, any vertex $w\in N_2$ and vertex $x\in N_1$ with $w$ and $x$ being at distance exactly two in $R_G$ are connected by a path in $G[N_1\cup N_2].$ This way we conclude that connectivity of $H$ implies that $G[N_1\cup N_2]$ is connected.

Property (2) follows by observing that $R_{G}/J(H)$ is a minor of $R_G/E(H),$ combined with the fact that, as shown above, $R_{G}/J(H)$ is the radial graph of $G/E_{H}(G)$ and that if the radial graph of a surface-embedded graph is planar, then the graph itself is planar.

	For Property (3), note that
	since each vertex in $v\in V(G)\setminus V(E_{H}(G))$ is in radial distance at least two from every vertex in $N_{2}$ and every vertex in $N_{2}$ is in radial distance at least two from any vertex $u$ in $N_{1},$ we get that the radial distance between $v$ and $u$ is at least four.
\end{proof}

\subsection{Maintaining the bound on treewidth}
\label{main_tw}

In order to prove that the contraction of the edges of the shortest radial paths given in the previous subsection does not yield big changes in the treewidth of the graph, we will use a result proved by Demaine, Hajiaghayi, and Kawarabayashi in \cite{DemaineHK11cont}.

This states that in a surface-embedded graph, if there is a large enough wall $W$ whose perimeter bounds a closed disk where $W$ is embedded, every collection of shortest radial paths with terminals outside this disk is guaranteed to not touch a sufficiently large wall in this disk.

Let $G$ be a graph that contains as a subgraph a wall $W$ whose boundary is a contractible cycle bounding a closed disk $\Delta$ where the entire $W$ is embedded. The graph $G\cap \Delta$ is called the \emph{compass} of $W$ in $G.$ The following result can be easily extracted from the proof of \cite[Theorem~6]{DemaineHK11cont}.
After stating the result, we give a very brief sketch of the proof of~\cite[Theorem~6]{DemaineHK11cont}.

%

%

%
\begin{proposition}
\label{dem_contr}
There is a function $\lambda:\Nbbb\to\Nbbb$ 
such that for every $r,k\in\mathbb{N}$ the following holds.
Let $G$ be a $\Sigma$-embedded graph that contains as a subgraph an $(r+\lambda(k))$-wall $W$ whose perimeter is a contractible cycle bounding a closed disk $\Delta$ where the entire $W$ is embedded.
If $\Pcal$ is a collection of $k$ shortest radial paths in $G$ whose endpoints are disjoint from the compass of $W,$
then there is an $r$-wall $W'$ embedded in $\Delta$ such that the vertex set of each path in $\Pcal$ is disjoint from the compass of $W'.$
Moreover, $\lambda$ is a linear function.
\end{proposition}

%

Let us sketch how~\cref{dem_contr} can be extracted from~\cite{DemaineHK11cont}. As described in~\cite[Subsection~4.2]{DemaineHK11cont}, by starting with a sufficiently large wall $W$ as above, one may first find a still sufficiently large wall $W_0$ whose compass is also disjoint from the radius-$2r$ radial neighborhoods of the endpoints of the paths in $\mathcal{P}$. Since $W_0$ is sufficiently large, by losing a constant factor, we can still find a large wall $W_1$ (inside the compass of $W_0$) whose compass is disjoint from $P_1$; see~\cite[Lemma~8]{DemaineHK11cont}. By repeating this argument recursively $k$ times, we obtain a wall $W_k$ whose compass is disjoint from all paths in $\mathcal{P}$, with a constant additive loss in $k$ in the size of the initial wall $W$.

We are now ready to show the next result which gives a linear-time reduction from planar sin gle-rooted to surface-embedded multi-rooted graphs.

\begin{lemma}
\label{trans_rad}
	There is an algorithm
	that, given a $\Sigma$-embedded graph $G$ and a set $B\subseteq V(G),$
outputs, in linear time, a collection $\Pcal$ of $\Ocal(|B|+\eg(G))$ many shortest radial paths in $G$ such that, given that $H=\cupall\Pcal,$ the following hold:
\begin{itemize}
	\item $B\subseteq V(E_{H}(G)),$
	\item $E_{H}(G)$ is connected in $G,$
	\item $G':=G/E_{H}(G)$ is planar, and
	\item $\tw(G)=\Ocal((|B|+1)^{3/2}\cdot(\eg(G)+1)^{5/2}\cdot \tw(G')).$
\end{itemize}
\end{lemma}

\begin{proof}
The algorithm invokes the algorithm of~\Cref{lem_contract_paths} with the radial graph $R_{G}$ of $G$ and the vertex set $B\subseteq V(G)$ as input; note that the radial graph of every $\Sigma$-embedded graph is a connected graph. Therefore, in linear time, we get a collection $\Pcal$ of $\Ocal(\eg(G))$ many shortest paths of $R_{G}$ such that if we set $H=\cupall \Pcal$ then $H$ is a connected subgraph of $R_{G}$ whose vertex set contains $B$ and $R_{G}/E(H)$ is planar.
Since $B\subseteq V(H)\cap V(G)$ and $V(H)\cap V(G)\subseteq V(E_{H}(G))$ (by definition), we get that $B\subseteq V(E_{H}(G)).$
By (1) of~\Cref{lem_prop_contr} we get that connectivity of $H$ implies that $E_{H}(G)$ is a connected set of edges in $G.$
Also, since $R_{G}/E(H)$ is planar, we have that $G/E_{H}(G)$ is also planar (due to (2) of~\Cref{lem_prop_contr}).

To show that if $\tw(G')\le t,$ then $\tw(G)=\Ocal((|B|+1)^{3/2}\cdot(\eg(G)+1)^{5/2}\cdot t),$ we argue as follows.
Let $l:=t+5$ and $d:=l+\lambda(|\Pcal|),$ where $\lambda$ is the function from~\Cref{dem_contr}.

Assume towards a contradiction that $\tw(G)>c\cdot d\cdot \sqrt{2|\Pcal|+1}\cdot (\eg(G)+1),$ where $c$ is the constant of~\Cref{thi_oplos_s}.
Then, $G$ contains as a subgraph a $d$-wall $W$ whose perimeter is a contractible cycle bounding a closed disk $\Delta$ where the entire $W$ is embedded such that $\Delta$ does not contain any of the endpoints of the paths in $\Pcal.$
Also, since $d=l+\lambda(|\Pcal|),$
by~\Cref{dem_contr}, we get that there is an $l$-wall $W'$ embedded in $\Delta$ such that no vertex of $V(H)\cap V(G)$ is a vertex of the compass of $W'.$
By (3) of~\Cref{lem_prop_contr}, we have that $V(E_{H}(G))$ is disjoint from the vertex set of the compass of an $(l-4)$-wall $\tilde{W}.$
Therefore, since $l=t+5,$ $\tilde{W}$ is a $(t+1)$-wall in $G/E_{H}(G),$ which implies that $\tw(G/E_{H}(G))\ge \tw(\tilde{W})>t,$ a contradiction.
\end{proof}


\subsection{Wrap-up: proof of \texorpdfstring{\cref{main_th}}{\cref{main_th}}}

We are now ready to prove \cref{main_th}.
\begin{proof}[Proof of~\Cref{main_th}]
We first apply \cref{trans_rad} and compute a collection $\Pcal$ of shortest radial paths in $G$ such that if $H:=\cupall\Pcal$ and $J:=E_{H}(G),$
then $J$ is a connected set of edges in $G$ such that $G/J$ is planar, $R(\mathbf{G})\subseteq V(J),$ and $\tw(\mathbf{G})=\Ocal((b+1)^{3/2}\cdot (g+1)^{5/2}\cdot \tw(G/J)).$
We next consider the reduced embedding pair $(G',u_{J})$ of $(G,J)$
obtained from $G$ by contracting all edges in $J$ (this way $u_{J}$ is defined as the result of these contractions) and then taking the $2$-edge-connected digon-completion of $G.$
We compute the global $c$-irrelevant set $I$ of $(G',u_{J})$ using the algorithm {\bf global\_trim}$(G',u_{J},c),$
which runs in time $\mathcal{O}_{c}(|G'|)=\mathcal{O}_{c}(|G|)$ as showed in~\cref{alg_basic_trim}.
Due to~\cref{eq_irrel}, $\mathbf{G}$ and $\mathbf{G}-I$
are $\Pi$-equivalent.
Also, by \cref{final_tw}, the treewidth of $G'-I$ is a linear function of $c$.
Since $G/J-I$ is a subgraph of $G'-I,$ then $\tw(G/J)\le \tw(G'-I)$ and therefore $\tw(\mathbf{G})=
\Ocal(c\cdot (g+1)^{5/2}\cdot (b+1)^{3/2}).$
\end{proof}

\section{Problems that have the insulation property}
\label{gen_esxopro}

Instead of enumerating problems that have the insulation property, we prefer to give two general families that contain most of them.
They express several containment and/or modification problems and can be all solved in linear time for graphs of bounded genus using our result. We need first some definitions.

A \emph{minor model} of a rooted graph $\bound{H} = (H,B_{H},\rho_{H})$ in a rooted graph $\bound{G} = (G,B_{G},\rho_{G})$
is a set $\Mcal=\{X_{v}\mid v\in V(H)\}$ of pairwise vertex-disjoint connected
vertex sets of $G$ where $|B_{H}|=|B_{G}|=:t,$ for every $i\in[t]$
$X_{\rho_{H}^{-1}(i)}\cap B_{G}=\{\rho_{G}^{-1}(i)\},$ and 
for every two vertices $v,u$ of $H$
\begin{eqnarray}
\{v,u\}\in E(H)\Rightarrow G[X_{v}\cup X_{u}] \mbox{~is connected.}\label{inlop} \end{eqnarray}
If we replace $\Rightarrow$ by $\Leftrightarrow$ in \eqref{inlop}
then we say that $\Mcal$ is an \emph{induced minor model} of $H$ in $G.$ We say that ${\bf H}$ is a \emph{minor} (resp. \emph{induced minor}) of ${\bf G}$
if there is a minor (resp. induced minor) model of ${\bf H}$ in ${\bf G}.$
 For every finite set ${\cal H}$ of rooted graphs and every integer $k,$ we define the following problem.
\medskip\medskip

\noindent{\sc $\Hcal$-Minor $k$-Deletion}\\
\noindent{\sl Input}: a rooted graph ${\bf G}=(G,B,\rho)$\\
\noindent{\sl Question}: is there a set $S\subseteq \overline{V}(\mathbf{G})$ where $|S|≤k$ and such that none\\ \phantom{{\sl Question}:} of the rooted graphs in $\Hcal$ is a minor of ${\bf G}-S$?
\medskip\medskip

We also denote by {\sc $\Hcal$-Induced Minor $k$-Deletion}
the problem defined as above if we use induced minors instead of minors.
When restricted to (rooted) graphs of bounded genus, both above problems, and all problems that can be seen as special cases of them, have the insulation property. We next give a brief explanation of how this follows from known combinatorial results on the existence of irrelevant vertices.

We first consider {\sc $\Hcal$-Minor $k$-Deletion} and we start with the case where $k=0.$ This 
expresses the {\sc Rooted Minor Containment} problem, asking whether ${\bf G}$ contains as a minor some 
of the rooted graphs in ${\cal H}.$
The insulation property for this problem on surface-embedded graphs follows from~\cite[Theorem~5.9]{BasteST19hittIV}, which in case of planar
graphs, can be seen as a direct consequence of \cite[Theorem~5.6]{BasteST19hittIV}. 
We stress that \cite[Theorem~5.6]{BasteST19hittIV} is, in turn, making use of the ``combing lemma''~\cite[Theorem~1.4]{GolovachST23comb}, which is based on the {linkage} theorem \cite{ReedRSS91find,Reed95rooted} (see \cite{AdlerKKLST17irre} for the case of planar graphs and in \cite{Mazoit13asin} for the case of surface-embedded graphs).
For surface-embedded graphs, a version of \cite[Theorem~5.6]{BasteST20hittI} can be proven for the induced minor setting as well by using the ``scattered version'' of the combing lemma~\cite[Corollary~3.6]{GolovachST23comb}, which in turn is based on the ideas of the ``induced linkage'' results of Kawarabayashi and Kobayashi in \cite{KawarabayashiK08thei}.
This implies that also the 
{\sc Rooted Induced Minor Containment} problem, which is {\sc $\Hcal$-Induced Minor $k$-Deletion} 
for $k=0,$ has the insulation property.
When $k≥1,$ the insulation property for {\sc $\Hcal$-Minor $k$-Deletion} 
follows from \cite[Lemma 16]{SauST21kapiI}, specialized on surface-embedded graphs.
By using the same arguments and having as departure point the induced minor version of \cite[Theorem~5.6]{BasteST19hittIV}, the insulation property follows for {\sc $\Hcal$-Induced Minor $k$-Deletion} as well.

Notice that {\sc $\Hcal$-Minor $k$-Deletion} contains most of the problems where 
an (almost) linear time algorithm implementation of the irrelevant technique is known.
For instance, {\sc Disjoint Paths} is derived when $k=0$ and ${\cal H}$ contains 
only the graph consisting of roots with a matching between them. Also, {\sc Rooted Minor Containment} follows if $k=0$ and ${\cal H}$ consists of the minor we are looking for.
Moreover, {\sc Planarization} follows if ${\cal H}$ contains the two Kuratowski graphs.
For the general case where ${\cal H}$ is a set of graphs, algorithms for the {\sc $\Hcal$-Minor $k$-Deletion} have been 
proposed in \cite{SauST20anfp} (running in cubic time) and \cite{MorelleSST24fast} (running in quadratic time).
For bounded genus graphs, this problem as well as its rooted and/or induced version can be solved 
in linear time, because of our results.

We wish to stress that in the induced minor case few results for the applicability of the irrelevant vertex technique are known. To our knowledge the only problem of this kind that has 
been examined is the {\sc Induced Cycle} problem for surface-embedded graphs in~\cite{KawarabayashiK10algor,KK09algori}. Interestingly, Fellows, Kratochvíl, Middendorf, and Pfeiffer in ~\cite{FellowsKMP95thecom} proved that there is a planar graph $H$
where asking for the induced minor containment of $H$ is an {\sf NP}-complete problem.
As this is the {\sc $\{H\}$-Induced Minor $0$-Deletion} problem, no polynomial 
algorithm may be expected in the induced setting in general graphs.
\medskip

At this point we would like to mention a family of problems with the insulation property that extends further the above framework.
It is the {\sc $\Hcal$-Minor $k$-Elimination Distance} problem, introduced by Bulian and Dawar~\cite{BulianD16grap,BulianD17fixe}
that is defined as {\sc $\Hcal$-Minor $k$-Deletion} where ${\Hcal}$
is a finite set of graphs and with the difference that instead of looking for 
a set $S\subseteq V(G)$ where $|S|≤k$ we ask that the tree-depth 
of ${\sf torso}(G,S)$ is at most $k.$\footnote{For a vertex set $X \subseteq V(G),$ the \emph{torso} of $X$ in $G,$ denoted by ${\sf torso}(G,X),$ is the graph obtained from the induced subgraph $G[X]$ by turning $N_{G}(V(C)),$ the set of vertices in $X$ adjacent to each connected component $C$ of $G - X,$ into a clique. The \emph{tree-depth} of a graph $G$ is the minimum height of a forest $F$ with the property that every edge of $G$ connects a pair of nodes that have an ancestor-descendant relationship to each other in $F.$}
 Using similar arguments as those in \cite[Lemma 16]{SauST21kapiI} in the case of {\sc $\Hcal$-Minor $k$-Deletion}, in \cite{MorelleSST24fast} it was proved that {\sc $\Hcal$-Minor $k$-Elimination Distance} also has the insulation property. In fact, the same arguments 
 can be applied in the more general setting where, instead of tree-depth, we consider any 
 minor-monotone graph parameter that is functionally bigger than treewidth.

\section{Open problems and research directions}
\label{sec_open}

In this paper we gave an algorithmic framework for the linear-time implementation of the irrelevant vertex technique for a wide family of problems, namely the problems that have the insulation property. 
As we already mentioned in the introduction, the applicability of the irrelevant vertex technique extends to more general problems 
that have the following ``weak'' version of the insulation property:
\begin{quote}
{\em There is a constant $c_{\Pi}$ so that for every $\Sigma$-embedded rooted graph ${\bf G}$
with a $c_{\Pi}$-railed nest $W,$ if $R({\bf G})\subseteq {\sf ext}(W)\cap V({\bf G})$ then \underline{some} vertex in ${\sf int}(W)\cap V({\bf G})$ is $\Pi$-irrelevant. }
\end{quote}
A typical problem with the above weak insulation property is the {\sc ${\cal H}$-Topological Minor Containment} problem where, given a graph $G,$ we ask whether it contains a subdivision of $H$ as a subgraph. This problem can be extended to the {\sc ${\bf H}$-Rooted Topological Minor Containment} problem if we correspond a bounded number of vertices of $G$ to each of the roots of ${\bf H}.$
More generally, most known problems that have the above 
weak insulation property are captured by the meta-algorithmic 
frameworks developed in \cite{GolovachST22model,FominGSST23comp,FominGST22anal,SchirrmacherSST24mode,SauST2024parame}.
These algorithmic meta-theorems apply for graphs 
of fixed Hadwiger number\footnote{The \emph{Hadwiger number} of a graph is the maximum size of a clique minor in it.}
and they are based on an extended notion of a railed-nest that is applicable 
to flat territories of the input graph instead of disk-embeddable ones. 
However, even in planar graphs, the currently fastest algorithm for such problems runs in quadratic time and is based on the repetitive finding (in linear-time) of an irrelevant vertex. To improve the algorithms produced by \cite{GolovachST22model,FominGSST23comp,FominGST22anal,SchirrmacherSST24mode,SauST2024parame} to sub-quadratic ones is an open challenge even for the case of planar graphs. An important step in this 
direction would be to prove that {\sc ${\bf H}$-Rooted Topological Minor Containment} can be solved in linear time in planar graphs.\medskip

\paragraph{Revealing and improving the parametric dependences.}
In our proof of the main result of the paper, that is \cref{main_th},
we did not make any effort to specify the dependencies in the $\Ocal_{c}$ notation on the running time of the algorithm.
The core of these dependencies resides 
in the proof of \cref{lemma_courcelle} that applies the optimization version of Courcelle's theorem (\cref{ow_dec_dource}) 
on $D_{\Ccal}$ whose treewidth is linearly bounded on $c,$ due to \Cref{tw_slice}. Here the optimization question \cref{ow_dec_dource} is asked to resolve is the following:
\begin{quote}
Given a graph $G,$ a set $R\subseteq V(G),$ a vertex $x\not\in R,$ and a non-negative integer $q,$ 
find the maximum size of a set $S\subseteq R$ 
such that for every $s\in S$ there are $q$ internally vertex disjoint 
paths in $G$ from $x$ to $s.$ 
\end{quote}
We claim that the above query
can be answered in time $2^{\Ocal(t\cdot \log t)}\cdot |G|,$ where $t=\tw(G)+q,$ using dynamic programming. 
This would give an algorithm for \cref{main_th} running 
in time $2^{\Ocal(c\cdot \log c)}\cdot |D|.$

We must stress here that the task of improving the parametric dependencies of algorithms for problems where the irrelevant vertex technique is applicable has been quite an active topic during the last years; see e.g.~\cite{ChoOhOh2023PlanarDisjointPaths} and~\cite{WlodarczykZehavi2023PlanarDisjointPaths} for some recent developments.
It is a challenging question whether the ideas in \cite{ChoOhOh2023PlanarDisjointPaths} and \cite{WlodarczykZehavi2023PlanarDisjointPaths} can be extended, perhaps combined with our technique, for wide families of problems having the insulation property.

\paragraph{Linear algorithms beyond surface embeddable graphs.}
Another direction for improvement is to extend \cref{main_th} beyond the class of bounded-genus graphs.
The target here would be again graphs of bounded Hadwiger number. 
In this direction
we believe that our technique is applicable in linear time if we 
are given the decomposition of the Graph Minors Structural Theorem (GMST) proved for such graphs by Robertson and Seymour in \cite{RobertsonS03GMXVI}. However the currently fastest algorithm for the computation of 
this decomposition is the one of Grohe and Kawarabayashi and Reed~\cite{GroheKR13asimpl} that runs in quadratic time. 
A linear time algorithm that outputs such a decomposition appears to be out of reach, for the moment.

\section*{Acknowledgements} We wish to thank the reviewers of the paper for their insightful comments on an earlier version of the paper.

%

\newpage

\appendix

\section{Appendix. Proof of~\texorpdfstring{\cref{segments}}{Lemma 4.3}}
\label{app_resp}

In the proof of~\cref{segments} we will use the fact that there is a exactly one root edge in the tree of the radial distance decomposition of every rooted embedding.

\begin{lemma}\label{lemma_deg1root}
	Let $(T,t_{0},\chi)$ be the radial distance decomposition of a rooted embedding $(G,u).$
	Then the root $t_{0}$ of $T$ has degree one in $T.$
\end{lemma}

\begin{proof}
	Recall that $\chi(t_{0})=\{u\}.$
	We show that the graph $R_{G}\setminus \{u\}$ is connected, which by the definition of the tree distance decomposition implies that $t_{0}$ has exactly one child in $T.$
	Since $G$ is a $2$-edge-connected $\Sbbb^{2}$-embedded graph, by~\Cref{obs_squareface} we have that every face of $R_{G}$ is incident to exactly four vertices. Let $E$ be the set of all edges of $R_{G}$ that are contained in faces of $R_{G}$ incident to $u,$ except from the edges that are incident to $u.$
	Observe that since $G$ is planar and every face of $R_{G}$ is incident to exactly four vertices of $R_{G},$ there is a cycle $C$ in $R_{G}$ whose edge set is a subset of $E.$ It is now easy to see that every path with vertices $f,u, f'$ in $R_{G}$ can be rerouted via $C$ to a path from $f$ to $f'$ that avoids $u.$
\end{proof}

\paragraph{Duals of surface embedded graphs.}
Let $\Sigma$ be a surface and let $G$ be a $\Sigma$-embedded graph.
The \emph{dual graph} of $G,$ denoted by $G^{*},$ is the graph whose vertex set is the set $F(G)$ of faces of $G$ and where two faces are adjacent if they share an edge of $G.$ For every $e\in E(G),$ we denote by $e^{*}$ the edge \emph{dual} to $e,$ which connects the two faces adjacent to $e$ in the embedding of $G.$
For every $f\in F(G),$ we use $f^{*}$ to denote the vertex of $G^{*}$ corresponding to the face $f$ of $G.$
The dual graph $G^{*}$ has a natural embedding in $\Sigma,$ where each vertex $f^{*}$ is assigned to a point $p_{f}$ in the interior of the face $f$ and for each edge 
$e\in E(G),$ incident to faces $f$ and $f',$ the dual edge $e^{*}$ is assigned to a curve that connects the points $p_{f}$ and $p_{f'}$ and crosses $G$ precisely at the edge $e.$
Given a set of edges $E\subseteq E(G),$
we use $E^{*}$ to denote the set $\{e^{*}\mid e\in E\}.$

\paragraph{Connected sets of faces.}
Let $G$ be a $\Sigma$-embedded graph and let
$F$ be a subset of its faces. We say that $F$ is {\em connected} in $G$ if for every two faces $f_{1}, f_{2}$ in $F$
there is a $V(G)$-avoiding arc starting
from a point in $f_{1}$ and finishing to a point in $f_{2}$ and not containing points 
from a face outside $F.$ Observe that $F\subseteq F(G)$ is connected in $G$ 
if and only if $G^{*}[F^{*}]$ is connected.
Keep in mind that the empty subset of $F(G)$ is connected in $G$ (as the definition is vacuously satisfied).

\paragraph{Prefixes and suffixes.}
Let $(T,t_{0},\chi)$ be the radial distance decomposition of a rooted embedding $(G,u).$
Given a node $t\in V(T),$ we define $\mathsf{suffix}_{T}(t)$ as the set of all nodes that are descendants of $t$ (including $t$), 
while we define $\mathsf{prefix}_{T}(t)$ as the subtree of $T$ that is obtained from $T$ after removing the children of $t$ and all their descendants.
We also define $\mathsf{prefix}_{\chi}(t)$ (resp. $\mathsf{suffix}_{\chi}(t)$) as the union of all $\chi(z)$ where $z$ is a node in $\mathsf{prefix}_{T}(t)$ (resp. $\mathsf{suffix}_{T}(t)$).
Given an edge $e=(t,t')$ of $T$ (in this case $t'$ is a child of $t$),
recall that we use $\mathsf{suffix}_{\chi}(e)$ to denote the union of all $\chi(z),$ where $z$ is either $t'$ or a descendant of $t'$ in $(T,t_{0}).$
Also, we use $\mathsf{prefix}_{\chi}(e)$ to denote the union of all $\chi(z),$ where $z$ belongs to the connected component of $T\setminus e$ that contains $t_{0}.$ 

In order to prove~\cref{segments}, we start by showing that the set of faces in the suffix of a vertex-face edge is connected.

\begin{lemma}\label{lem_suffix}
	Let $(T,t_{0},\chi)$ be the radial distance decomposition of a rooted embedding $(G,u).$
	For every vertex-face edge $e=(v,f)\in E(T),$
	the set $F(G)\cap \mathsf{suffix}_{\chi}(e)$ is connected in $G.$
\end{lemma}

\begin{proof}
We use $Z$ as a shortcut for $\mathsf{suffix}_{\chi}(f).$
Keep in mind that $f$ is a node in a face layer of $T,$ the set $N_{Z}$ of all vertices of $R_{G}$ that are adjacent to vertices from $Z$ but do not belong to $Z$ are vertices of $G,$ and $R_{G}[Z]$ is connected.
We set $F=Z\cap F(G)$ and show that $F$ is connected in $G.$
 	Let $g,g'$ be two faces of $F.$ It is enough to prove that there is a path connecting $g$ and $g'$ in $G^{*}[F^{*}].$ Since $Z$ is connected, there exists a path $P$ in $R_{G}[Z]$ whose vertices are $g_{0}=g, u_{1}, g_{1}, \dots, u_{m-1}, g_{m-1}, u_{m}, g_{m}=g',$ starting from $g$ and finishing at $g'$; here, $g_{i},i\in[0,m]$ correspond to faces of $G$ and $u_{i}, i\in[m]$ to vertices of $G.$
 	Also, since $N_{Z}\subseteq V(G),$ for every $i\in[m],$ both $u_{i}$ and all neighbors of $u_{i}$ in $R_{G}$ are in $Z.$
	Observe that for every $u_{i},$
	the set of all faces of $G$ incident to $u_{i}$ is connected in $G.$
 	Thus, there exists a path in $G^{*}[F^{*}]$ between $g_{j-1}$ and $g_{j}$ for every $j\in[m]$ and therefore also a path connecting $g$ and $g'$ in $G^{*}[F^{*}].$
\end{proof}

We next show that 
the set of faces
in the prefix of a vertex-face edge is connected.

\begin{lemma}
	\label{granting}
	Let $(T,t_{0},\chi)$ be the radial distance decomposition of a rooted embedding $(G,u).$
	For every vertex-face edge $e=(v,f)\in E(T),$ the set $F(G)\cap \mathsf{prefix}_{\chi}(e)$ is connected in $G.$
\end{lemma}

\begin{proof}
	Let $E$ be the set of all vertex-face edges $\bar{e}$ of $T$ such that $F(G)\cap\mathsf{prefix}_{\chi}(\bar{e})$ is not connected in $G.$
	Consider such an edge $e'=(v',f')\in E$ such that $v'$ has minimum distance (among all other edges in $E$) from the root $t_{0}$ of $T.$
	We can assume that $v'\neq t_{0},$ since otherwise by~\Cref{lemma_deg1root} $f'$ is the only child of $v'$ and the empty set is trivially a connected set of faces in $G.$

	Let $e=(v,f)$ be the vertex-face edge of $T$ such that $f$ is the parent of $v'$ in $T.$
	By minimality of $e',$ we have that $F(G)\cap\mathsf{prefix}_{\chi}(e)$ is connected in $G.$
	We will argue that $F(G)\cap\mathsf{prefix}_{\chi}(e')$ is also connected in $G,$ arriving to a contradiction to the choice of $e.$ We prove this gradually, by showing that $F(G)\cap\mathsf{prefix}_{\chi}(f),$ then $F(G)\cap \mathsf{prefix}_{\chi}(v'),$ and finally $F(G)\cap \mathsf{prefix}_{\chi}(e')$ is connected in $G.$\medskip

	\noindent\emph{Claim 1:} The set $F(G)\cap\mathsf{prefix}_{\chi}(f)$ is connected in $G.$\medskip

	\noindent\emph{Proof of Claim 1:}
 We use $F$ to denote $F(G)\cap\mathsf{prefix}_{\chi}(f).$
 Note that $\mathsf{prefix}_{\chi}(e)\subseteq \mathsf{prefix}_{\chi}(f).$
	To show that $F$ is connected in $G,$ it suffices to prove that for every face $g\in\mathsf{prefix}_{\chi}(f)\setminus \mathsf{prefix}_{\chi}(e),$
	there is a path in $G^{*}[F^{*}]$ from the vertex $g^{*}$ of $G^{*}$ to some vertex $\tilde{g}^{*}$ of the dual $G^{*}$ corresponding to a face $\tilde{g}\in \mathsf{prefix}_{\chi}(e).$
		
	Let $u$ be a vertex of $G$ that is in $\chi(v)$ and is incident to both the face $g$ and some face in $\mathsf{prefix}_{\chi}(e).$
	Notice every face incident to $u$ is in $\mathsf{prefix}_{\chi}(f).$
	Therefore the set of faces incident to $u$ is partitioned into two sets, one consisting of faces in $\mathsf{prefix}_{\chi}(e)$ and the other consisting of faces in $\mathsf{prefix}_{\chi}(f)\setminus \mathsf{prefix}_{\chi}(e).$
	This implies the existence of the claimed path.\hfill$\diamond$
	\medskip

	We next show that $F(G)\cap\mathsf{prefix}_{\chi}(v')$ is also connected. Before proceeding to the proof, we note that for every edge $(t,t')$ of $T,$ $\mathsf{suffix}_{\chi}(t')=\mathsf{suffix}_{\chi}((t,t')).$
	\medskip

	\noindent\emph{Claim 2:} The set $F(G)\cap\mathsf{prefix}_{\chi}(v')$ is connected in $G.$\medskip

	\noindent\emph{Proof of Claim 2:}
	For every vertex-face edge $e''=(v'',f'')$ of $T$ where $v''$ is a child of $f,$ we denote by $F''$ the set of all faces of $G$ in $\mathsf{prefix}_{\chi}(f)\cup\mathsf{suffix}_{\chi}(e'').$
	
	For every such $e'',$ we aim to show that $F''$ is connected in $G.$ To see this, first
	observe that both $F(G)\cap \mathsf{prefix}_{\chi}(f)$ and $F(G)\cap \mathsf{suffix}_{\chi}(e'')$ are connected in $G$ (because of Claim 1 and~\Cref{lem_suffix}, respectively).
	Therefore to show that $F''$ is connected in $G,$ it suffices to show that given faces $g,g'$ of $G$ where $g$ belongs to $\mathsf{prefix}_{\chi}(f)$ and $g'$ belongs to $\mathsf{suffix}_{\chi}(e''),$ there is a path in the dual $G^{*}$ consisting only of vertices of $G^{*}$ that correspond to faces in $F''.$
	Let $u$ be a vertex of $G$ that is in $\chi(v'')$ and is incident to a face in $\mathsf{prefix}_{\chi}(f)$ and a face in $\mathsf{suffix}_{\chi}(e'').$
	Notice that the faces, call them $F_{u},$ of $G$ that are incident to $u$ are also faces of $F''.$
	Therefore the set $F_{u}$ is partitioned into two sets, one consisting of faces 
	of $\mathsf{prefix}_{\chi}(f)$ and one consisting of faces in $\mathsf{suffix}_{\chi}(e'').$
	This implies the existence of a path in $G^{*}$ between any two faces in $F_{u}$ using only vertices corresponding to faces in $F'',$ which together with the connectivity of $\mathsf{prefix}_{\chi}(f)$ and $\mathsf{suffix}_{\chi}(e'')$ implies that $F''$ is connected in $G.$

	The proof of the claim follows by observing that the fact that for every vertex-face edge $e''=(v'',f'')$ of $T$ where $v''$ is a child of $f,$ $F(G)\cap (\mathsf{prefix}_{\chi}(f)\cup\mathsf{suffix}_{\chi}(e''))$ is connected in $G$ implies that $F(G)\cap (\bigcup_{e''}\mathsf{suffix}_{\chi}(e''))$ is connected in $G,$ thus proving that $F(G)\cap\mathsf{prefix}_{\chi}(v')$ is connected in $G.$\hfill$\diamond$
\medskip

We conclude the proof of the lemma by showing that $F(G)\cap \mathsf{prefix}_{\chi}(e')$ is connected in $G.$
Let $z$ be a child of $f$ in $T.$
We use $F$ to denote the set $F(G)\cap \mathsf{prefix}_{\chi}(\{t,z\}),$ $A$ to denote the set $F(G)\cap \mathsf{suffix}_{\chi}(z),$ and $B$ to denote the set $F(G)\cap \mathsf{prefix}_{\chi}(x).$
Keep in mind that $F=F(G)\setminus A$ and $B\subseteq F.$
		
Suppose towards a contradiction that $F$ is not connected in $G.$
Let $C_{1},\ldots,C_{r}, r\geq 2$ be the connected components of $G^{*}[F^{*}]$ and let $F_{1},\ldots,F_{r}$ be the sets of faces corresponding to the vertex sets of $C_{1},\ldots,C_{r},$ i.e., for every $i\in[r],$ $F_{i}\coloneqq \{f\in F(G)\mid f^{*}\in V(C_{i})\}.$
Observe that since $G$ is $2$-edge-connected, $G^{*}$ is connected.
Therefore, for every $i\in[r],$ there is an $f_{i}\in F_{i}$ and an $f_{i}'\in A$ such that $f_{i}$ and $f_{i}'$ are incident faces in $G.$
Let $e_{i}$ be an edge of $G$ that is incident to both $f_{i}$ and $f_{i}'$ and let $v_{i}$ be some endpoint of $e_{i}.$
Notice now that $v_{i}\in\chi(x)$ and therefore, by definition of the tree distance decomposition, there is a face $\tilde{f}_{i}$ of $G$ that belongs to $B$ and is incident to $v_{i}.$
The latter implies that for every $i\in[r],$ there is a $\tilde{f}_{i}\in B\cap F_{i},$ a contradiction to the assumption that $B$ is connected in $G.$
\end{proof}

We are now ready to prove~\Cref{segments} using~\Cref{granting,lem_suffix}.

\begin{proof}[Proof of~\Cref{segments}]
	Let $e=(v,f)$ be a non-root vertex-face edge of $T.$
	By \Cref{granting} and~\Cref{lem_suffix}, the sets $F(G)\cap \mathsf{prefix}_{\chi}(e)$ and $F(G)\cap\mathsf{suffix}_{\chi}(e)$ are both connected in $G.$
	Let $F_{\rm pre}\coloneqq F(G)\cap \mathsf{prefix}_{\chi}(e)$ and $F_{\rm suf}\coloneqq F(G)\cap \mathsf{suffix}_{\chi}(e).$
	Notice that $\{F_{\rm pre},F_{\rm suf}\}$ is a partition of the faces of $G.$
	Since $G$ is $2$-edge-connected,
	for every edge of $G,$ there are exactly two faces of $G$ that are incident to this edge.
	Let $Z$ be the set of all edges of $G$ that are incident to a face from $F_{\rm pre}$ and a face from $F_{\rm suf}$ and let $V_{Z}$ be the set of endpoints of the edges of $Z.$
	It is easy to see that $V_{Z}\subseteq\chi(v).$	
	Connectivity of both $F_{\rm pre}$ and $F_{\rm suf}$ in $G$ and planarity of $G$ imply that the graph $(V_{Z},Z)$ is isomorphic to a cycle.
\end{proof}

\end{document}